\documentclass[11pt,a4paper,onecolumn,aps,nofootinbib,showpacs,floatfix]{revtex4}

\usepackage[T1]{fontenc}
\usepackage[cp1250]{inputenc}

\usepackage{graphicx}
\usepackage{placeins}
\usepackage{dsfont}
\usepackage{amsmath,amssymb,graphics}

\usepackage{color}
\usepackage{bm}
\begin{document}

\title{Empirical symptoms of catastrophic bifurcation transitions \\ on financial markets: A phenomenological approach}
\author{M. Koz{\l}owska}
\email{marz.kozlowska@poczta.onet.pl}
\affiliation{Institute of Experimental Physics, Faculty of Physics \\ University of Warsaw,
Ho\.za 69, PL-00681 Warsaw, Poland}
\author{T. Gubiec}
\email{Tomasz.Gubiec@fuw.edu.pl}
\affiliation{Institute of Experimental Physics, Faculty of Physics \\ University of Warsaw,
Ho\.za 69, PL-00681 Warsaw, Poland}
\author{T. R. Werner}
\email{Tomasz.Werner@fuw.edu.pl}
\affiliation{Institute of Theoretical Physics, Faculty of Physics \\ University of Warsaw,
Ho\.za 69, PL-00681 Warsaw, Poland}
 \author{M. Denys}
\email{Mateusz.Denys@fuw.edu.pl}
\affiliation{Institute of Experimental Physics, Faculty of Physics \\ University of Warsaw,
Ho\.za 69, PL-00681 Warsaw, Poland}
\author{A. Sienkiewicz}
\email{asien@okwj.edu.pl}
\affiliation{Institute of Experimental Physics, Faculty of Physics \\ University of Warsaw,
Ho\.za 69, PL-00681 Warsaw, Poland}
\author{R. Kutner}
\email{Ryszard.Kutner@fuw.edu.pl (for_the_correspondence)}
\affiliation{Institute of Experimental Physics, Faculty of Physics \\ University of Warsaw,
Ho\.za 69, PL-00681 Warsaw, Poland}
\author{Z. Struzik}
\email{zbigniew.struzik@gmail.com}
\affiliation{The University of Tokyo, 7-3-1 Hongo, \\ Bunkyo-ku, Tokyo 113-0033, Japan \\ and \\
RIKEN Brain Science Institute, \\ 2-1 Hirosawa, Wako-shi 351-0198, Japan}






\begin{abstract}
The principal aim of this work is the evidence on empirical way that catastrophic bifurcation breakdowns or
transitions, proceeded by flickering phenomenon, are present on notoriously significant and unpredictable  financial
markets. Overall, in this work we developed various metrics associated with catastrophic bifurcation transitions, in
particular, the catastrophic slowing down (analogous to the critical slowing down). All these things were considered
on a well-defined example of financial markets of small and middle to large capitalization. The catastrophic
bifurcation transition seems to be connected with the question of whether the early-warning signals are present in
financial markets. This question continues to fascinate both the
research community and the general public. Interestingly, such early-warning signals have recently been identified
and explained to be a consequence of a \emph{catastrophic bifurcation transition} phenomenon observed in multiple
physical systems, e.g. in ecosystems, climate dynamics and in medicine (epileptic seizure and asthma attack). In the
present work we provide an analogical, positive identification of such phenomenon by examining its several different
indicators in the context of a well-defined daily bubble; this bubble was induced by the recent worldwide financial
crisis on typical financial markets of small and middle to large capitalization.
\end{abstract}


\maketitle


\section{Introduction}\label{section:introduct}

Discontinuous phase transitions in complex systems (much as in liquid-gas systems) together with critical phenomena
are topics of canonical importance in statistical thermodynamics \cite{DS0,SCN,LB,BaPo,DGM,HH,LWCBT}. During the evolution
of the complex systems we observed various catastrophic breakdowns proceeded by flickering phenomenon. This type of
evolution is an example of a generic problem how small changes can lead to dramatic consequences. Overall, in this
work we developed various metrics associated with catastrophic bifurcation transitions. All these things were
considered on a well-defined example of notoriously significant and unpredictable financial markets
\cite{MS,BMR,DiSo,AJK,MaSo,JJH,JF} (and refs. therein).

The significant contribution to explanation of mechanism of financial market evolution, in particular the
settlement of whether the early-warning signals are visible there, is a generic challenge of our work.
Notably, the problem of whether the early-warning signals in the form of a \emph{critical slowing down} phenomena
(observed in multiple physical systems \cite{LB,SCN,LWCBT}) are present on financial markets was clearly formulated in ref.
\cite{SBBB}. Nowadays, as well critical as catastrophic slowing downs (the latter systematically considered in this
work) seem to be the most refined indicators of whether a system is approaching a critical threshold or
a bifurcation catastrophic tipping point, respectively \cite{GJ,MQBR,CB,CBCKP,BCS}.

Our contribution can open possibilities for numerous applications, for instance, to forecasting, market risk
analysis and financial market management. Econophysics -- the physics of financial markets --- offers
particularly promising approaches to our challenge and therefore it constitutes a principal framework
of our work \cite{Wiki}.

The basic term of our work, i.e., the term 'complex system' defines a system composed of a large number
of interconnected entities -- degrees of freedom, often open to its environment, self-organizing its internal
structure and dynamics, expressing novel macroscopic or emergent properties. The paramount property of complex
system is its indivisibility. For this reason, reductionist approaches fail in the case of complex systems
which must be considered holistically.

Complex systems play an increasing role in majority of scientific disciplines. Notable examples can be drawn
from biology and biophysics (biological networks, ecology, evolution, origin of life, immunology, neurobiology,
molecular biology, etc), geology and geophysics (plate-tectonics, earthquakes and volcanoes, erosion and landscapes,
climate and weather changes, environment evolution, etc.). Besides their essential role in the characterization
of a range of phenomena in physics and chemistry, complex systems are now indispensable in economy (covering
econometrics and financial markets) as well as in social sciences (including cognition, opinion dynamics, distributed
learning, interacting agents, etc.).
%
%

The cross-disciplinary approach enabled the discovering of several features and stylized facts of financial market's
complexity. For instance, the approach stimulating our present work comes in part from ecology \cite{SBBB,GJ,DSDS},
where sometimes ecosystem undergoes a \emph{catastrophic regime shift} (in the sense of R\'{e}ne Thom catastrophe
theory \cite{GJ}), over relatively short period of time. This regime shift seems to be one of the most sophisticated
because of bifurcation existence. This means that the catastrophe or tipping points \cite{DSDS,AJ} exist, at which
a sudden shift of the system to a contrasting regime may occur\footnote{For instance, the sudden shifts (or jump
discontinuities) of magnetization plotted versus magnetic field was already found, at critical fields, in our early
physical work \cite{KBK}, where we studied the influence of lattice ordering on diffusion properties.}. Furthermore,
before reaching a tipping point, there is also a possibility of the system transition to such a state, where system
again continues its evolution gradually. This is a subcatastrophic bifurcation transition. Anyway, both kind of
transitions, i.e., the catastrophic and subcatastrophic ones, are discontinuous that is, they are the analog of the
first-order phase transitions \cite{DS0}.

One of the most important attainment of the catastrophe theory in the context of economics seems to be an introducing
the complexity into it. For instance, already elementary catastrophes introduced a nonlinear topological complexity
which was utilized within the different economical sectors \cite{ECZ}-\cite{JBR}.

There is a well-known controversy, concerning two-state transitions on financial markets, which is the significant
inspiration of our work. Namely,
Plerou, Gopikrishnan and Stanley \cite{PGS1,PGS2} observed a two-phase behavior on financial markets by using
empirical transactions and quotes within the intra-day data for 116 most actively traded US stocks, for the two-year
period 1994-1995. By examining the fluctuation of volume imbalance that is, by using some
conditional probability distribution of the volume imbalance, they found the change of this distribution from
uni- to bimodal one. That is, the market shifted from equilibrium to out-of-equilibrium
state, where these two different states were interpreted as distinct phases.

In contradiction, Potters and Bouchaud \cite{PB} pointed out that two-phase behavior of the above mentioned
conditional distribution is a direct consequence of generic statistical properties of traded volume and not real
two-phase phenomena.

Nevertheless, two-phase phenomenon was again examined for financial index DAX by Zheng, Qiu and Ren by using
minority games and dynamic herding models \cite{ZQR}. They found that this phenomenon is a significant characteristics
of financial dynamics, independent of volatility clustering.

Moreover, Jiang, Cai, Shou and Zhou observed the bifurcation phenomenon for Hang-Seng index \cite{JCZZ}
as non-universal one, which requires specific conditions.

However, Matia and Yamasaki in his work \cite{MY} on trading volume indicated that bifurcation phenomenon is an
artifact of the distribution of trade sizes which follows a power-law distribution with exponent belonging to the
L\'evy stable domain.

Very recently, Filimonov and Sornette \cite{FS} proved that the trend switching phenomena in financial
markets considered by Preis et al. \cite{TP,PSS,PS,PTS,SBFHMKPP,PTSE,PTSHE} has
a spurious character. They argued that this character stems from the selection of price peaks that imposes
a condition on the studied statistics of price change and of trade volumes that skew their distributions.

In the present work we consider the bifurcation phenomenon from complementary point of view because, we focus our
attention on its unconditional catastrophic properties. The main goal of our complementary work is to provide some
empirical facts for possible existing of catastrophic bifurcation transitions (CBT) of stock markets of small and
middle to large capitalization. If several effects found in this work would be considered separately one could
suppose that they have different sources. However, the fact that they appeared together indicates that their origin
is in the catastrophic bifurcation transition.

The classification of crises as bifurcation between a stable regime and a novel regime provides a first step towards
identifying signatures that could be used for prediction \cite{DS0} (and refs. therein). For instance, the problem
of the tipping points existence in financial markets is a heavily researched area. This is because the
discovery of predictability leads to its elimination according to the one of the fundamental financial market
paradigm. This paradigm states that as profit can be made (for instance, from predictability), a financial market
gradually annihilates such an arbitrage opportunity.


%
%

\begin{figure}
\begin{center}
\includegraphics[width=160mm,angle=0,clip]{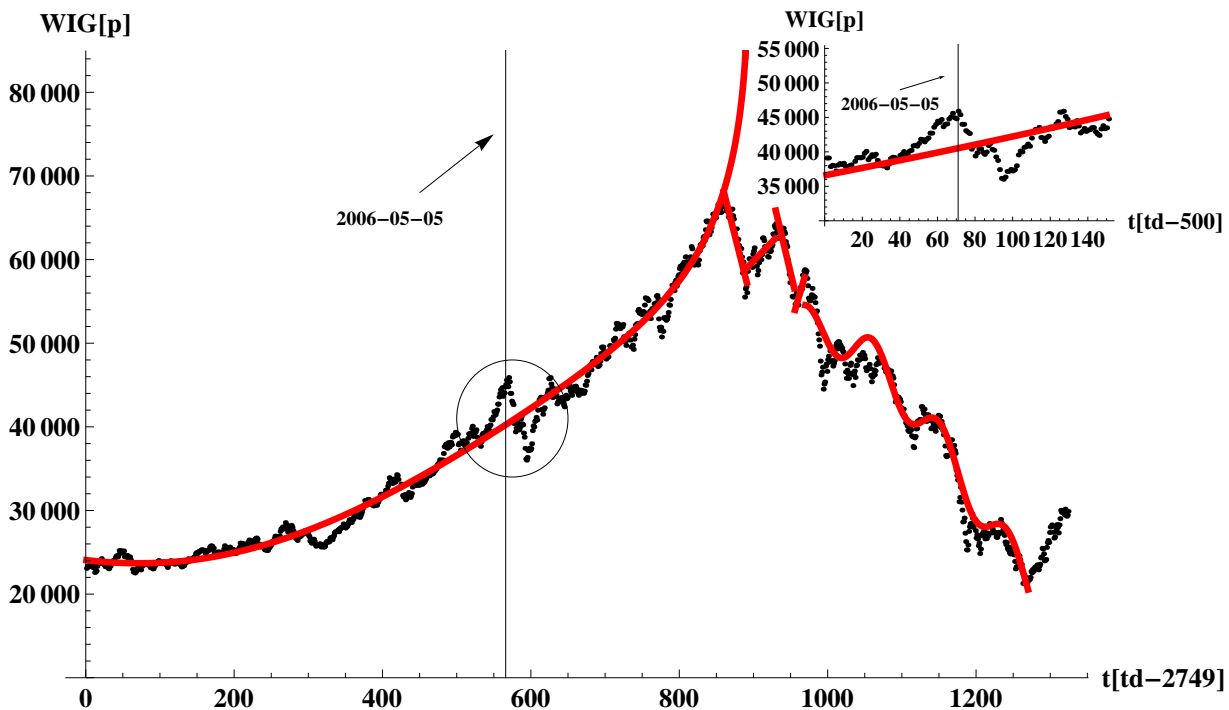}
\caption{The latest well formed peak of WIG (the erratic curve) beginning at 2004-02-06
(the $2480^{th}$ trading day (td) of the Warsaw Stock Exchange) and ended at 2009-05-18 (the $4075^{th}$ td).
The maximum of empirical data is located at 2007-07-06 (the $3340^{th}$ td) while the corresponding theoretical
maximum at 2007-08-22 (the $3372^{nd}$ td). The
solid curve is our theoretical long-term (multi-year) trend \cite{KK} fitted to the hossa (left-hand side of the
peak). The dashed vertical line denotes the position of the local peak's maximum placed at 2006-05-05. This is the
most pronounced local peak defined relatively to the trend, which was presented separately (in the upper right corner
of the figure) by the smaller plot.}
\label{figure:WIG_C_24_09_2010}
\end{center}
\end{figure}

\begin{figure}
\begin{center}
\includegraphics[width=160mm,angle=0,clip]{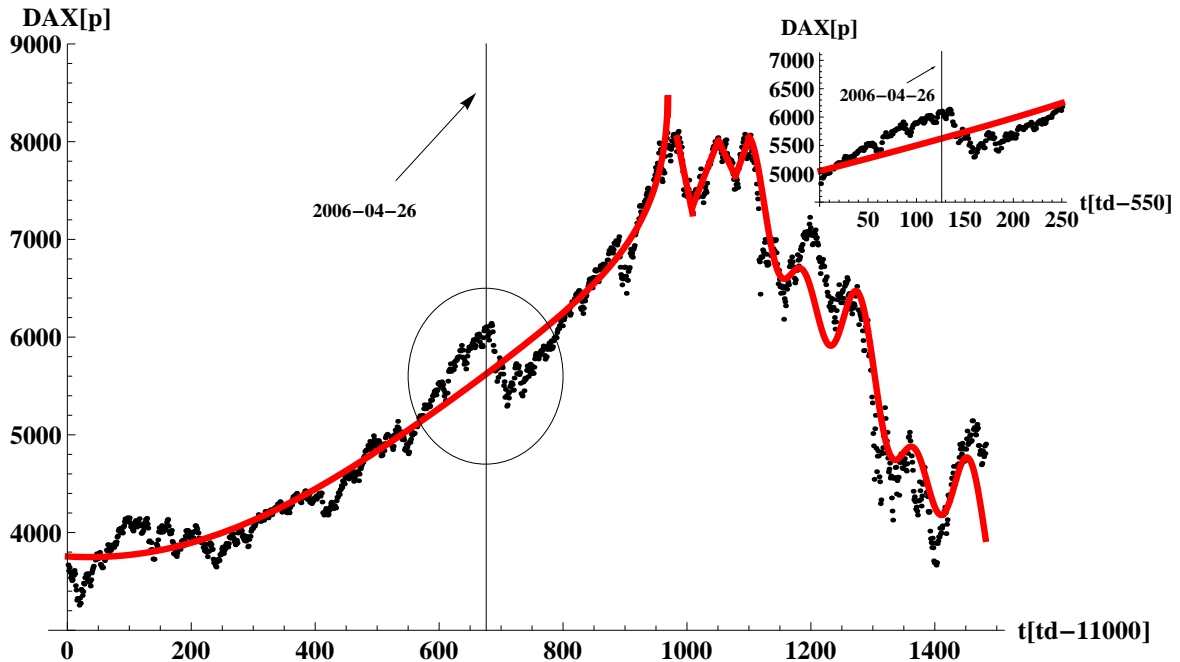}
\caption{The latest well formed peak of DAX (the erratic curve) beginning at 2003-09-04
(the $11001^{st}$ trading day (td) of the Frankfurter Wertpapierb\"orse) and ended at 2009-07-01 (the $12482^{nd}$ td).
The maximum of empirical data is located at 2007-07-13 (the $11985^{th}$ td) while the corresponding theoretical
maximum at 2007-07-17 (the $11989^{th}$ td). The
solid curve is our theoretical long-term (multi-year) trend \cite{KK} fitted to the hossa (left-hand side of the
peak). The dashed vertical line denotes the position of the local peak's maximum placed at 2006-04-26. This is the
most pronounced local peak defined relatively to the trend, which was presented separately (in the upper right corner
of the figure) by the smaller plot.}
\label{figure:DAX_rys1_NEW}
\end{center}
\end{figure}

\begin{figure}
\begin{center}
\includegraphics[width=160mm,angle=0,clip]{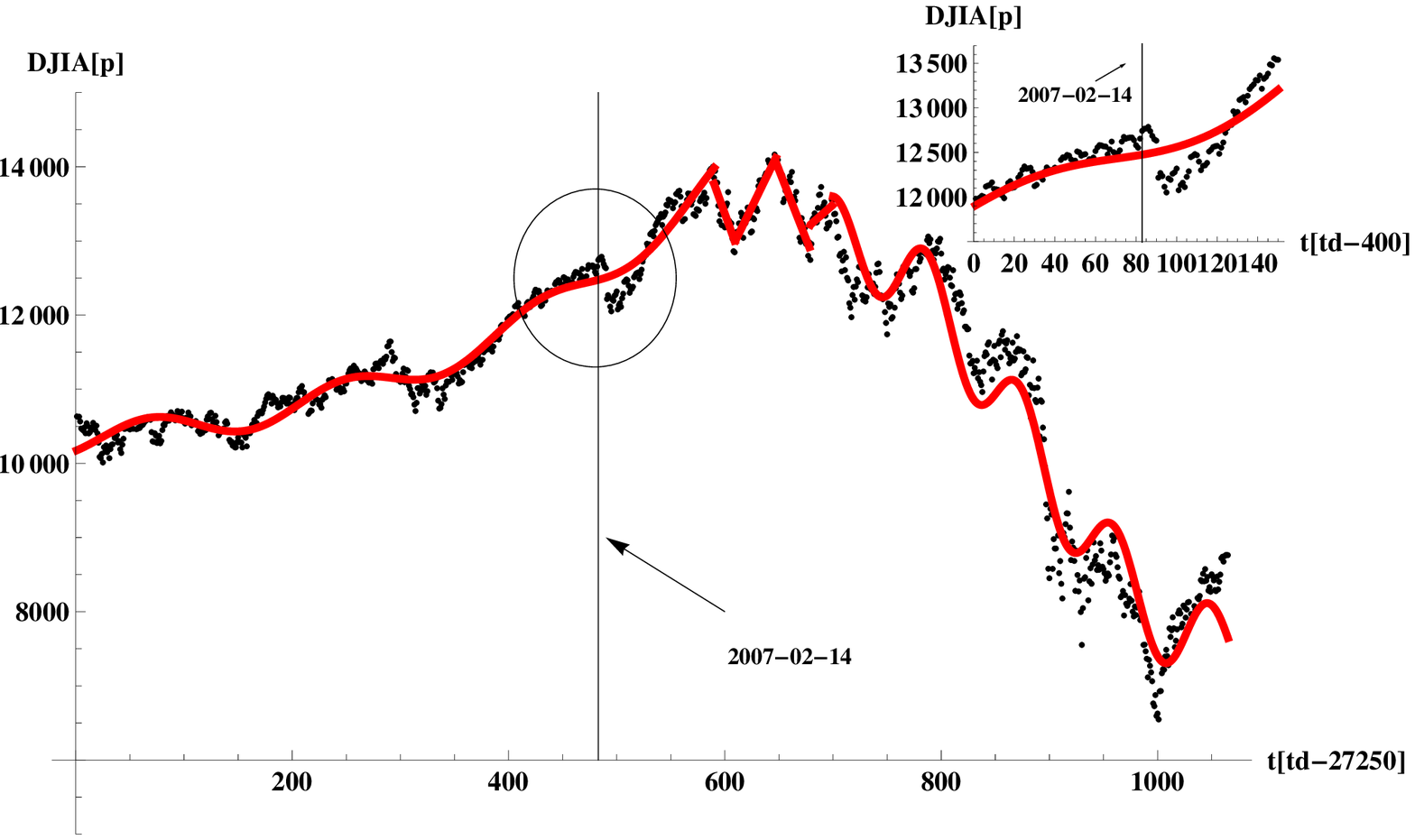}
\caption{The latest well formed peak of DJIA (the erratic curve) beginning at 2005-03-16
(the $27251^{st}$ trading day (td) of the New York Stock Exchange) and ended at 2009-06-09 (the $28315^{th}$ td).
The maximum of empirical data is located at 2007-10-09 (the $27896^{th}$ td) while the corresponding theoretical
maximum at 2007-09-12 (the $27877^{th}$ td). The
solid curve is our theoretical (long-term multi-year) trend \cite{KK} fitted to the hossa (left-hand side of the
peak). The dashed vertical line denotes the position of the local peak's maximum placed at 2007-02-14. This is the
most pronounced local peak defined relatively to the trend, which was presented separately (in the upper right corner
of the figure) by the smaller plot.}
\label{figure:DJIA_rys1_NEW}
\end{center}
\end{figure}

In this article we identify catastrophic bifurcation by verifying a list of its basic indicators, herein for
WIG\footnote{The index WIG (Warszawski Indeks Gie{\l}dowy) is the main one of the Warsaw Stock Exchange, which is of
a small size.}, DAX, and DJIA daily speculative bubbles on the Warsaw Stock Exchange, Frankfurter Wertpapierb\"orse,
and New York Stock Exchange, respectively. That is, we mainly consider stock market speculative bubbles of recent
worldwide financial crisis of small and middle to large capitalization (cf. Figures
\ref{figure:WIG_C_24_09_2010}-\ref{figure:DJIA_rys1_NEW}).


In this work we pay our attention to the analysis of daily financial market data by considering them from the
interdisciplinary point of view. We suppose that daily data is the most significant as they contain some reminiscences
both from the high- as well as from the low-frequency trading. That is, the data have an intermediate character
containing some informations both on the intra-day trading as well as on the less frequent longer-term inter-day one.
Moreover, the detrending procedure of the daily data is better established than for the intra-day data because for
the latter well known intra-day patterns exist. Both bullish and bearish sides of considered peaks are detrended
herein by using generalized exponential (or Mittag-Leffler function) decorated by some oscillations (for details
see Appendix \ref{section:MLF}) because such a function slightly better fits peaks considered in this work than
commonly used log-periodic function.

The content of the paper is as follows. In Section \ref{section:hypothesis} we consider the structural transformation
between graphs (trees), which constitutes environment for bifurcation transitions. The trees were calculated
by MST technique and found as quite different before and during the recent worldwide financial crisis. Section
\ref{section:Aedata} is devoted systematic empirical analysis of daily data originating from three typical
stock markets of small, intermediate and large capitalization. In Section \ref{section:This} we  explain how linear
and nonlinear indicators arise when system approaches the catastrophic bifurcation threshold. Section
\ref{section:conclus} contains concluding remarks.



\section{Structural phase transitions on financial markets}\label{section:hypothesis}

Barely since about two decades, physicists have intensively studied the structural (or topological) properties of
complex networks \cite{DGM} (and refs. therein). They discovered that in the most real graphs small and finite loops
are rare and insignificant. Hence, it was possible to assume their architectures as locally treelike; the property
which has been extensively exploited. It is also surprise how well this simplification works in the case of numerous
loopy and clustered networks. Therefore, we decided for the Minimal Spanning Tree (MST) technique as particularly
useful, canonical tool of a graph theory \cite{BeBo} being a correlation based network without any loop
\cite{RNM,BCLMVM,MS,BLM,VBT,KKK,TMAM,TCLMM} (and refs. therein).

We consider a simple dynamics of the empirical network. That is, the network consists of the fix number of vertices
where only distances between them can vary in time. Obviously, during the network evolution some of its edges can
disappear while others can be born. Hence, the number of edges is a non--conserved quantity continuously varying in
time likewise, for instance, their mean length and a mean occupation layer \cite{RNM,BLM,BR,TSC}.

We applied the MST technique to find the transition of complex network during its evolution from the hierarchical tree
representing the stock market structure before the recent worldwide financial crash \cite{DiSo}, to star like tree
representing the market structure already comprising the crash. Subsequently, we found the transition
from the latter tree to the hierarchical one decorated by several local star like trees (hubs) representing the market
structure after the worldwide financial crash. These two types of transitions were found, for instance, on the Warsaw
Stock Exchange (WSE), for the network of that 274 companies which are present on WSE all the duration time round \cite{SGKS}.
Analogous results we obtained for the German (Frankfurt) Stock Exchange (FSE) (herein, for the network of that
564 companies which are present on FSE all the duration time round) \cite{WSGKS,WSGKS1}.

We suppose that our results can serve as an empirical foundation for the modeling of the dynamic structural phase
transitions and critical phenomena on financial markets \cite{WH}.

\subsection{Initial results and discussion}

\begin{figure}
\begin{center}
\includegraphics[width=150mm,angle=0,clip]{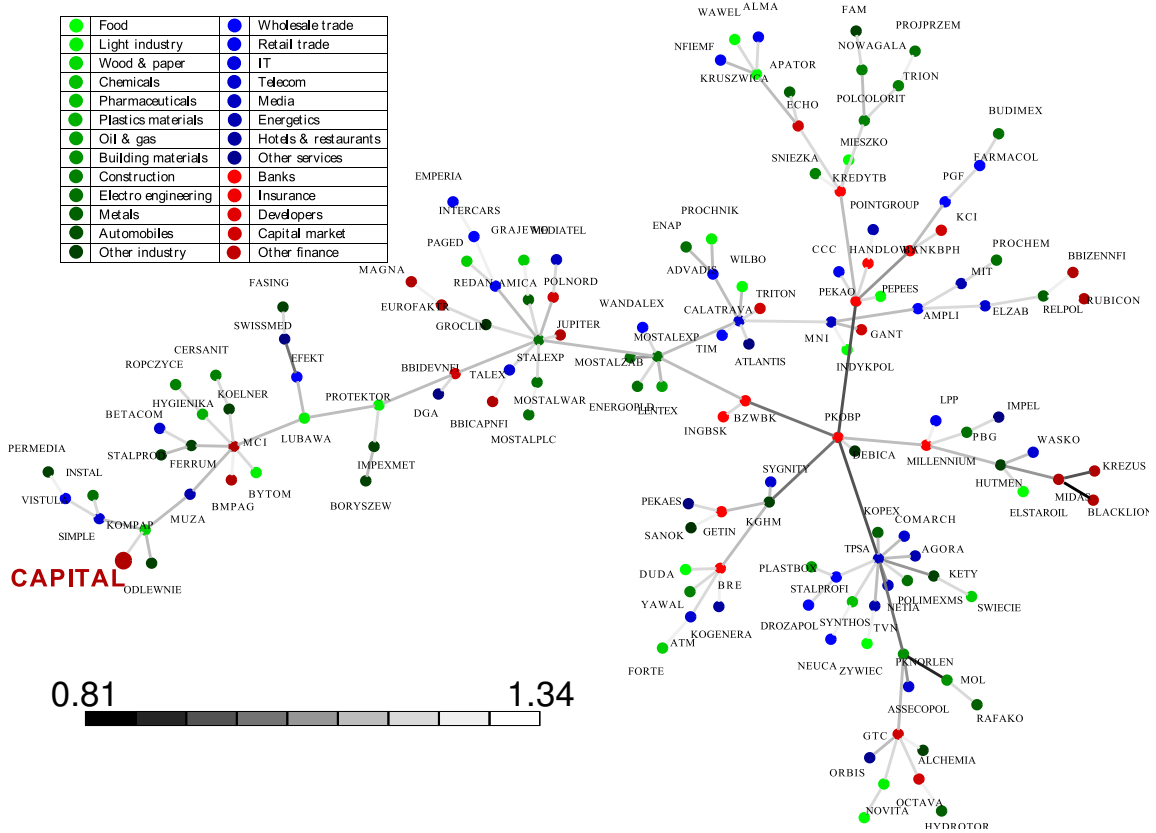}
\caption{The hierarchical Minimal Spanning Tree associated with the Warsaw Stock Exchange for the long enough
duration time from 2005-01-03 to 2006-03-09 that is, before the worldwide crash. The companies were marked by the
coloured circles (see the legend). We pay our attention for the CAPITAL financial company, which plays
a central role within MST shown in Figure \ref{figure:20080812_asien}. When the tint of grey of the link
between two companies is greater then the cross-correlation between them is also greater while the distance between
them is shorter (cf. the corresponding scale enclosed there). However, the geometric distances between companies,
shown in the Figure by the lengths of straight line segments are arbitrary otherwise, the tree would be much less
readable.}
\label{figure:20060309_asien}
\end{center}
\end{figure}

The initial state (graph or complex network) of Warsaw Stock Exchange is shown in Figure \ref{figure:20060309_asien}
in the form of the hierarchical MST\footnote{For construction of MST we used here the Prim's algorithm
\cite{PA}, which is quicker than the Kruskal's one \cite{PA,JK}, particularly for $N\gg 1$. Both algorithms are most
often used in this context.}.
This graph was calculated for sufficiently long duration time, from 2005-01-03 to 2006-03-09, where the worldwide
financial crash was still absent \cite{DiSo}.

\begin{figure}
\begin{center}
\includegraphics[width=150mm,angle=0,clip]{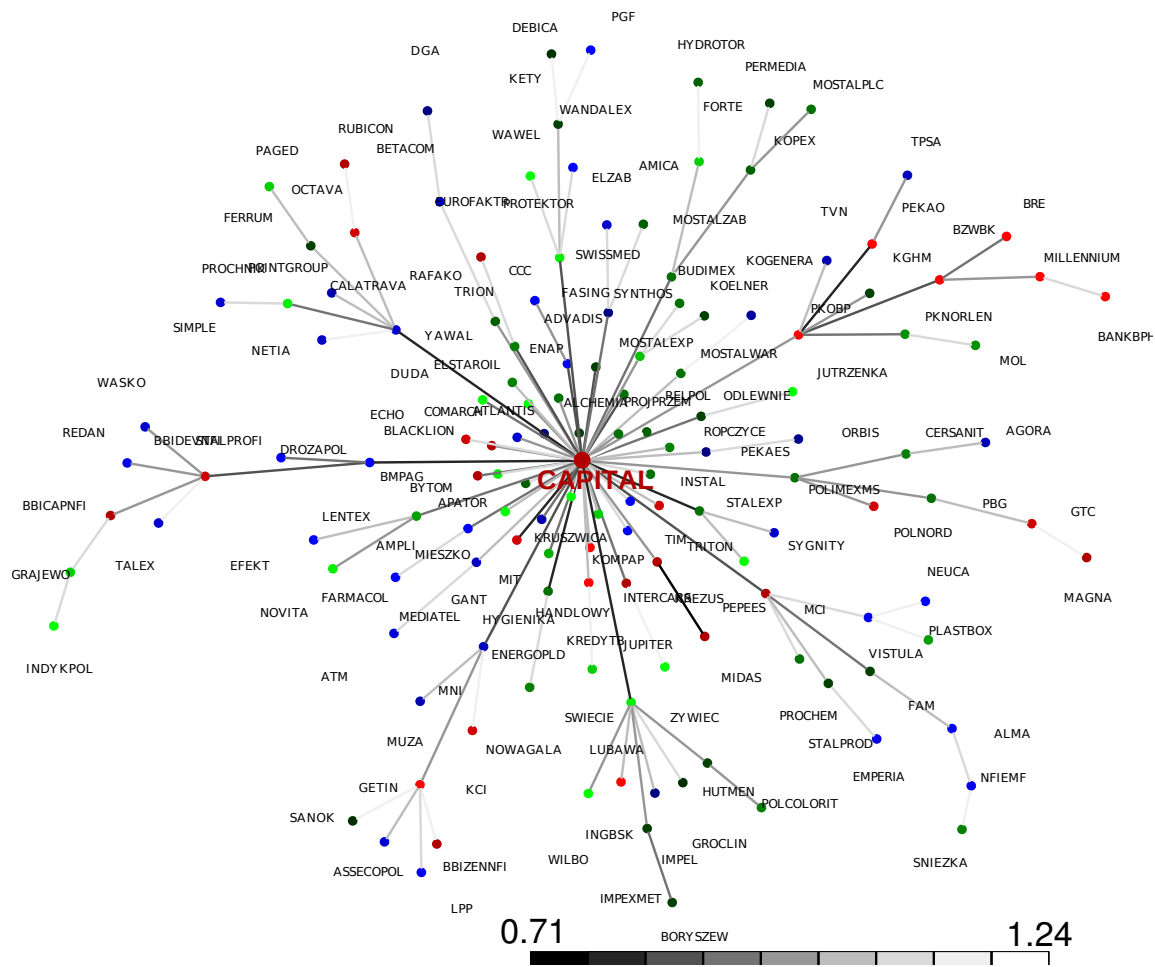}
\caption{The superstar like graph (or superhub) of the Minimal Spanning Tree for the Warsaw Stock Exchange from
2007-06-01 to 2008-08-12 which comprises the worldwide financial crash. Now, the CAPITAL company became a dominated
hub (or superhub being a giant component), i.e. the central company of the market.}
\label{figure:20080812_asien}
\end{center}
\end{figure}

We pay our attention mainly for the CAPITAL financial company\footnote{The full name of this
company is CAPITAL Partners. It is present on the Warsaw Stock Exchange from 20th October 2004. The capital investment
in various assets and investment advisement are main purposes of this company activity.}, which is a suburban company
for the most duration time round. However, it becomes a central company for the tree presented in Figure
\ref{figure:20080812_asien} that is, it is a central company for the duration time from 2007-06-01 to 2008-08-12,
which already comprises the worldwide financial crash.
In other words, for the latter duration time the CAPITAL company is represented by the vertex, which has much larger
number of edges (or it is of much larger degree) than any other vertex (or company) that is, it becomes a dominated
hub (or superhub being a giant component). Notably, the company Salzgitter AG -- Stahl und Technologie plays, on the
German (Frankfurt) Stock Exchange, the role analogous to the CAPITAL Partners one.

Indeed in the way described above, the transition between two structurally (or topologically)
different states of stock exchange is realized. That is, we observed the transition from hierarchical
tree (consisting of hierarchy of local stars or hubs) to the superstar like tree (or superhub).

\begin{figure}
\begin{center}
\includegraphics[width=160mm,angle=0,clip]{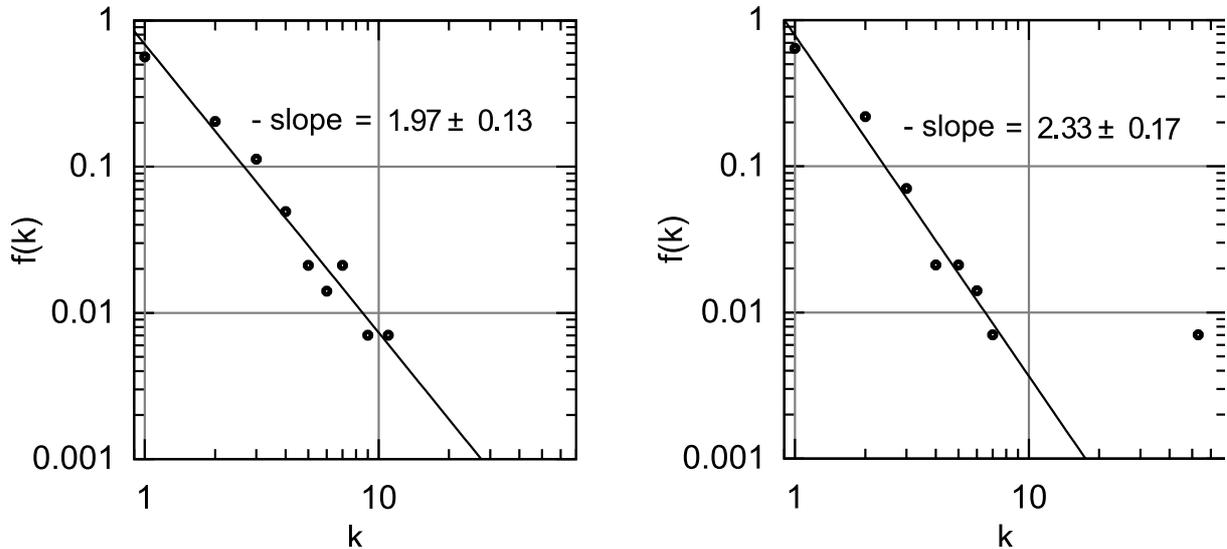}
\caption{The comparison of power-law distributions $f(k)$ vs. $k$ (where $k$ is the vertex degree) for the hierarchical
Minimal Spanning Tree shown in Figure \ref{figure:20060309_asien} and the superstar like tree shown in
Figure \ref{figure:20080812_asien}. One can observe that for the latter tree there is a single vertex herein
(rhs plot), which has degree much larger (about 50) than any other vertex. Indeed, this vertex represents
the CAPITAL company which seems to be a super-extreme event or a dragon king \cite{DSDS,WGKS,AJK,MaSo} being
a giant component \cite{DGM}.}
\label{figure:spok_nie_spok}
\end{center}
\end{figure}

Furthermore, in Figure \ref{figure:spok_nie_spok} we compared distributions of vertex degrees for both trees.
Although these distributions are power--laws, we cannot say that herein we deal with the Albert--Barab{\'a}si (AB) like
complex network with their rule of preferential linking of new vertices \cite{AB}. This is because for both our trees
the power--law exponents are distinctly smaller than 3 (indeed, exponent equals 3 characterizes the AB network), which
is a typical observation for many real complex networks \cite{DGM}. More precisely, the slope (equals $-1.97\mp 0.13$)
of the lhs plot in Figure \ref{figure:spok_nie_spok} also characterizes complex networks of e--mails \cite{EMB}.
The slope (equals $-2.33\mp 0.17$) of the rhs plot in this Figure is also characteristic for the complex network
of actors (where also superstars are present) \cite{WS,ASBS}.

Remarkable that rhs plot in Figure \ref{figure:spok_nie_spok} makes possible to consider tree presented in
Figure \ref{figure:20080812_asien} also as a hierarchical MST decorated by a dragon king. Due to our opinion, the
appearance of such a dragon king is a signature of a crash.

It seems to be an interesting project to find a proper local dynamics (perhaps nonlinear) for our network.
The more so, the single vertex (representing the CAPITAL company) is located far from the straight line (in log-log
plot) and can be considered as a temporal outstanding, super-extreme event or a dragon king \cite{DSDS,WGKS,AJK,MaSo},
which condenses the most amount of edges (or links).

\begin{figure}
\begin{center}
\includegraphics[width=160mm,angle=0,clip]{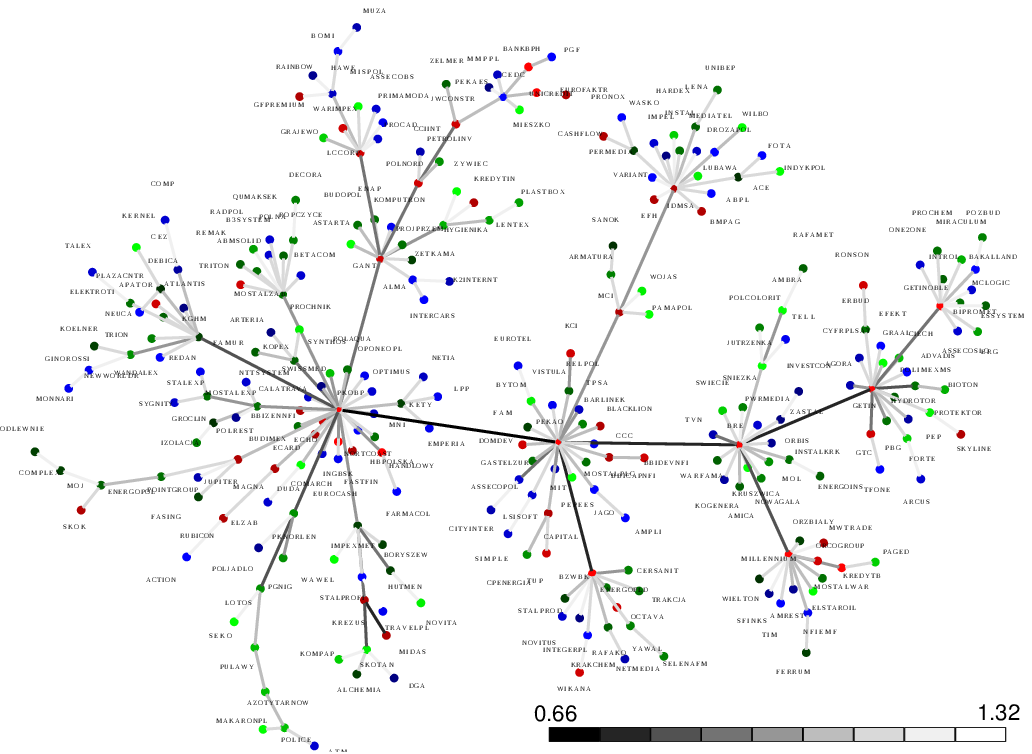}
\caption{The hierarchical graph of the Minimal Spanning Tree decorated by several local star like trees for the Warsaw
Stock Exchange for the long enough duration time from 2008-07-01
to 2011-02-28 that is, placed after the worldwide financial crisis. The companies were marked by coloured circles (see
the legend in Figure \ref{figure:20060309_asien}). Apparently, the CAPITAL financial company is no more a central hub.
If the tint of grey of the link between two companies is greater then the cross--correlation between them is also
greater while the distance between them is shorter. However, the geometric distances between companies, shown in the
Figure by the length of straight lines are arbitrary otherwise, the tree would be much less readable.}
\label{figure:20080701-20110228_New}
\end{center}
\end{figure}

For completeness, the MST was constructed for WSE for the third range of duration time that is, from 2008-07-01 to
2011-02-28, i.e. placed after the worldwide financial crash (cf. Figure \ref{figure:20080701-20110228_New}).

\begin{figure}
\begin{center}
\includegraphics[width=160mm,angle=0,clip]{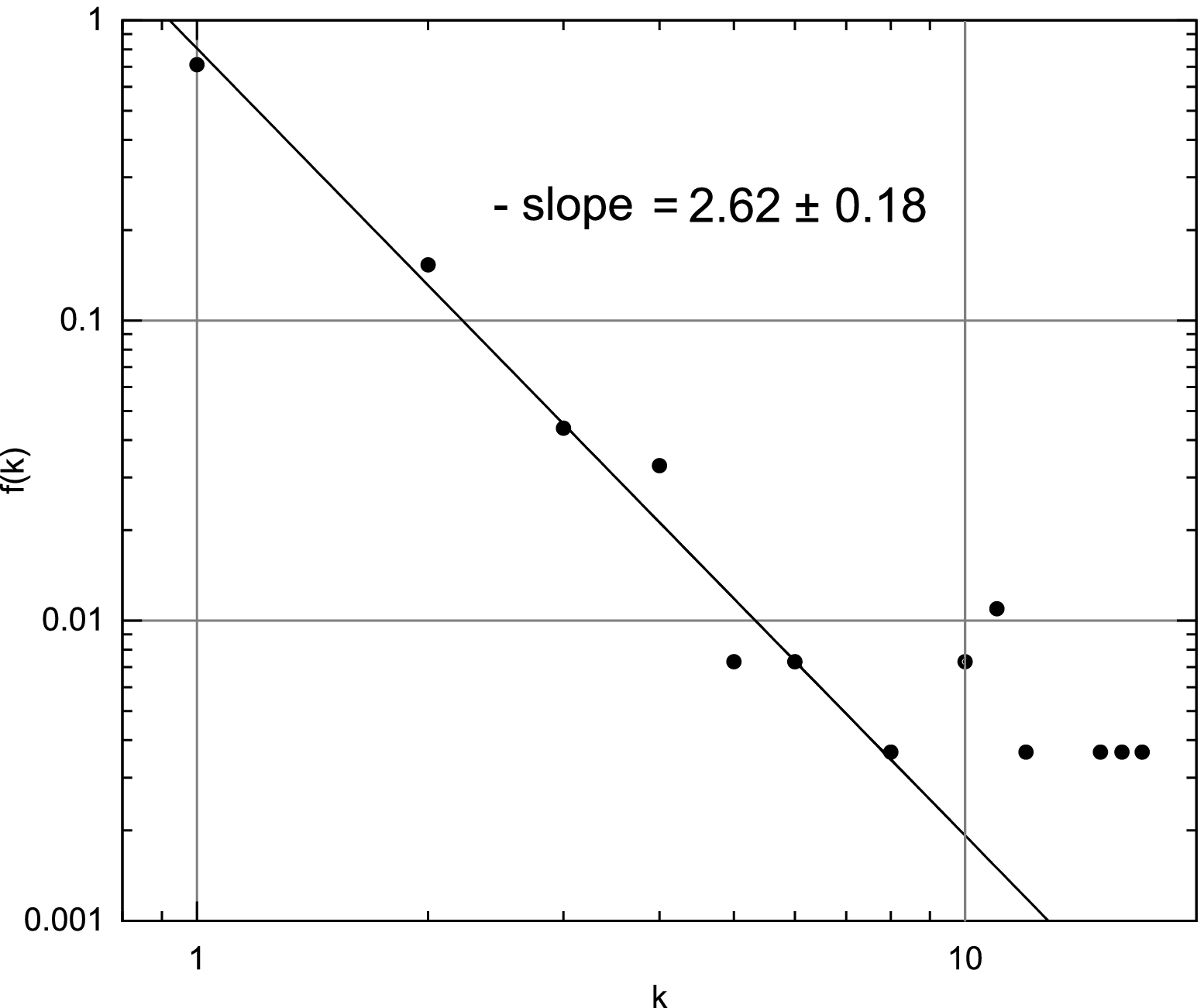}
\caption{The power--law distribution $f(k)$ vs. $k$ (where $k$ is the vertex degree) for the structured
graph of the MST shown in Figure \ref{figure:20080701-20110228_New}. Six points (associated with several different
companies) apeared above the power law. This means that several large hubs appeared instead of a single superhub.}
\label{figure:214_New}
\end{center}
\end{figure}

It is interesting that several new hubs appeared while the single superhub (superstar) disappeared (as it became
a usual hub). This means that the structure (or topology) of the network is varied during its evolution over the
market crash. This is also well confirmed by the plot in Figure \ref{figure:214_New}, where several points
(represented large hubs but not superhubs) are placed above the power law. Apparently, this power law is defined by
the slope equals $-2.62 \mp 0.18$ and also cannot be considered as AB complex network. Rather, it is analogous to the
internet, which is characterized by almost the same slope \cite{FFF,CCGJSW}.

\begin{figure}
\begin{center}
\includegraphics[width=160mm,angle=0,clip]{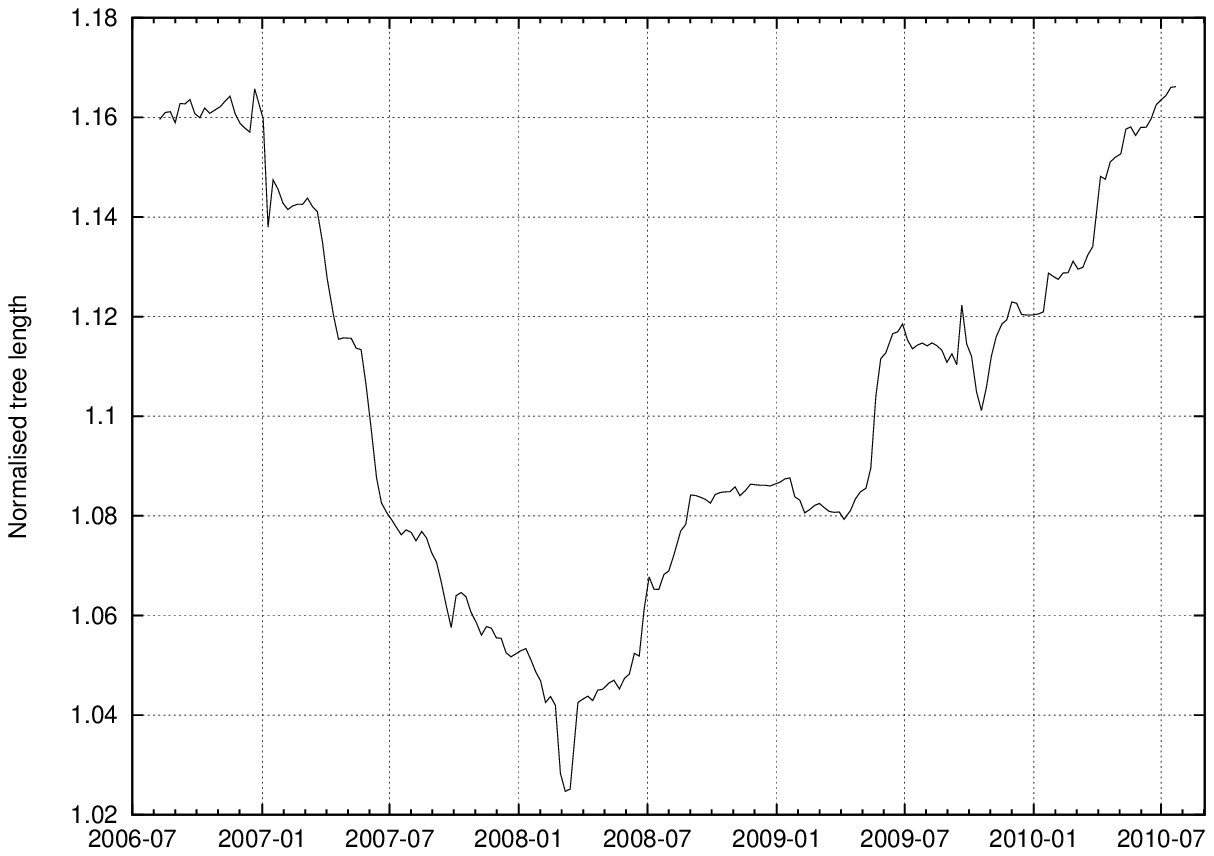}
\caption{Normalized length of the MST tree vs. time (counted in trading days, td).
Apparently, the well defined absolute minimum of the curve is located at the beginning of 2008. This localization
(inside the duration time from 2007-06-01 to 2008-08-12 comprising the crisis) together with the
corresponding length so close to 1.0 confirm the existence of a central dominated hub (or superhub).}
\label{figure:tree_length}
\end{center}
\end{figure}

\begin{figure}
\begin{center}
\includegraphics[width=160mm,angle=0,clip]{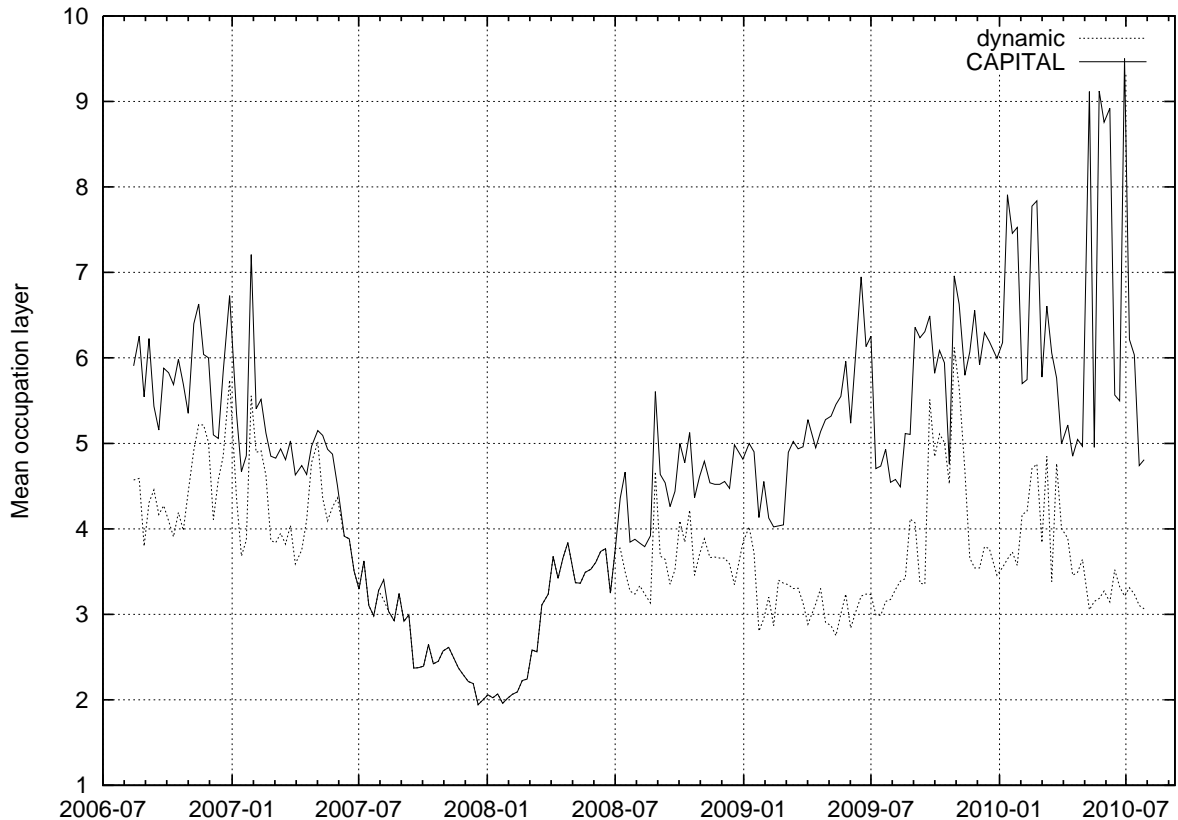}
\caption{Mean occupation layer for the MST tree vs. time (counted in trading days, td), where the CAPITAL company
was assumed as a central hub (the solid curve). For comparison, the result based on the central temporal hubs (having
temporary the largest degrees, the dashed curve) was obtained. Indeed, the latter approach is called the dynamic one.
Apparently, the well defined absolute minimum, common for both curves, is placed at the beginning of 2008
(inside the duration time from 2007-06-01 to 2008-08-12).}
\label{figure:mean_layer}
\end{center}
\end{figure}

Above given considerations were confirmed by plots shown in Figures \ref{figure:tree_length} and
\ref{figure:mean_layer}, where well defined absolute minimums at a beginning of 2008 (inside the duration time from
2007-06-01 to 2008-08-12) are clearly shown.
As usual, by the normalized length of the MST tree it is simply understood the average length of the edge (directly)
connecting two vertices\footnote{Obviously, the edge between two vertices is taken into account if any connection
between them exists.}. The normalization was chosen so to obtain length equals 1 for pure star like tree.
Apparently, this normalized length vs. time has absolute minimum close to 1 at the beginning of 2008 (cf. Figure
\ref{figure:tree_length}), while for other moments much shallower (local) minimums are observed. Furthermore, by
applying the mean
occupation layer defined, as usual, by the mean number of subsequent edges connecting a given vertex of a tree with
the central vertex (being herein the CAPITAL company), we obtained very similar result (cf. solid line in Figure
\ref{figure:mean_layer}). For comparison, the result based on the other central temporal hubs (having temporary the
largest degrees) was also obtained (cf. the dashed curve in Figure \ref{figure:mean_layer}). This approach is called
the dynamic one. Fortunately, all above used approaches give fully consistent results.

The existence of the absolute minimum (shown in Figure \ref{figure:mean_layer}) for the CAPITAL company and
simultaneously, existence of the absolute minimum shown in Figures \ref{figure:tree_length} (to good approximation)
at the same moment, confirms the existence of the star like MST structure centered indeed around the CAPITAL company
as a superhub.

\subsection{Initial conclusions}

In this Section we studied the empirical evolving correlated network associated with Warsaw Stock
Exchange\footnote{We also studied the Franfurt Stock Exchange. Because obtained results are analogous to that found
for WSE, we skipped them.}. The analytical treatment of such a network is still a challenge.

Our work supplies empirical evidence that there is the dynamic topological (structural) phase transition inside the
time range dominated by a crash. Before and after this range superhub disappeared and we observed pure hierarchical
MST and hierarchical MST decorated by several hubs (but not superhub), respectively. That is, our results consistently
confirm the existence of a dynamic structural phase transitions:
\begin{eqnarray}
\mbox{phase of hierarchical tree $\Rightarrow $ phase of star like tree $\Rightarrow $} \nonumber \\
\mbox{$\Rightarrow $ phase of hierarchical tree decorated by several local star like trees}.
\nonumber
\end{eqnarray}

One of the most significant observation contained in this work comes from plots in Figures \ref{figure:spok_nie_spok}
and \ref{figure:214_New}. Namely, exponents of all degree distributions are smaller than 3, which means that all
variances of vertex degrees diverge\footnote{It seems that the degree distribution exponent even smaller than 2 could
happened for MST obtained for the duration time from 2005-01-03 to 2006-03-09 (cf. the lhs plot in Figure
\ref{figure:20060309_asien}).}. This result indicates that herein we deal with criticality that is, the scenario
of our network evolution takes place within some scaling region \cite{DGM,DS0,BaPo,HH} containing a critical phenomena.
Although our situation is more complicated (as power laws here are either decorated by hubs (including the case of
a superhub) or governed by exponent slightly smaller than 2), we suppose that phenomenological theory of cooperative
phenomena in networks proposed by Goltsev et al. \cite{GDM} could be a promising first attempt.

Alternative view for our results could consider the superhub phase as a temporal condensate \cite{DGM}. Hence, we
can reformulate above mentioned phase transitions as representing the dynamic transition from excited phase into the
condensate and next the transition outside of the condensate again to some excited phase.

We can summarize this work by the conclusion that it seems promising to study in details above considered
phase transitions, which yet defines the basis for understanding of a stock market evolution as a whole.

\section{Systematic analysis of empirical data}\label{section:Aedata}

\subsection{Signals and detrended signals}\label{section:Sdsignal}

The systematic analysis of empirical data we perform (as we already indicated in Section \ref{section:introduct}), for
the bubbles of WIG, DAX and DJIA indices concerning the recent worldwide financial crisis (cf. Figures
\ref{figure:WIG_C_24_09_2010}-\ref{figure:DJIA_rys1_NEW}).
That is, the bubbles (hossas or booms) represent the left-hand side of the corresponding peaks which seems to be the
typical of stock exchanges of small to large capitalization. It is remarkable how awfully similar are shapes of WIG
and DAX peaks. This is a generic property of European stock market corresponding peaks.

\begin{figure}
\begin{center}
\includegraphics[width=120mm,angle=0,clip]{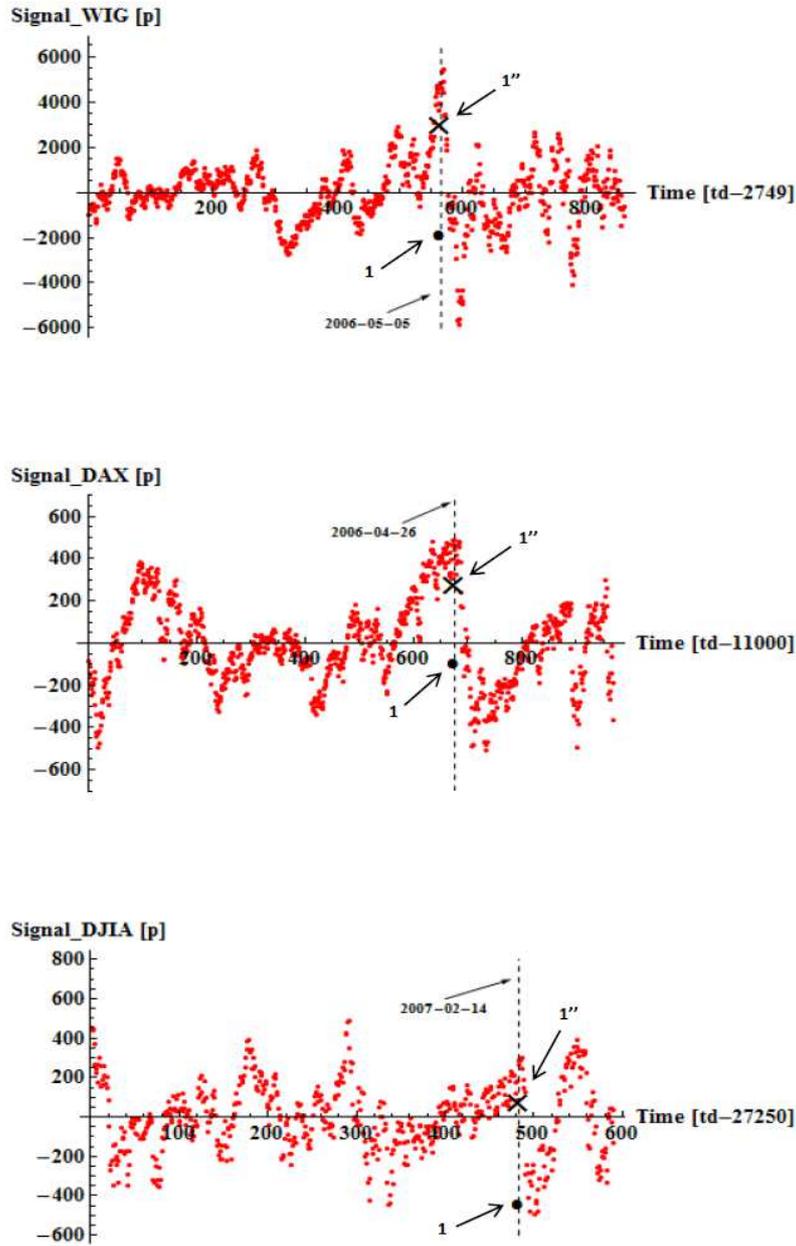}
\caption{Detrended indices (signals) WIG, DAX, and DJIA constituting the basis for further considerations. By dashed
also denoted by the vertical dashed lines in three plots in Figures \ref{figure:wariancja_MA_WIG_DAX_DJIA},
\ref{figure:WariancjaL_WDD}, \ref{figure:Autocorrel}, \ref{figure:Recovery}, \ref{figure:skewness_MA}, and
\ref{figure:Skosnosc_col}. Furthermore, the positions of spikes' variance were also denoted in Figures
\ref{figure:Noise_Lside} and \ref{figure:fixedpoint_WIG} - \ref{figure:fixedpoint_DJIA}. Points denoted by
1 (black circle) and 1'' (cross) mean mechanical equilibrium points (considered in details
in Section \ref{section:ECBT}), which define an empirical catastrophic bifurcation transition (cf. again Figures
\ref{figure:fixedpoint_WIG} - \ref{figure:fixedpoint_DJIA}).}
\label{figure:SzumL}
\end{center}
\end{figure}

The bubbles were
quite well approximated by our (deterministic) long-term (multi-year) trend (cf. Equation (1.2) in \cite{MKRK}).
By subtractions this trend from empirical time series (i.e., from the daily time series of indices) we
obtained a noisy, at least partially detrended quick-changing signal as a remainder of the bubbles
(cf. Figure \ref{figure:SzumL}).
Indeed, this signal is the main subject of our analysis particularly, the most visible downcast ones marked by dashed
vertical lines. These objects are located around $566^{th}$, $676^{th}$ and $483^{rd}$ trading days of the WIG, DAX
and DJIA bubbles, respectively. Precisely, these characteristic trading days are the positions of the variance spikes'
centers, well seen in Figure \ref{figure:wariancja_MA_WIG_DAX_DJIA}. Note that DJIA largest spike appeared only at
2007-02-14 when the \emph{subprime} market was definitely broken down that is, the latest from these three indices.

\subsection{Noise and its distributions}\label{section:noise}

\begin{figure}
\begin{center}
\includegraphics[width=150mm,angle=0,clip]{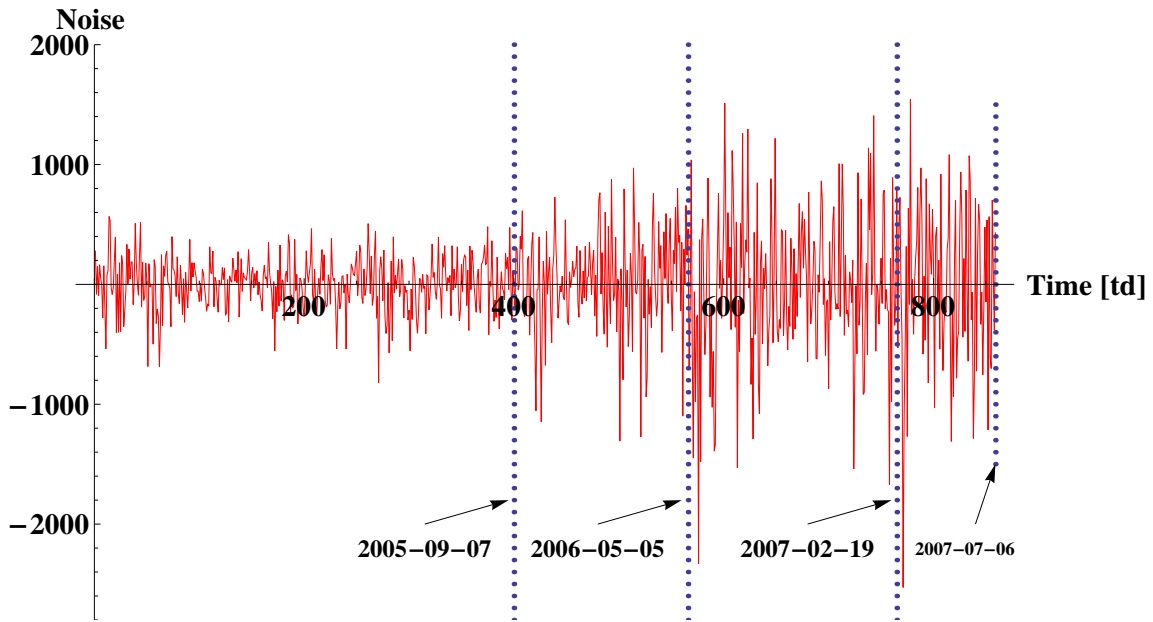}
\caption{The noise (increments) of detrended signal for WIG running from 2004-02-06 to 2007-07-06 that is, increments
of the signal shown in the upper plot in Fig. \ref{figure:SzumL}.}
\label{figure:Noise_Lside}
\end{center}
\end{figure}

For completeness, besides the signal we analyze its noise (increments), for example, for WIG (cf. Figure
\ref{figure:Noise_Lside}) from 2004-02-06 to 2007-07-06 that is, for the left side of the peak shown in Figure
\ref{figure:WIG_C_24_09_2010}.
One can observe that before 2005-09-07 the amplitude of the noise is distinctly smaller than that after 2006-05-05
(the region inbetween one can consider as an intermediate one).

\begin{figure}
\begin{center}
\includegraphics[width=150mm,angle=0,clip]{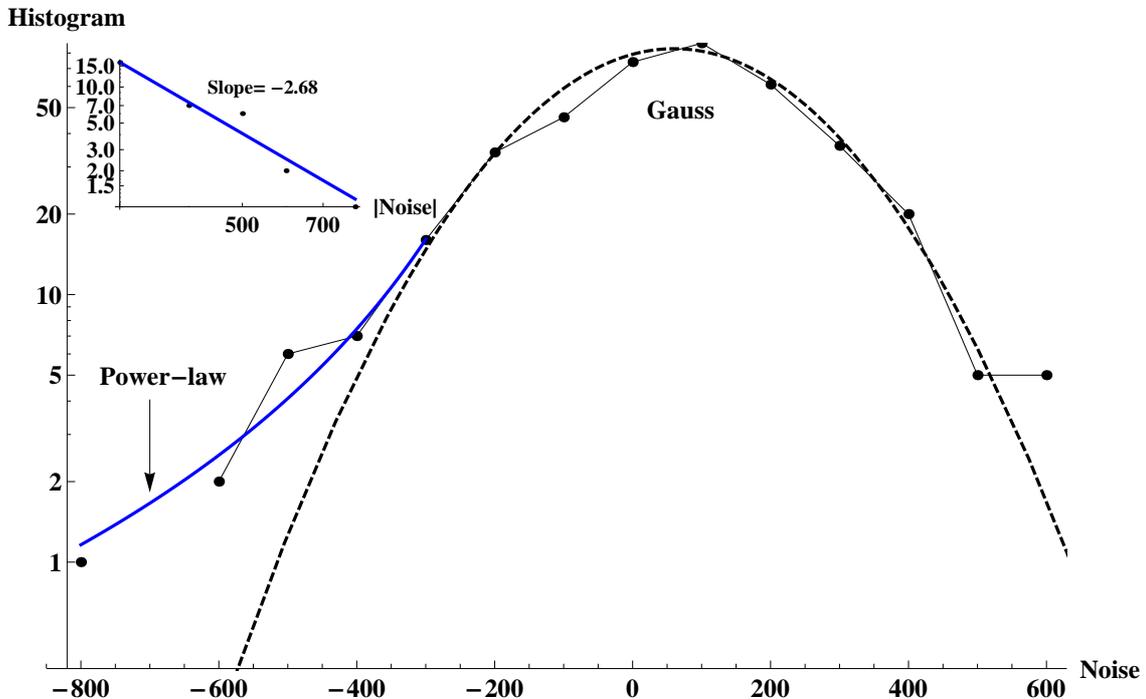}
\caption{Slightly asymmetric histogram of empirical increments consisting of the central Gaussian and poorly formed
left-sided power-law (for the negative increments). This histogram composed empirical data from 2004-02-06 to
2005-09-07. The inset presents the power-law in the log-log scale
(for convenience, the sign of the argument was changed there from the negative to positive one)}
\label{figure:Histogram_noise_WIG_do400_Lside1_2}
\end{center}
\end{figure}

For better insight into the structure of the noise the histograms of the noise were prepared for two above mentioned
regions (that is, before and after 2005-09-07). In Figure \ref{figure:Histogram_noise_WIG_do400_Lside1_2} the
semi-logarithmic plot of the histogram for the noise for the first region is presented. As it is seen, it consists
of well formed central Gaussian part and poorly formed power-law with slope equals -2.68 for relatively large
absolute values of negative increments.

\begin{figure}
\begin{center}
\includegraphics[width=150mm,angle=0,clip]{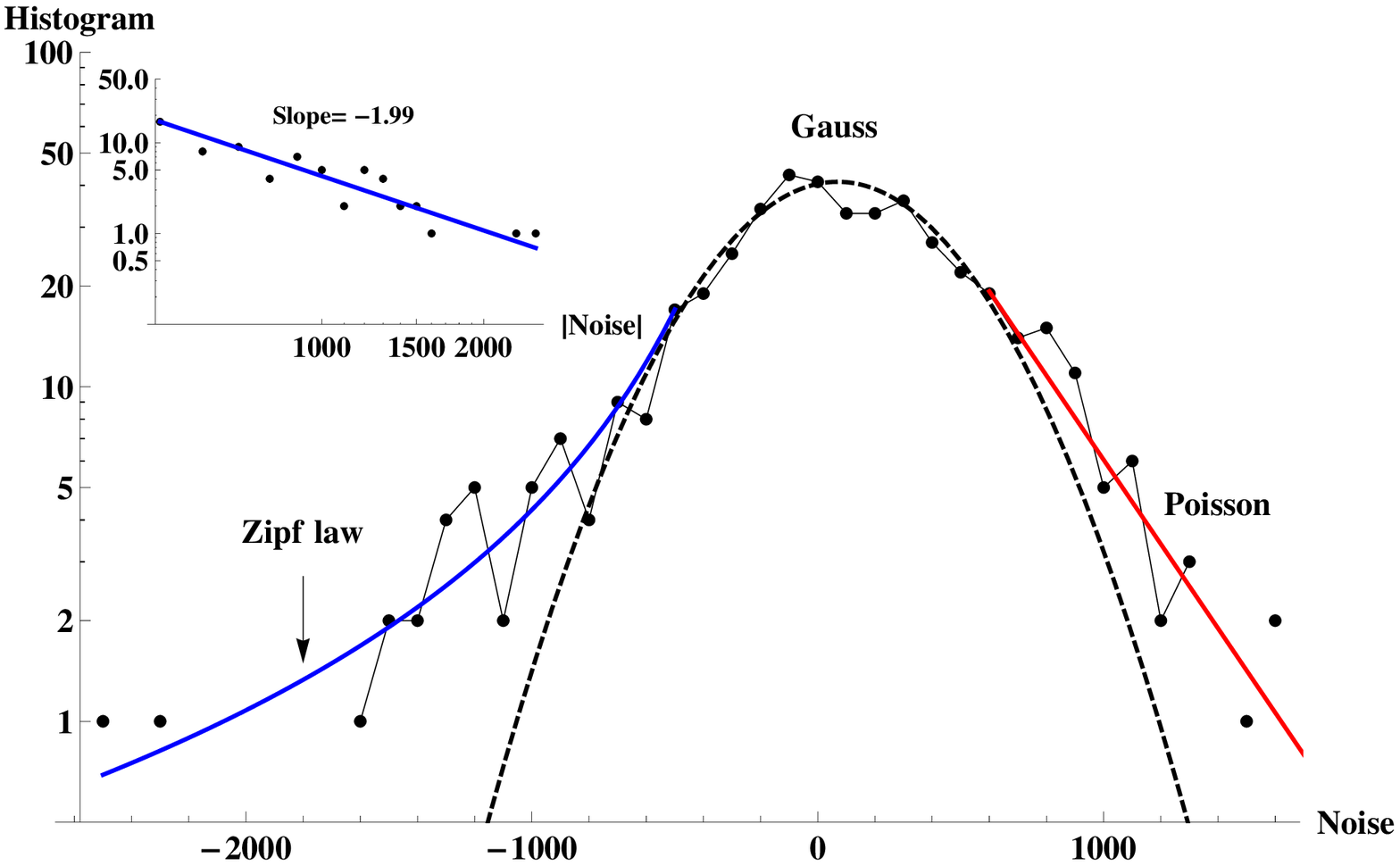}
\caption{Strongly asymmetric histogram of increments consisting of the central Gaussian, the left-sided Zipf-law
and right-sided Poisson (exponential). This histogram composed empirical data from 2005-09-08 to 2007-07-06 that is,
it also contains noise from the intermediate region. The inset presents the Zipf law in the log-log scale
(for convenience, the sign of the argument was changed there from the negative to positive one.}
\label{figure:Histogram_noise_WIG}
\end{center}
\end{figure}

In Figure \ref{figure:Histogram_noise_WIG} the analogous histogram was shown for the second region. Again, the
central part of the histogram is Gaussian and left part is better formed power-law with slope equals -1.99.
However, the right part of the histogram is well fitted by the exponential (Poisson) distribution.

\begin{figure}
\begin{center}
\includegraphics[width=150mm,angle=0,clip]{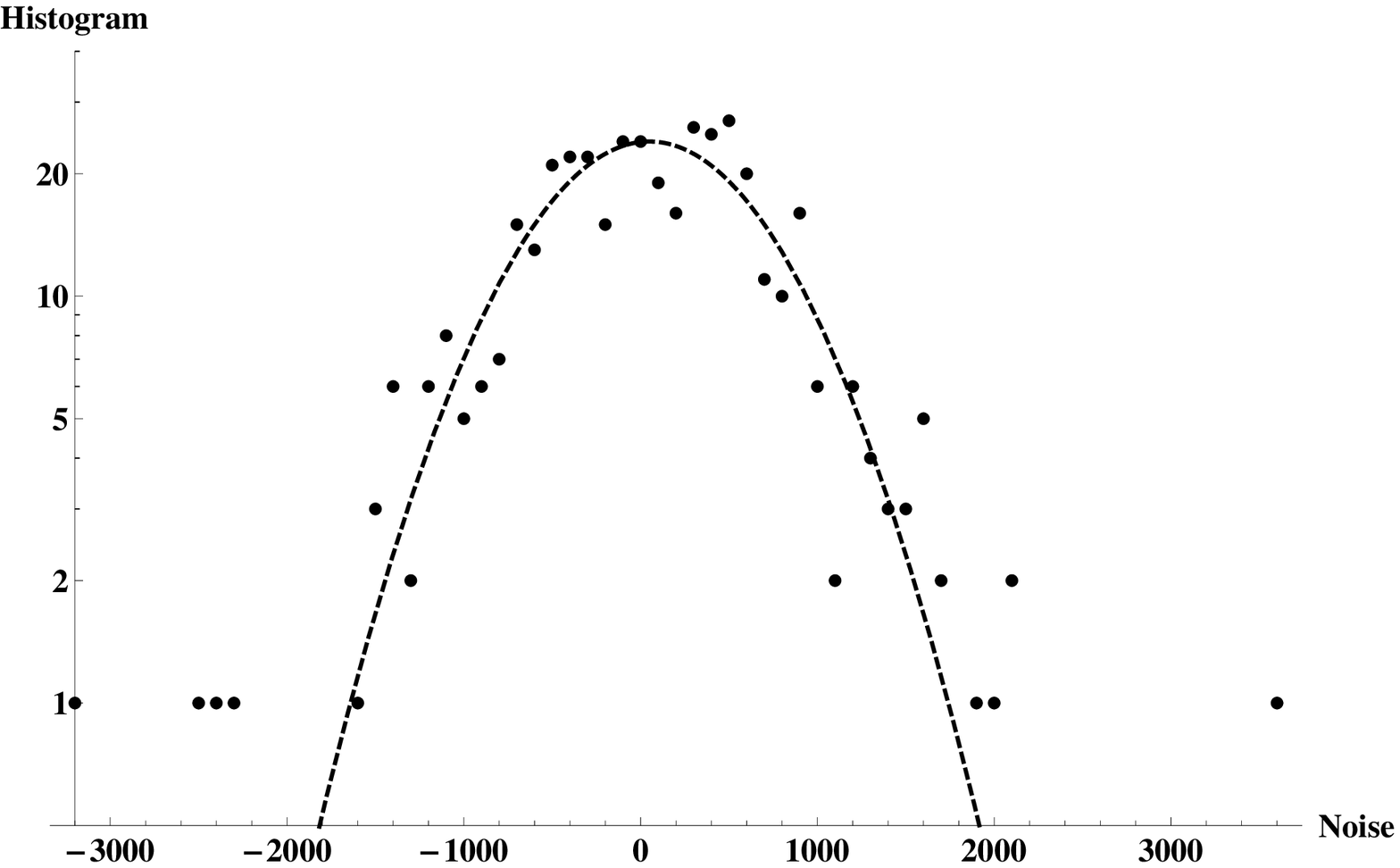}
\caption{The histogram of the noise obtained from the right side of the peak (shown in Fig.
\ref{figure:WIG_C_24_09_2010}) consisting of the Gaussian (the dashed curve) fitted to the empirical data points.}
\label{figure:Histogram_WIG_right1}
\end{center}
\end{figure}

For completeness, in Figure \ref{figure:Histogram_WIG_right1} the histogram of the noise of detrended signal
for the right part of the peak was shown. As it is seen, the Gaussian distribution satisfactorily fits the empirical
data, except few singlets outside the distribution.

\begin{figure}
\begin{center}
\includegraphics[width=150mm,angle=0,clip]{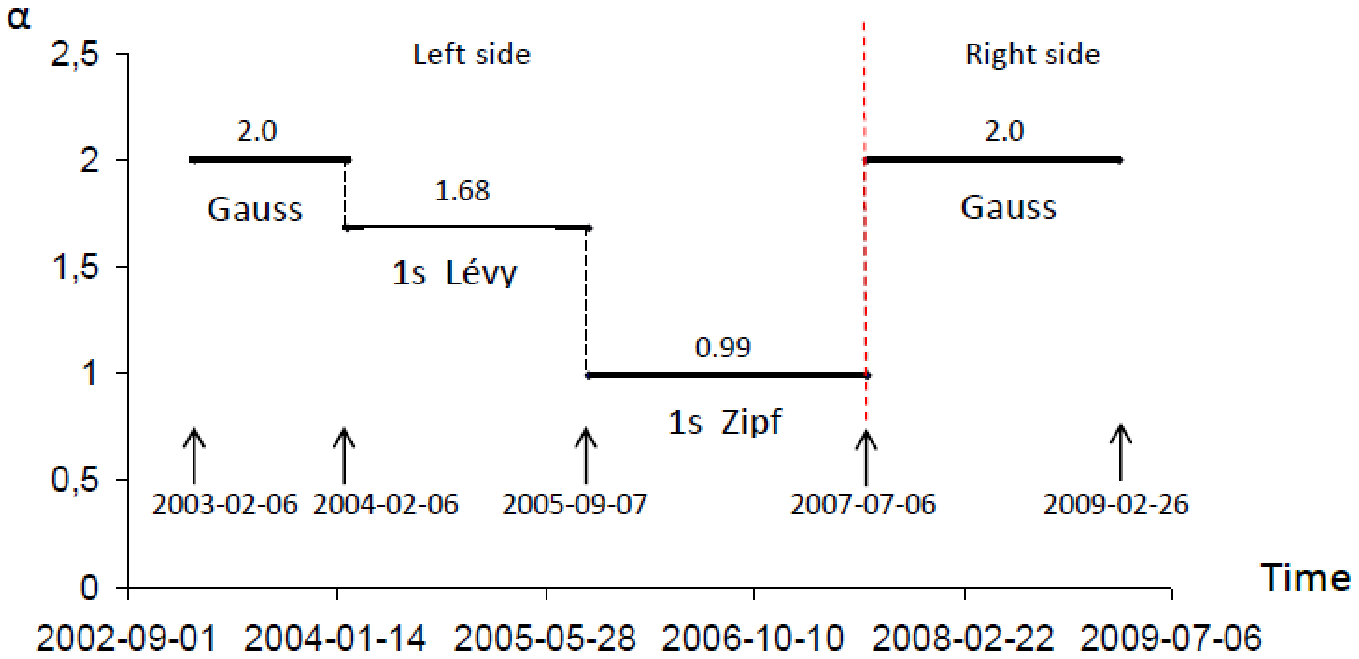}
\caption{The schematic plot of the L\'evy exponent $\alpha $ vs. time (counted in months). The discontinuous transition
of exponent $\alpha $ from the Zipf value to the Gaussian one at the threshold (marked by the dashed vertical line),
where hossa changes to bessa, is well seen.}
\label{figure:alpha_time}
\end{center}
\end{figure}

To summarize results presented in Figures
\ref{figure:Histogram_noise_WIG_do400_Lside1_2}-\ref{figure:Histogram_WIG_right1} we plotted in Figure
\ref{figure:alpha_time} the one-sided L\'evy exponent $\alpha $ vs. time (counted in months).
All values of this exponent was obtained by the corresponding fits presented in
Figures \ref{figure:Histogram_noise_WIG_do400_Lside1_2}-\ref{figure:Histogram_WIG_right1}. The jump of the exponent
$\alpha $, from the Zipf value before the threshold (marked by the dashed vertical line) to the Gauss one after
this threshold, is clearly shown.

Importantly, in Appendix \ref{section:scalrels} we proposed a time-dependent distribution of increments in the scaling
form (\ref{rown:propgatorsd}). That is, we proposed distribution obtained in our earlier work \cite{RK} (and refs.
therein) for the Continuous-time Weierstrass Flights model, where spatio-temporal coupling was used. Hence, we found
the searched distribution of increments (independently of any time interval) in the form (\ref{rown:border}), where
$\alpha =-2/\eta $ and $\eta =-2.02$ (as here $\alpha =-0.99$). Further in the text we exploit Expression
(\ref{rown:propgatorsd}) in a way fully consistent with all our empirical data.

\subsection{Usual variance of detrended signal}\label{section:uvds}

\begin{figure}
\begin{center}
\includegraphics[width=90mm,angle=0,clip]{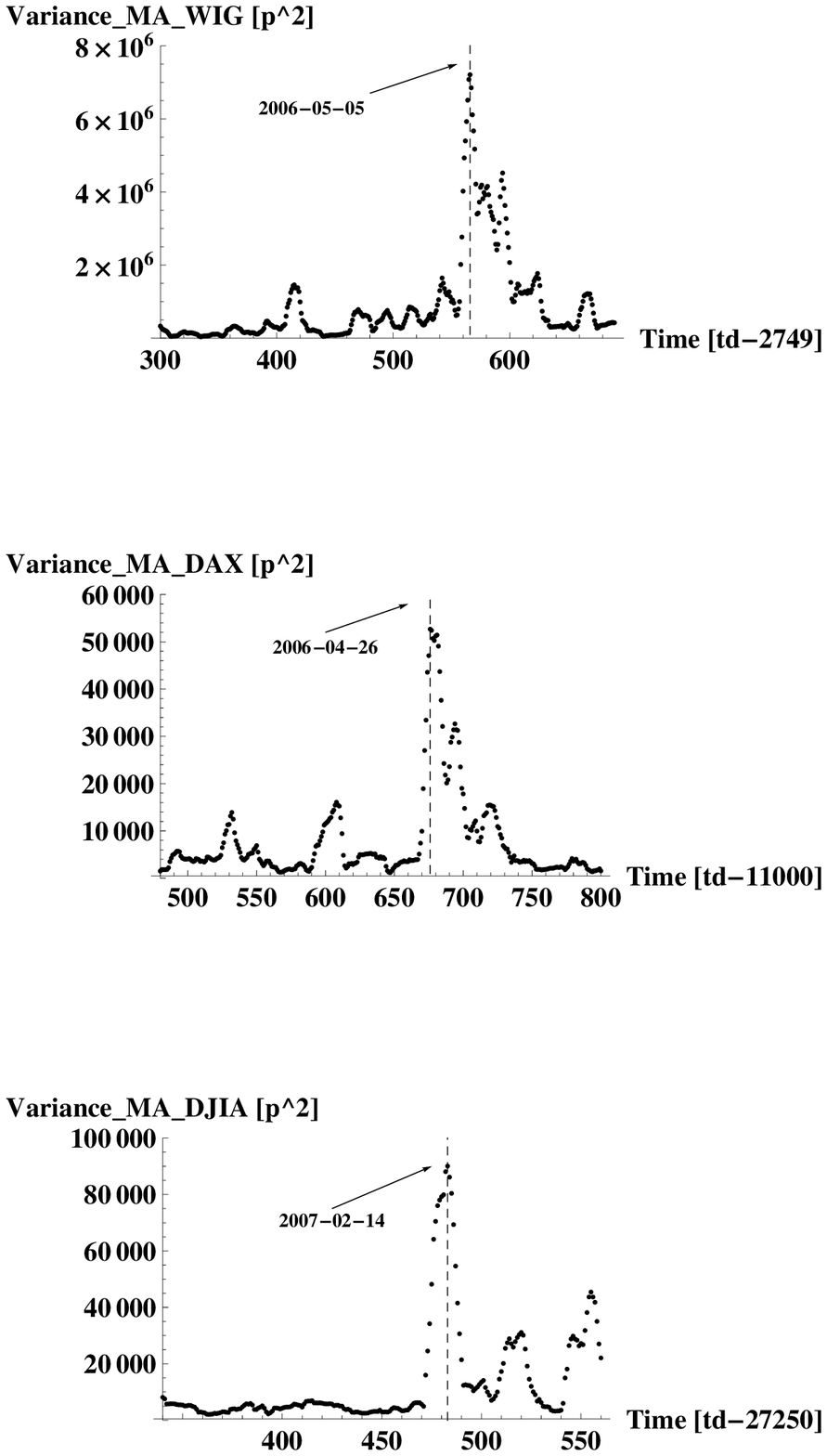}
\caption{Plots of  usual variance estimators of detrended indices (signals) WIG, DAX and DJIA (these signals were
shown in the corresponding three plots in Figure \ref{figure:SzumL}). Here, time ranges from
2005-04-15 to 2006-11-15 for WIG (upper plot), from 2005-08-18 to 2006-10-19 for DAX (middle plot), and from
2006-07-21 to 2007-05-09 for DJIA (lower plot).
The vertical dashed lines denote the positions of spikes' centres.}
\label{figure:wariancja_MA_WIG_DAX_DJIA}
\end{center}
\end{figure}

For our three different signals the time dependence of sufficiently sensitive estimators of the usual variance,
defined within the moving (or scanning) time window of a one month width (or twenty trading days\footnote{Twenty
trading days is considered as one trading month keeping Central Bank (riskless) reference return fixed.}), is shown
in Figure \ref{figure:wariancja_MA_WIG_DAX_DJIA}. That is, we obtained these estimators by the corresponding
separate scannings of empirical time series. These scannings were made by using (mentioned above) time window of
the fixed width and also fixed scanning time step (again of one trading month). Indeed, within this window the
variance estimator was calculated for each temporal position of the time window.

Notably, the variance estimators of signals suddenly strongly increase in the range of downcasts (marked by circles
in Figures \ref{figure:WIG_C_24_09_2010} - \ref{figure:DJIA_rys1_NEW}), creating local peaks of these estimators in
the form of spikes (cf. three plots in Figure \ref{figure:wariancja_MA_WIG_DAX_DJIA}). The centers of these spikes are
indicated in the plots by vertical dashed lines. The existence of the spike is the one of the significant empirical
symptom of catastrophic or even critical slowing downs. Further in the text we call these spikes the
\emph{catastrophic spikes}.

Apparently (cf. Figure \ref{figure:wariancja_MA_WIG_DAX_DJIA}), the catastrophic spikes are preceded by well formed
local peaks of variance estimators having much smaller amplitude. This behavior clearly manifests
a \emph{flickering phenomenon} \cite{SBBB}. This can happen, for instance, if the system enters the intermediate
bistable (bifurcation) region placed between two tipping points. Then the system is stochastically moving down and
up either between the basins of attraction of two alternative attractors or between attractor and repeller; both
possibilities are defined by stable/stable or stable/unstable pairs of equilibrium states. Such
a behavior can be also considered as an early-warning indicator. Moreover, the flickering of the variance estimator
(although less intensive) together with intermittencies shrinking in time, are observed even for earlier time
intervals (cf. the upper plot in Figure \ref{figure:wariancja_MA_WIG_DAX_DJIA}). This flickering phenomenon is
considered in details in Section \ref{section:AR}.

\subsection{Accumulative variance of detrended signal}

\begin{figure}
\begin{center}
\includegraphics[width=97mm,angle=0,clip]{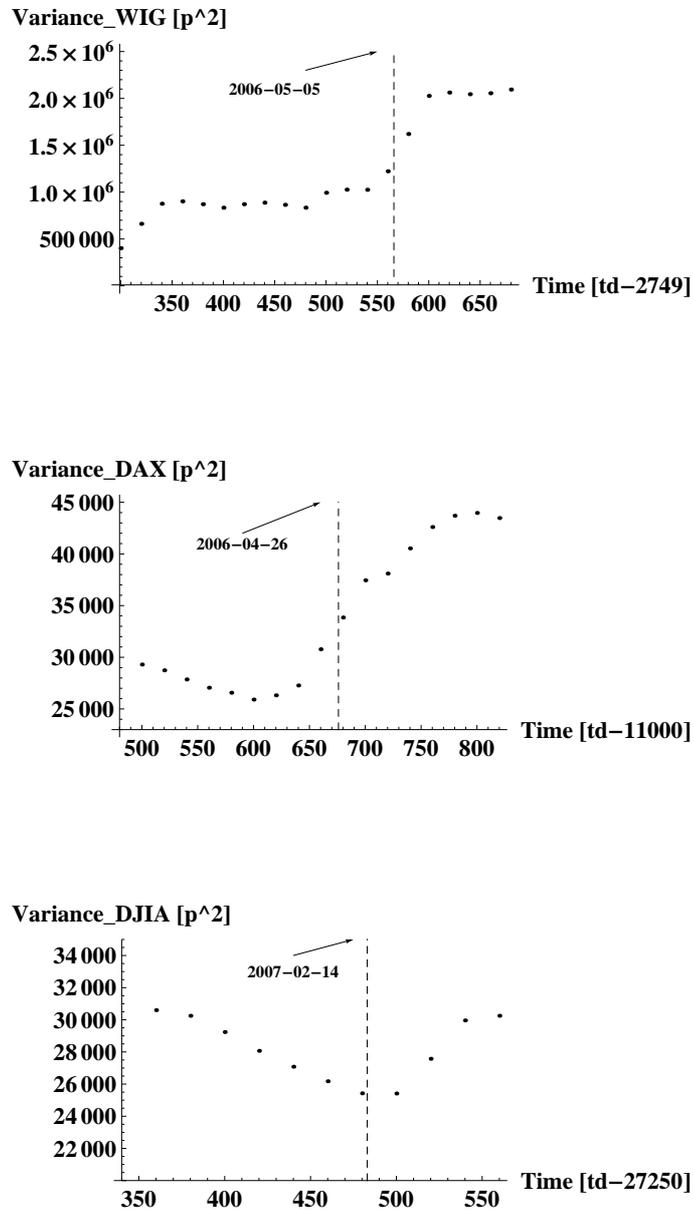}
\caption{The bimodal accumulative variance estimators of detrended indices (signals) WIG, DAX and DJIA concerning
the same time ranges as shown in plots in Figure \ref{figure:wariancja_MA_WIG_DAX_DJIA}. The vertical dashed lines
denote the position of the catastrophic spikes or the place of the greatest amplitude of the noise (cf.
Figure \ref{figure:SzumL}).}
\label{figure:WariancjaL_WDD}
\end{center}
\end{figure}

In Figure \ref{figure:WariancjaL_WDD} the accumulative variance (ACV) estimators of our three detrended signals
(WIG, DAX, and DJIA) were shown. These variance estimators were presented for the same time ranges as those used in
Figure \ref{figure:wariancja_MA_WIG_DAX_DJIA}. They are much more stable quantities than the
usual variance estimators and less sensitive to details of time series, which is a feature useful
for study both catastrophic and critical properties.

The ACV estimators show herein that our systems are bimodal ones in the vicinity of the catastrophic spikes. That is,
the state characterized by the lower value of the accumulative variance estimator is well separated from the state
of the higher value of this estimator. That is, ACV estimator is (almost) constant directly before and after as time
$t$ passes the relatively narrow intermediate bistable region where ACV estimator rapidly rises. For instance, for WIG
this region extends from about $t_{down}=520$ to $t_{up}=590$ trading days. Indeed, for time earlier than $t=t_{down}$
the system occupies the lower attractor while above $t_{down}$ the ACV estimator sharply increases to the upper
attractor. This increase of ACV estimator provides an advance warning that the system approaches a regime shift, being
a substantial reorganization of a financial market toward a herden collective or coherent phenomena.

It was shown recently \cite{BCS}, that the variance estimator increasing near the threshold should be expected in
wide range of social and ecological systems with multiple attractors. In this work we empirically prove that also
financial markets can obey this feature.

In the subsequent stages we are interested in the time-dependence of the coefficient $AR(1)$ and the lag-1
autocorrelation function, $ACF(1)$, mainly within the ranges defined by the catastrophic spikes.

\subsection{Recovery rate}\label{section:AR}

\begin{figure}[htb]
\begin{center}
\includegraphics[width=150mm,angle=0,clip]{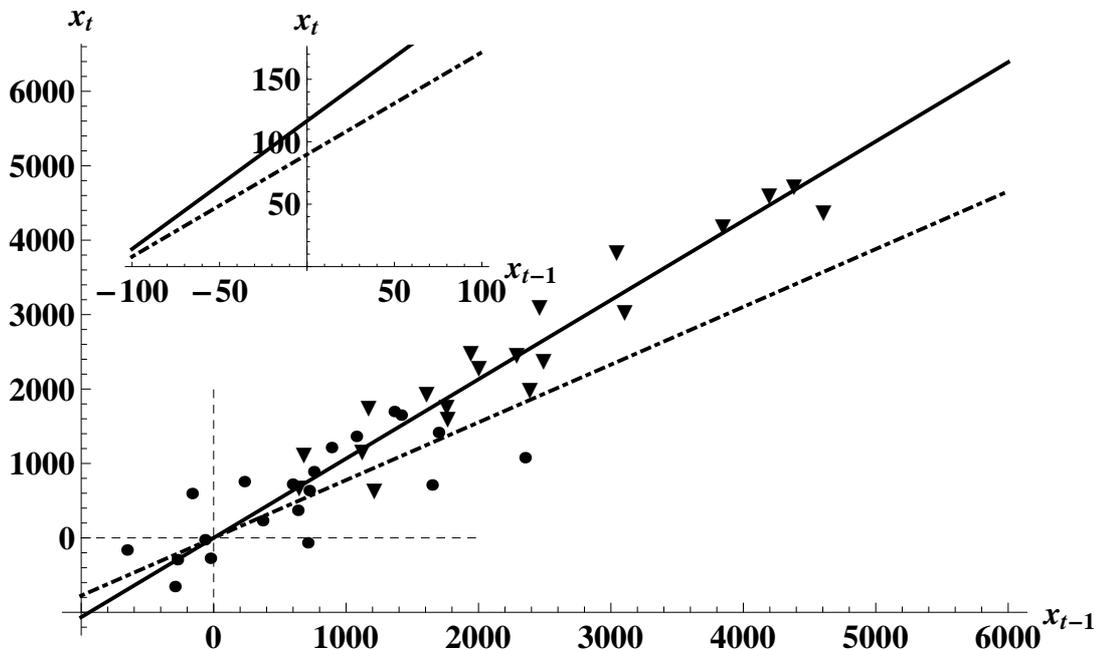}
\caption{Plot, for instance, of the WIG detrended successive signals $x_t$ vs.
$x_{t-1}$ for twenty pairs (or for one month) ranging from $t-1=521$ to $540$ (black circles and fitted dotted-dashed
straight line) as well as from $t-1=541$ to $560$ (inverted triangles and fitted solid straight line) time
steps. Apparently, the slopes of fitted curves almost equals $0.6$ and $1.0$, respectively. These results give
$-\lambda \approx 0.4$ and $-\lambda \approx 0.0$, respectively (see also plot
in Figure \ref{figure:Recovery}). Furthermore, different values of the shift coefficient $b$, although relatively
small, are well distinguishable in the inset plot.}
\label{figure:Szum_sz_541_560}
\end{center}
\end{figure}

In Figure \ref{figure:Szum_sz_541_560} was plotted, as a typical behavior, the WIG current signal $x_{t}$ against
the preceding signal $x_{t-1}$.
Two plots of significantly different sets of empirical data were shown there. Each set consists
of twenty successive pairs of detrended signal (process) $(x_{t-1},x_t)$ extended from $t=522$ to $t=541$ trading
days (full circles) and from $t=542$ to $t=561$ trading days (inverted triangles), respectively. The slopes of
straight lines fitted separately to both data sets give two different values of $AR(1)$ coefficient. That is, these
slopes give values of coefficient $AR(1)=1+\lambda $, where $\lambda $ is a derivative of the nonlinear drift term,
$f(x_t;P)$ (where $P$ is a driving or control parameter), over the signal variable, $x_t$, present in the Langevin
Equation (\ref{rown:fodeqex}). This equation is discretized and linearized in the vicinity of the catastrophic
bifurcation point (cf. Section \ref{section:dvequ}). Furthermore, both different values of the shift coefficient $b$,
although relatively small, are well distinguishable in the inset plot.

\begin{figure}[htb]
\begin{center}
\includegraphics[width=86mm,angle=0,clip]{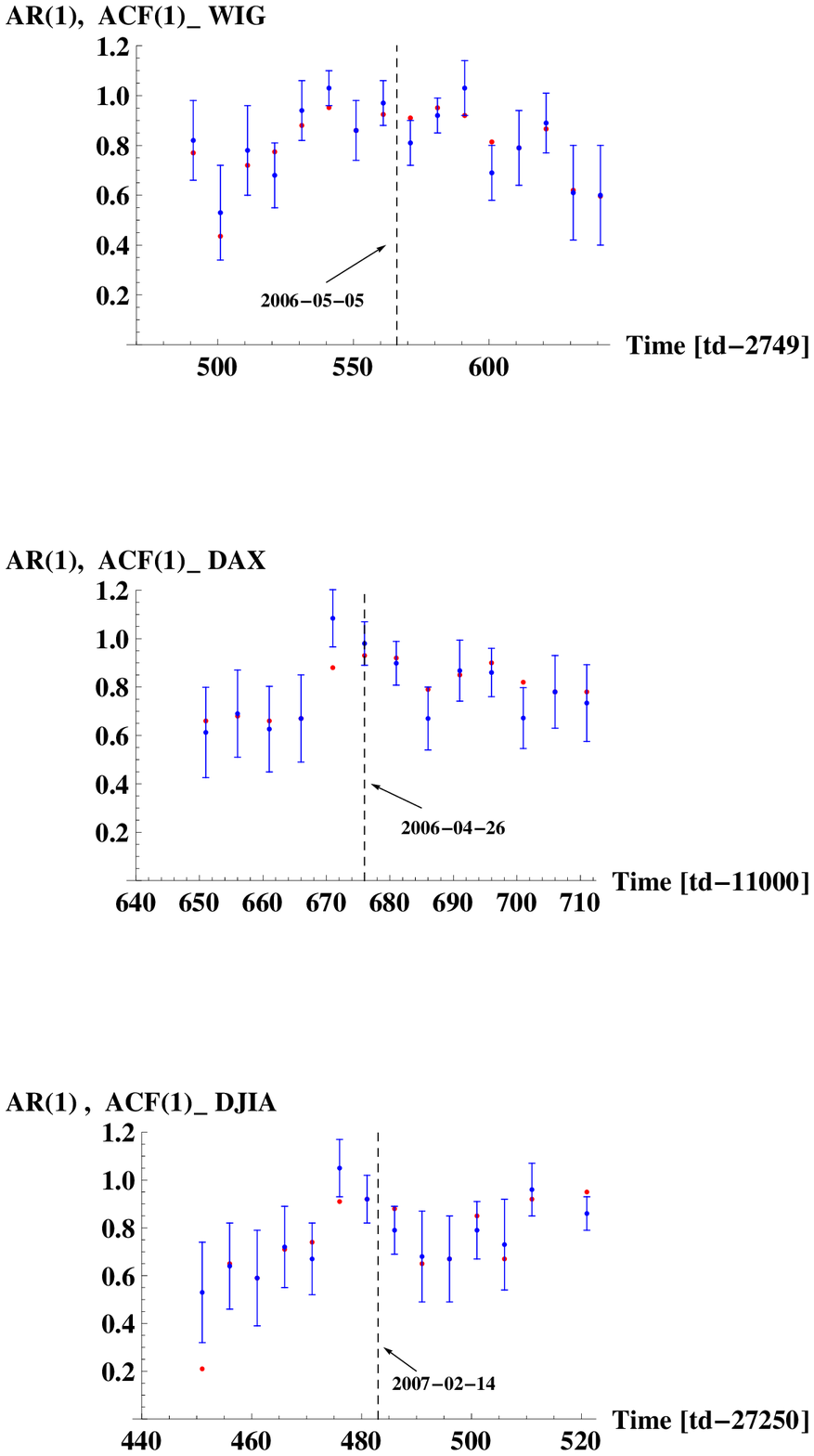}
\caption{Coefficient AR(1) (blue circles with bars) together with lag-1 autocorrelation function (ACF(1), red
circles) vs. time. These quantities were calculated for 16 successive time intervals each consisting of 20 points
and ranging from time interval $[481,500]$ to $[641,660]$ one. These intervals cover the catastrophic bifurcation
threshold. As it is seen, both quantities have similar dependence on time (they are concave) indicating that their
local maximum is broaded and placed somewhere inside the time interval $[541,600]$ (or within one quarter). The
vertical dashed line denotes the position of the varianes spikes' centres.}
\label{figure:Autocorrel}
\end{center}
\end{figure}

\begin{figure}[htb]
\begin{center}
\includegraphics[width=90mm,angle=0,clip]{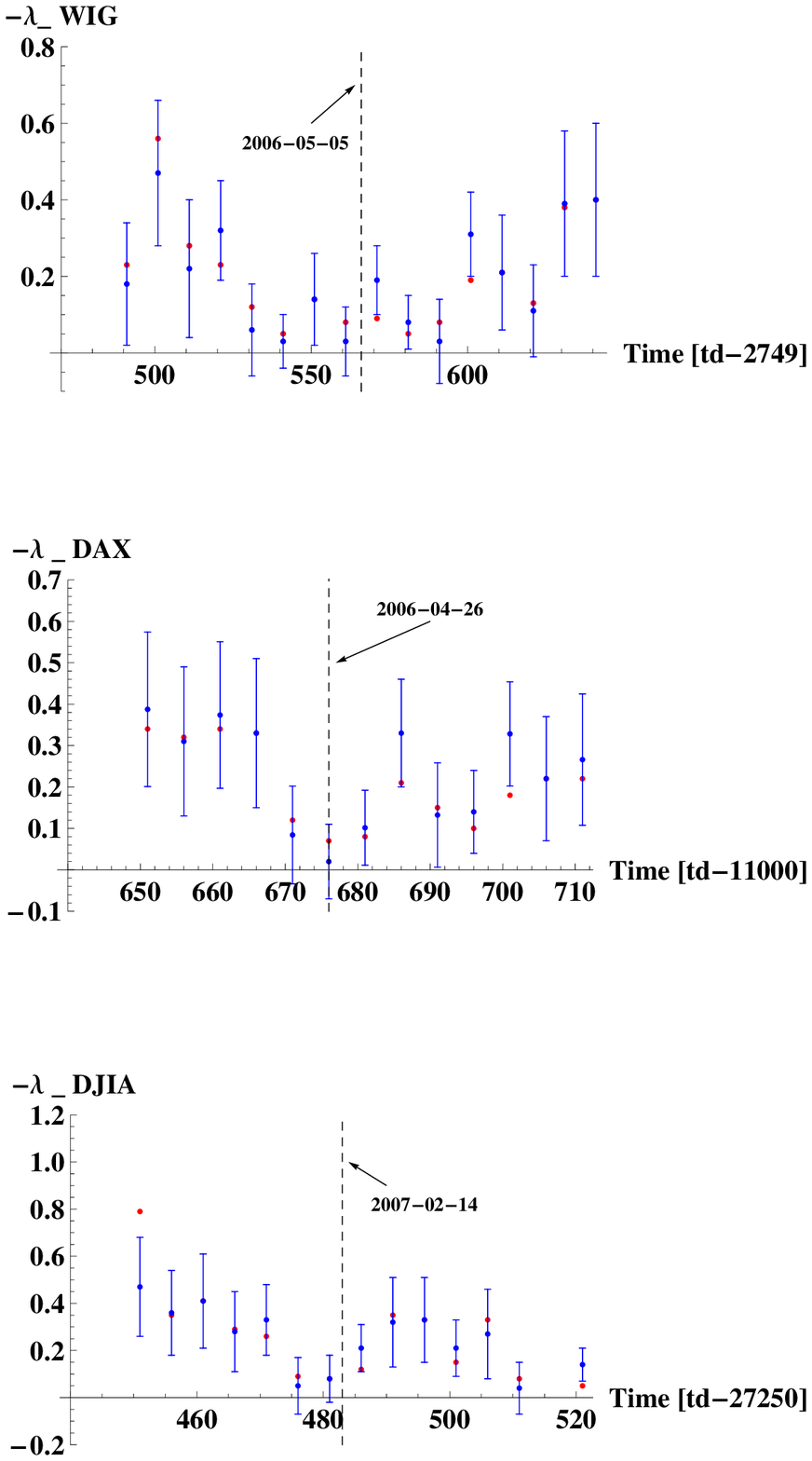}
\caption{The recovery rate $-\lambda (\geq 0)$
calculated by using two different formulas: (i) $-\lambda  = 1-AR(1)$ (full blue circles with bars)
and (ii) $-\lambda = 1-ACF(1)$ (full red circles), where  plots of AR(1) and ACF(1) vs. time were
presented in Fig. \ref{figure:Autocorrel}. Hence, both curves have similar shape in time (they are concave) having
broaded minima located at the same time interval as corresponding maxima shown in Fig. \ref{figure:Autocorrel}.
The vertical dashed lines denote the position of the spikes' centers.}
\label{figure:Recovery}
\end{center}
\end{figure}

In Figures \ref{figure:Autocorrel} and \ref{figure:Recovery} coefficients $AR(1)$ and $-\lambda $ were presented,
respectively.
It is remarkable that all our empirical data give $-1<\lambda \le 0$. Such a restricted range of
$\lambda $ is the crucial result, which is the one of necessary requirement (or signature) of existence the
catastrophic bifurcation transition.

Apparently, from fits mentioned above, the larger $AR(1)$ coefficient, almost equals $1.0$, is present for the latter
set of data placed near to the catastrophic bifurcation threshold (marked by the dotted vertical
straight line plotted in these figures). For the former set, placed one month earlier, we obtained much lower quantity
$AR(1)\approx 0.60$, as it should be.


In Section \ref{section:kp1oats} we proved that both for infinite and finite time series (which in all our cases
consists of 20 elements) $ACF(h)$ is expressed by the formula given in the second row in (\ref{rown:covar}).
In fact, we study $ACF(1)=AR(1)=1+\lambda $ by the method alternative to that used in Section
\ref{section:AR} to analysis of the coefficient $AR(1)$. Namely, we applied an usual estimator, $ACF_{EST}(1)$, of
$ACF(1)$ for a given month (which is our time window where $\lambda $ is constant),
\begin{eqnarray}
ACF_{EST}(1)=\frac{1}{Var(x_t)}\frac{1}{20}\left[\left(\sum_{t=0}^{19}x_t\, x_{t+1}\right)
-\frac{1}{20}\left(\sum_{t=0}^{19}x_t\right)\left(\sum_{t=0}^{19}x_{t+1}\right)\right].
\label{rown:estimat}
\end{eqnarray}
Indeed, this estimator is plotted in Figure \ref{figure:Autocorrel}, approaching (to good approximation)
its maximal value equals 1.0 in the vicinity of the threshold (marked by the dashed vertical line). This result,
together with the corresponding one for coefficient $AR(1)$, are the basic achievements of our work. The results
presented in Figure \ref{figure:Recovery} for recovery rate already follows from them.

Our approach is justified by assuming that $\lambda $ is a piecewise constant
function of time, i.e., it is a fixed quantity for each set of twenty data points; the same we assume for the shift
coefficient $b$ considered below. Hence, $\lambda $ and $b$ are slowly varying function of time (counted in months)
in comparison with process $x_t$ (counted in days). They play a crucial role in our considerations.

\subsection{Empirical catastrophic bifurcation transitions}\label{section:ECBT}

\begin{figure}
\begin{center}
\includegraphics[width=160mm,angle=0,clip]{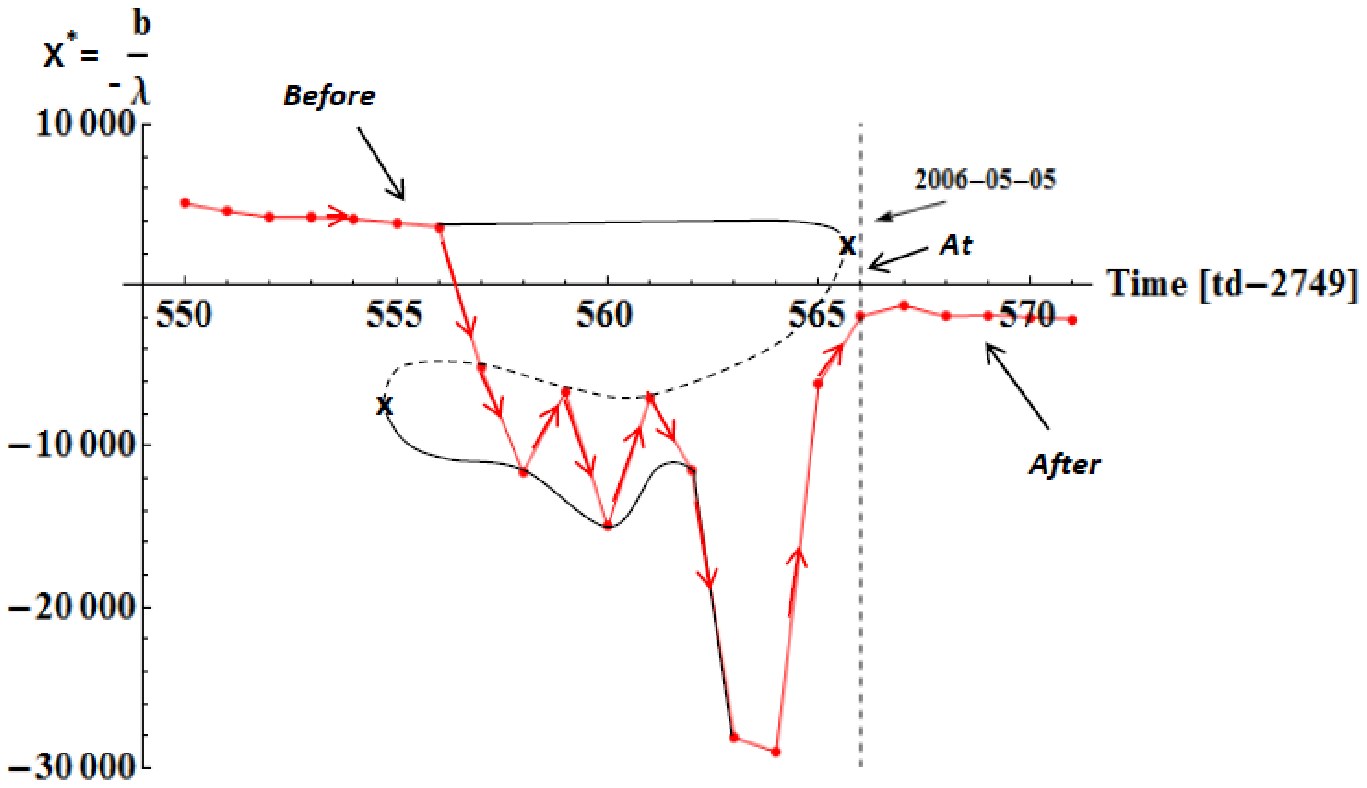}
\caption{Empirical curve (which consists of points joined by segments of lines with arrows), representing the mechanical
equilibrium states
defined by $x^*(=-b/\lambda )$ vs. time (counted in trading days, td), where $b$ and $\lambda $ were obtained from
empirical data for WIG (cf. Figures \ref{figure:Szum_sz_541_560} and \ref{figure:Recovery}). The flickering phenomena,
present prior to the negative spike are well seen before catastrophic bifurcation threshold. This threshold is localized
at 2006-05-05 and marked by the dashed vertical line. The upper segment of the backward folded curve (the solid one
ended by the right tipping point denoted by character x) is identified as sequence of stable mechanical equilibrium
states $x_{1''}^*$, the segment placed between two tipping points (the dashed curve) as a sequence of unstable
mechanical equilibrium states $x_{1'}^*$, and the lower segment (also solid curve) placed after the left tipping point
(again denoted by character x) is also identified as a sequence of stable mechanical equilibrium states, here $x_{1}^*$
(cf. Figures \ref{figure:BeforeCatBif}-\ref{figure:Zbiorczy_CatBif} in Section \ref{section:This} for details). Notably
that dashed curve (located in bistable region) is, obviously, smoothly plotted between two tipping points and over the
empirical points (dots).}
\label{figure:fixedpoint_WIG}
\end{center}
\end{figure}

\begin{figure}
\begin{center}
\includegraphics[width=155mm,angle=0,clip]{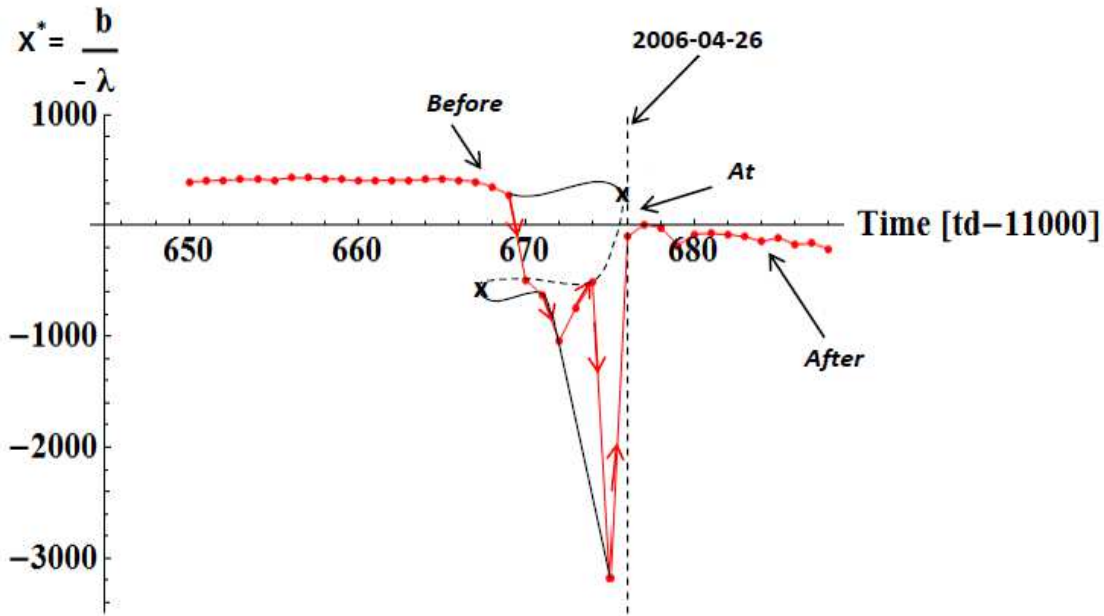}
\caption{Empirical curve (of points joined by segments of lines and arrows) being a sequence of
mechanical equilibrium states defined by $x^*(=b/|\lambda |)$ vs. time (counted in trading days, td), where $b$ and
$\lambda $ were obtained from empirical data for DAX (cf. Figures \ref{figure:Szum_sz_541_560} and \ref{figure:Recovery}).
The flickering phenomena, present prior to the negative spike are well seen before catastrophic bifurcation threshold.
This threshold is localized at 2006-04-26 and marked by the dashed vertical line. The upper segment of the backward
folded curve (the solid one ended by the right tipping point denoted by character x)), placed before the tipping point
and pointed by arrow termed `Before', is identified as a sequence of stable mechanical equilibrium states
$x_{1''}^*$. The segment placed between two tipping points (the dashed curve) is identified as sequence of unstable
mechanical equilibrium states $x_{1'}^*$. Finally, the lower segment (also solid curve) placed after the left tipping
point (again denoted by character x) is identified as a sequence
of stable mechanical equilibrium states, here $x_{1}^*$ (and pointed after the vertical dashed line by arrow termed
`After'). Note that the arrow termed `At' points possible catastrophic bifurcation threshold (cf. Figures
\ref{figure:BeforeCatBif}-\ref{figure:Zbiorczy_CatBif} for details). Notably that dashed curve (located in bistable
region) is, obviously, smoothly plotted between two tipping points and over the empirical points (dots).}
\label{figure:fixedpoint_DAX}
\end{center}
\end{figure}

\begin{figure}
\begin{center}
\includegraphics[width=160mm,angle=0,clip]{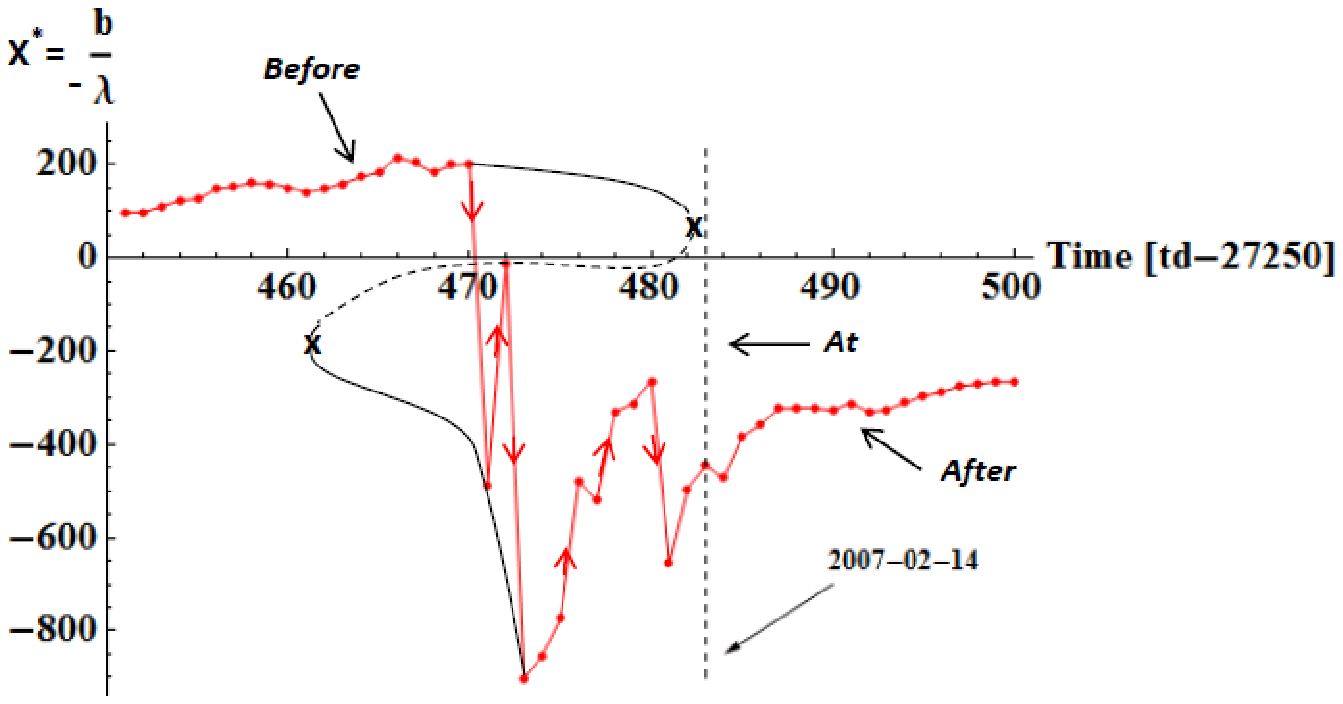}
\caption{Region of bistability spanned between two tipping points (marked by two characters x). All points represent
empirical data for DJIA.
Apparently, this region is analogous to those shown already in Figures \ref{figure:fixedpoint_WIG} and
\ref{figure:fixedpoint_DAX}.}
\label{figure:fixedpoint_DJIA}
\end{center}
\end{figure}

The shift coefficient $b$ relates to recovery rate $\lambda $
and root $x^*$ by equality $b=-\lambda x^*$ (see the second equation in (\ref{rown:fodeqex})). Hence, we derived
roots $x_j^*,\, j=1'', 1', 1$, and successively plotted their dependence on time in Figures
\ref{figure:fixedpoint_WIG}-\ref{figure:fixedpoint_DJIA}. Sufficiently far from the catastrophic bifurcation threshold
the spontaneous (maybe an accidental) reduction of error bars of the curve $x^*$ vs. time $t$ is observed together
with smoothing out of this curve. The significant jump of this curve is seen only in the nearest vicinity of
this threshold. We suppose that both these empirical facts have rather universal character as they are well seen for
typical stock markets of small, middle and large capitalization.

The so called, flickering phenomenon (defined already in Section \ref{section:uvds}) between stable roots
$x_{1''}^*$ and $x_1^*$ are well seen on these figures around the negative catastrophic spikes.
The flickering phenomenon can mainly appear within the bifurcation region where unstable intermediate state
($x_{1'}^*$) makes bounce of the system (between the above mentioned stable roots) easier. Indeed, this bounce
is able to significantly increase the variance (this effect we discuss in Section \ref{section:kp1oats}). We can
consider the flickering phenomenon as a possible precursor of the bifurcation catastrophic transition.

The key results shown in Figures \ref{figure:fixedpoint_WIG}-\ref{figure:fixedpoint_DJIA} constitutes basis for further
considerations because they suggest that bistabilities on financial markets can exist.

\subsection{Nonlinear indicator: skewness}

\begin{figure}[htb]
\begin{center}
\includegraphics[width=101mm,angle=0,clip]{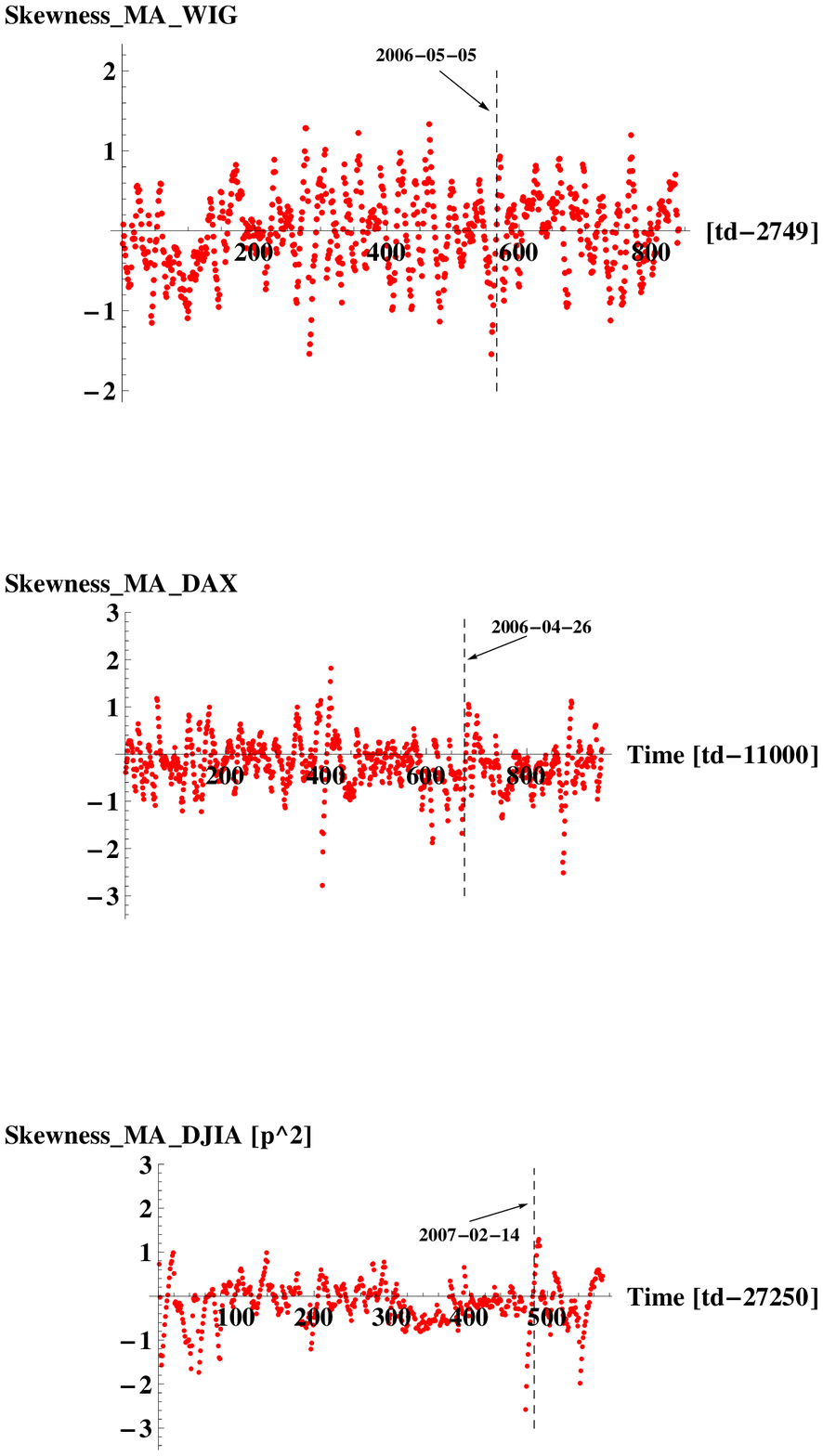}
\caption{Plot of the skewness where no systematic up-and-down asymmetry for WIG, DAX, and DJIA is seen for the first
glance. The vertical dashed lines denote the position of the variance spikes' center.}
\label{figure:skewness_MA}
\end{center}
\end{figure}

\begin{figure}[htb]
\begin{center}
\includegraphics[width=92mm,angle=0,clip]{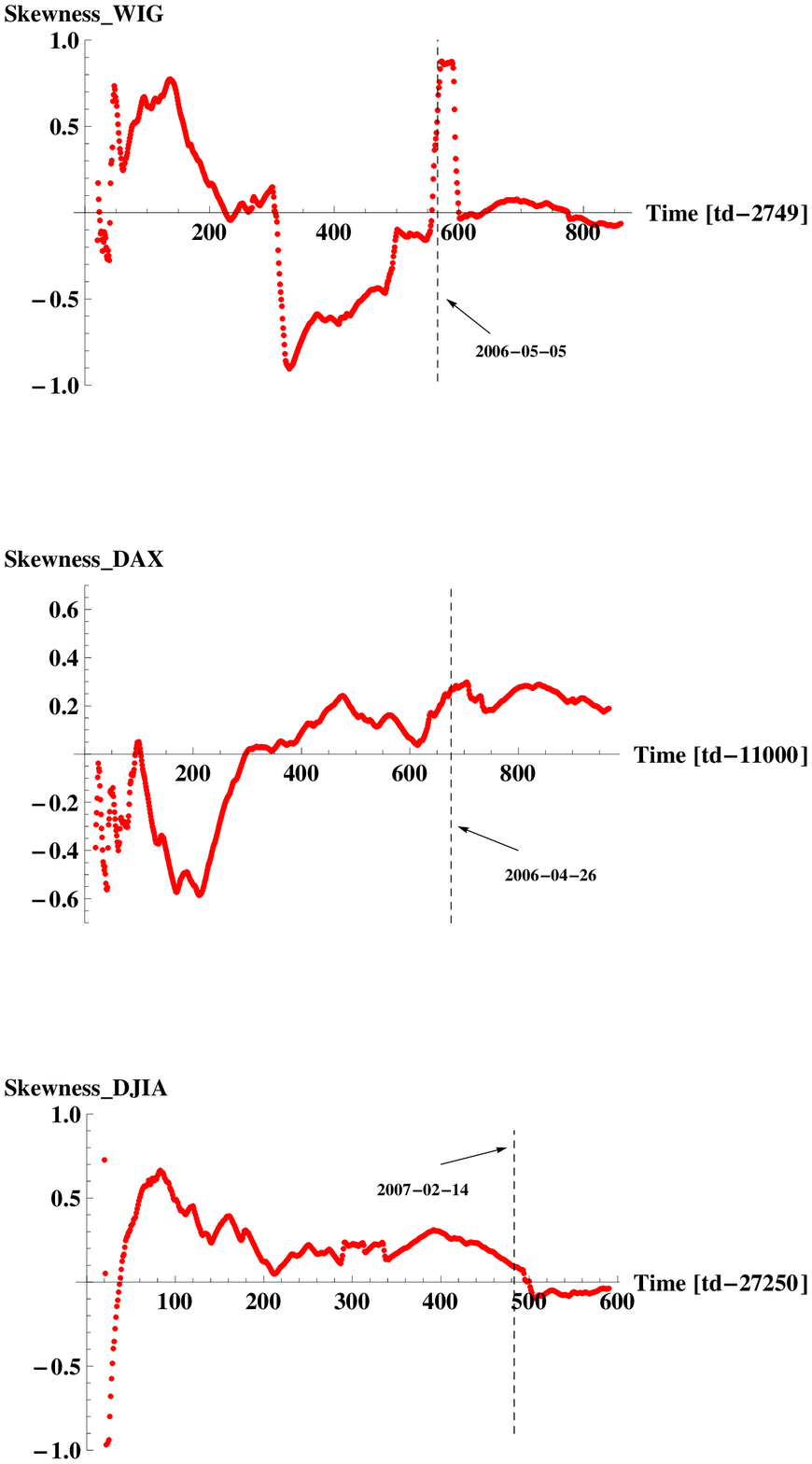}
\caption{Plot of the accumulative skewness where a sudden significant increase within the range of the spike is
well seen for WIG. For DAX the nonvanishing accumulative skewness is also observed within the range of the
spike. In fact, for DJIA the accumulative skewness vanishes within the range of the spike which means that the
catastrophic bifurcation transition has in this case a linear character.
}
\label{figure:Skosnosc_col}
\end{center}
\end{figure}

In Figure \ref{figure:skewness_MA} the time dependence of the skewness of detrended signal is shown. Relatively
large changes in skewness are observed. To verify whether these changes are statistically relevant we
calculated a cumulative skewness. This, more stable and less sensitive to unimportant details, quantity was plotted in
Figure \ref{figure:Skosnosc_col}. It presents a strong increase within the range of the spike. This
supplies a significant issue that consideration of empirical data within a linearized model, given by the latter
equation in (\ref{rown:fodeqx}), is insufficient \cite{GJ}.

\subsection{Periodograms}

\subsection{Periodograms of signals}\label{section:Pofs}

\begin{figure}
\begin{center}
\includegraphics[width=160mm,angle=0,clip]{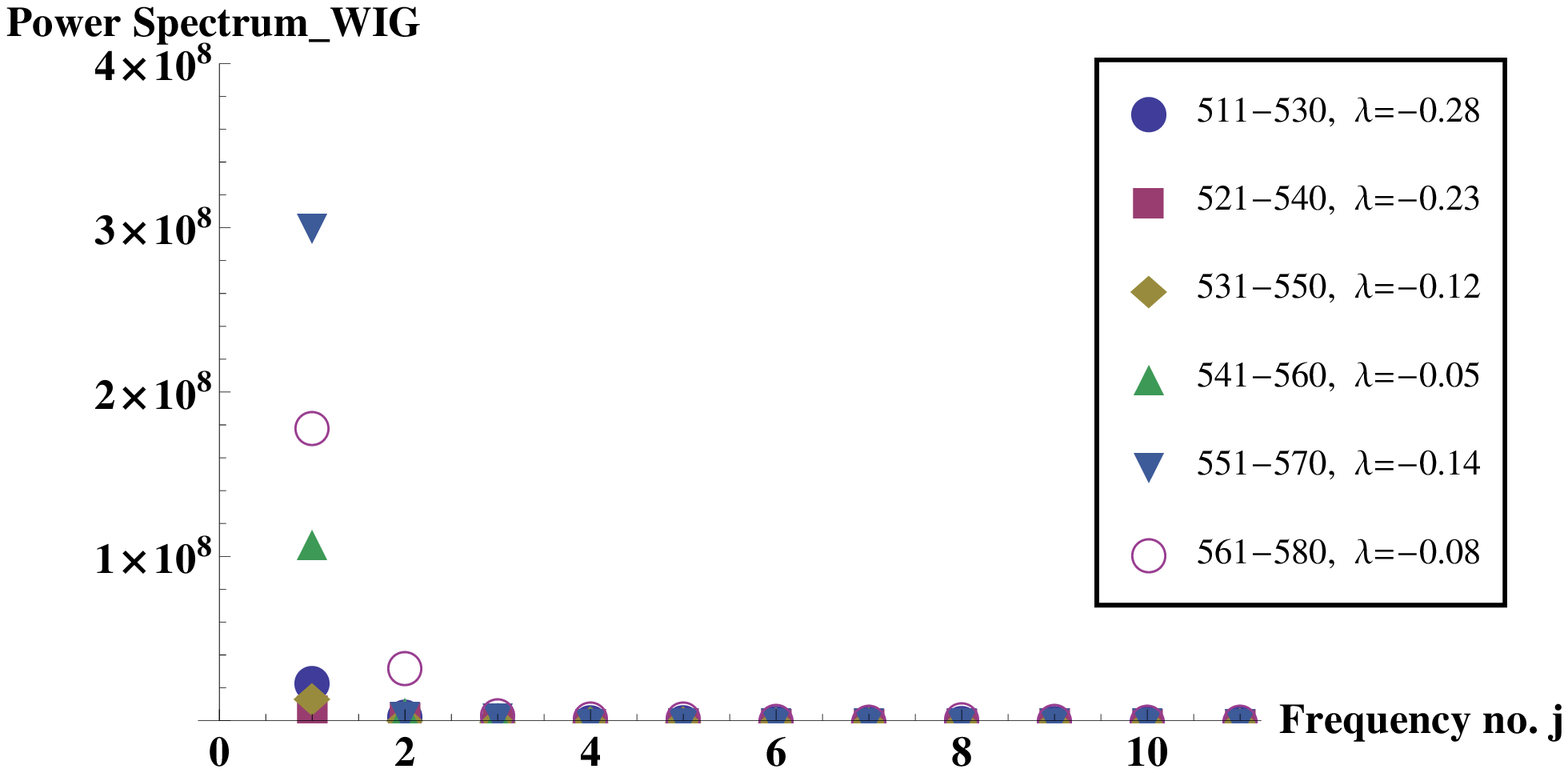}
\caption{Typical empirical periodograms, for instance, of WIG detrended signals calculated for six successive
overlapping time intervals (each consisting of 20 empirical data points). The first interval begins at 511 data point
and ends at 530 one (cf. blue circles). The inverted triangles and white circles were calculated for the last but one
and the latter time intervals ranging from 551 to 570 and from 561 to 580 data points, respectively. Indeed, these
intervals cover the catastrophic bifurcation threshold (placed at $566^{th}$ trading day). The increasing part of the
periodogram is placed at a vanishing frequency $\omega _1=0$ in agreement with suggestion given by Formula
(\ref{rown:PSsmall}). This is the one of the significant feature of the catastrophic bifurcation transition. The plot
is presented here for frequency numbers $j=1,2,\ldots ,T/2+1=11$ (it is sufficient due to symmetry of periodogram).
Apparently, recovery rate $-\lambda (>0)$ correspondingly decreases when it is reaching the catastrophic bifurcation
threshold (see also the upper plot in Figure \ref{figure:Recovery}). In fact, the difference between two periodograms
at $j=1$ representing inverted triangles and white circles can be considered as some error bar.}
\label{figure:periodogram_WIG}
\end{center}
\end{figure}

It is well known that for the case of finite length time series, the power spectrum ($PS(\omega )$, cf. Section
\ref{section:kp1oats}) is replaced by its estimator, i.e., by periodogram \cite{WAF,BD,KS} defined as
\begin{eqnarray}
I(\omega _j)=T^{-1}\left| \sum _{t=1}^T x_t \, exp (-i \, t \omega _j)\right| ^2,\; j=1,2, \ldots , T,
\label{rown:periodogr}
\end{eqnarray}
where $x_t$ is a detrended signal, $i$ is an imaginary unit and frequency $\omega _j=\frac{2\pi }{T}(j-1)$ (where
$j$ we name the frequency number). This enable the observation of a distinct increase of periodogram,
$I(\omega _1=0)=T^{-1}\mid \sum _{t=1}^T x_t \mid ^2$, that is, its strong
increase at $\omega _1 =0$ in comparison with remained values of the periodogram $I(\omega _j),\, j\ge 2$,
(cf. Figure \ref{figure:periodogram_WIG}; results for DAX and DJIA are analogous).
This means that we are able
to conclude about reddened of power spectra when discretized time-interval (consisting herein of $T=20$ empirical
data points) reach the catastrophic bifurcation transition finally containing it. More precisely, in Figure
\ref{figure:periodogram_WIG} the last two time intervals (extended from 551 to 570 and 561-580 trading days)
contain the catastrophic bifurcation threshold. Therefore, periodograms for these intervals increase at a vanishing
frequency $\omega _1=0$, as it was suggested by Formula (\ref{rown:PSsmallf}). In fact, the difference between two
periodograms at $j=1$ representing inverted triangles and white circles can be considered as some error bar.

Namely, for $\mid \omega \mid \ll 1$ we obtain from Expression
(\ref{rown:PSomeg}) that
\begin{eqnarray}
PS(\omega )\approx -\frac{\lambda \, (2+\lambda )}{\lambda ^2-(1+\lambda)\, \omega ^2}\approx -\, \frac{2\lambda }{\lambda ^2-\omega ^2},
\label{rown:PSsmall}
\end{eqnarray}
where the latter Expression in (\ref{rown:PSsmall}) was obtained at additional condition $\mid \lambda \mid \ll 1$.

By assuming that $\lambda $ vanishes slower than $\omega $ (at least in the nearest vicinity of the catastrophic
bifurcation threshold), we obtain from Equation (\ref{rown:PSsmall}) that
\begin{eqnarray}
PS(\omega )\approx -\frac{2}{\lambda }.
\label{rown:PSsmallf}
\end{eqnarray}

Expression (\ref{rown:PSsmallf}) says that as system approaches catastrophic bifurcation transition the singularity
(which for empirical data manifests by a distinct but finite spike) shifts systematically to lower frequencies as then
$\lambda $ vanishes. Indeed, this is well known reddened of the power spectrum. In other words, the significant
increase of the peridogram shown by inverted triangle in Figure \ref{figure:periodogram_WIG} at $\omega =0$ (or
$j=1$), is found as time-interval was shifted toward the catastrophic bifurcation transition. This increase is
a "finger print" of the catastrophic bifurcation transition.

\begin{figure}
\begin{center}
\includegraphics[width=140mm,angle=0,clip]{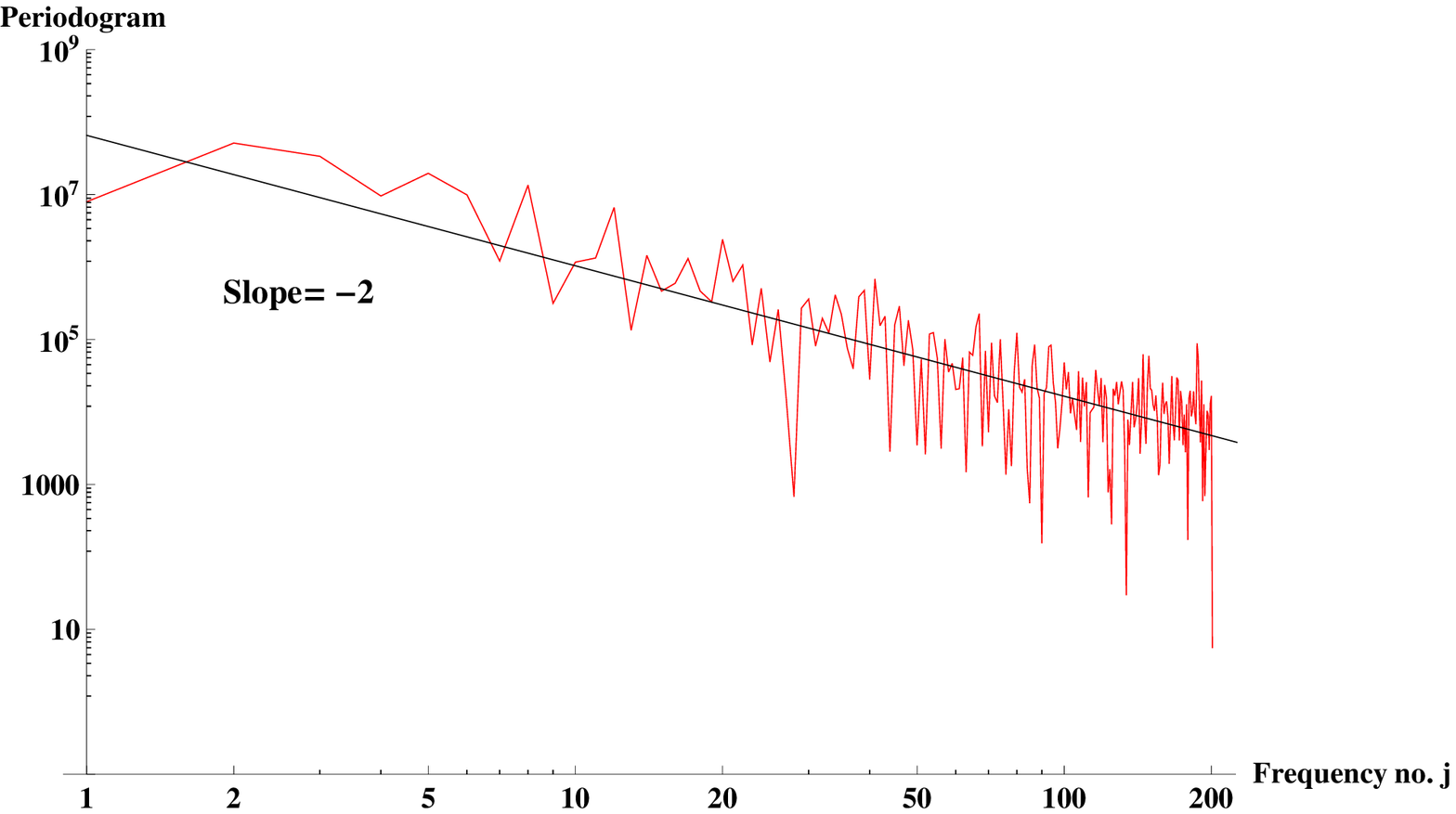}
\caption{Empirical periodogram of WIG's detrended signal $x_t,\; t=1,2,\ldots ,T=400$ that is, periodogram calculated
from 1 to 400 points $x_t$. The slope of the plot (in log-log scale) is well established.}
\label{figure:Power_spectrum_sygnal_1_400}
\end{center}
\end{figure}

\begin{figure}
\begin{center}
\includegraphics[width=150mm,angle=0,clip]{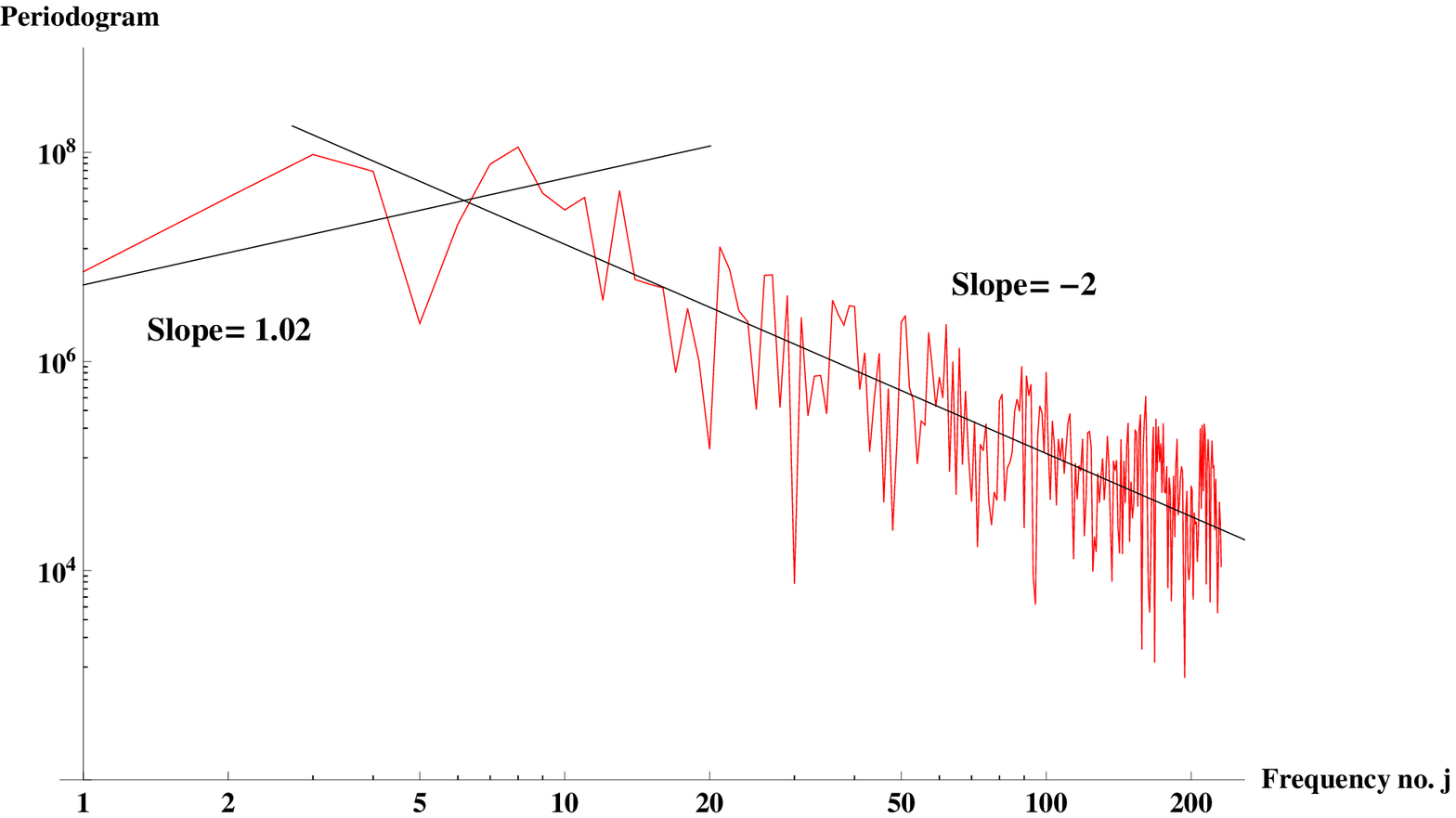}
\caption{Empirical periodogram of WIG's detrended signal $x_t,\; t=401,402,\ldots ,T=859$ that is, periodogram
calculated from 401 to 859 empirical points $x_t$). The slopes of the plot is well established both for small and
other frequencies.}
\label{figure:Power_spectrum_sygnal_400_859_polowa}
\end{center}
\end{figure}

It should be admitted that both empirical periodograms for detrended signals that is, the first one for time series
ranging from 1 to 400 (see Figure \ref{figure:Power_spectrum_sygnal_1_400} for details) and the second one from 401
to 859 (see Figure \ref{figure:Power_spectrum_sygnal_400_859_polowa} for further details) give (in log-log plot) slopes
consistent with distributions shown in plots in Figures \ref{figure:Histogram_noise_WIG_do400_Lside1_2} and
\ref{figure:Histogram_noise_WIG} (cf. Appendix \ref{section:scalrels}), respectively. These results are also
consistent with corresponding ones considered in Section \ref{section:Pons} below.

\subsection{Periodograms of noises}\label{section:Pons}

\begin{figure}
\begin{center}
\includegraphics[width=150mm,angle=0,clip]{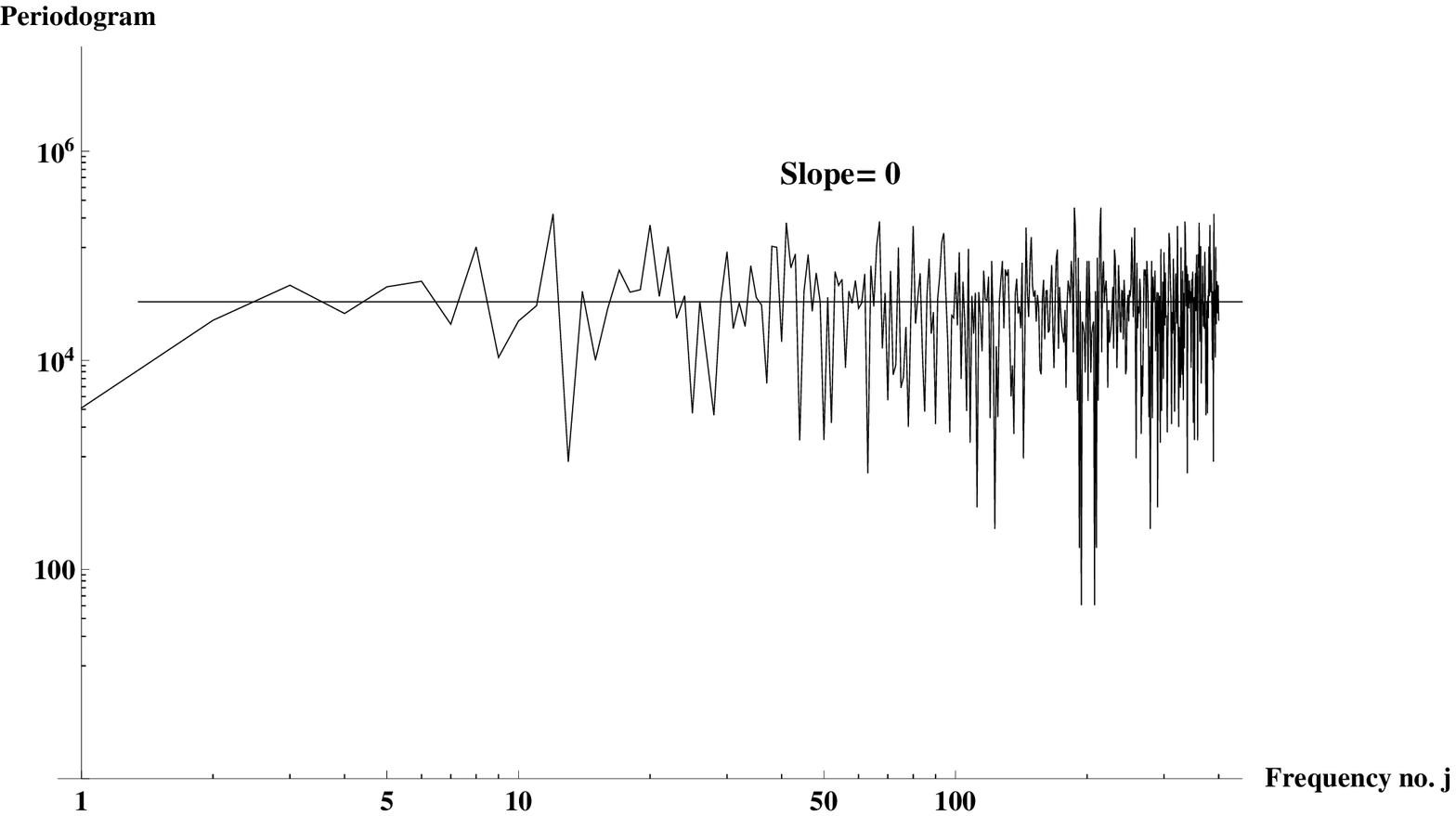}
\caption{Empirical periodogram of noise (increments) of the WIG's detrended signal calculated for T=399 that is,
calculated from 1 to 399 pair of points (defining $\Delta x_t$). The expected white noise (the horizontal part of
the plot) is well established.}
\label{figure:Periodogram_noise_WIG_do400}
\end{center}
\end{figure}

\begin{figure}
\begin{center}
\includegraphics[width=150mm,angle=0,clip]{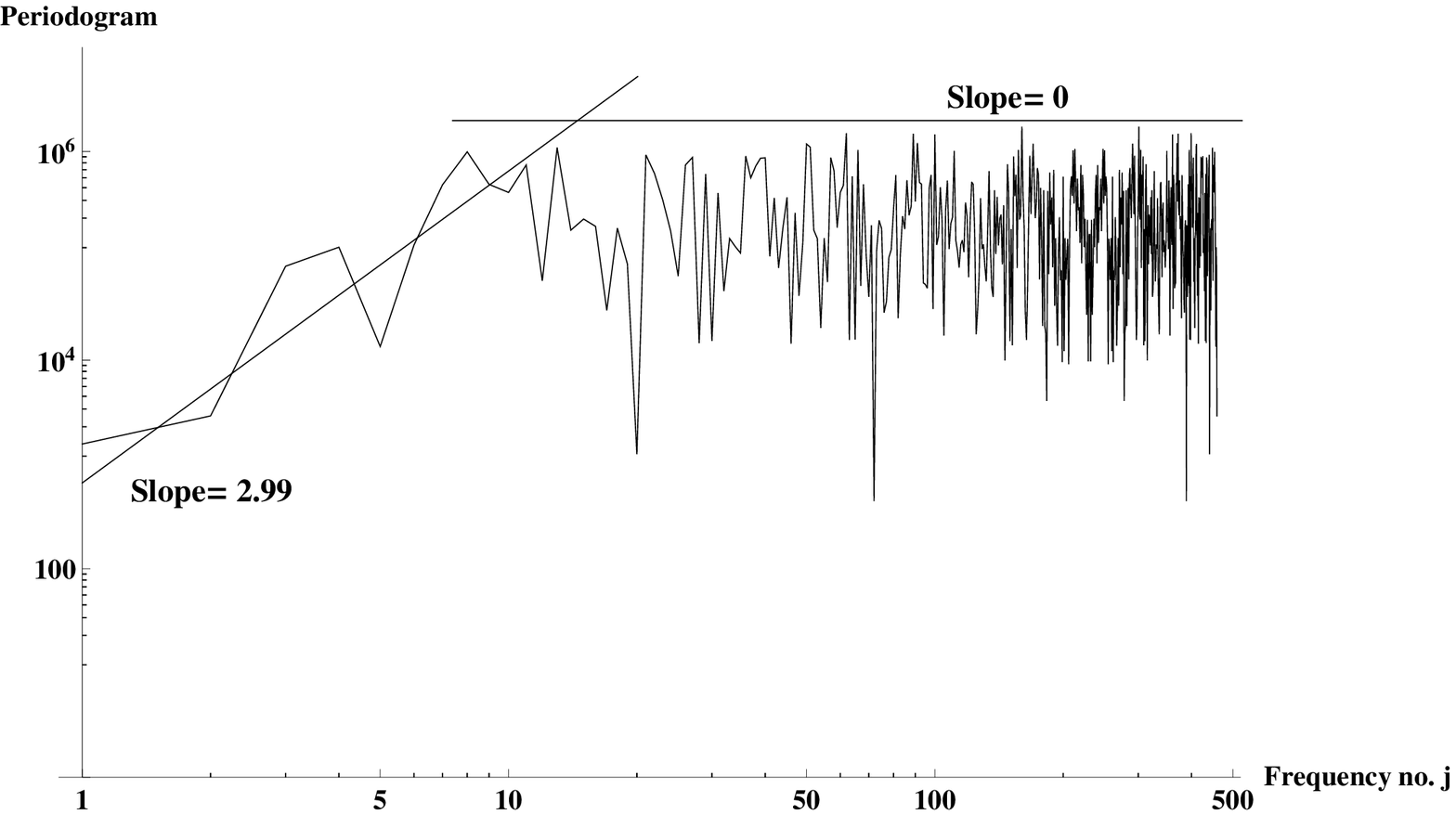}
\caption{Empirical periodogram of noise (increments) of the WIG's detrended signal calculated for T=458 that is,
calculated from 400 to 858 pair of points. The power-law for small frequencies (i.e., for $1\leq j\leq 14$) and white
noise for intermediate and large frequencies (i.e., for $j\geq 15$) are well seen.}
\label{figure:Periodogram_noise_WIG_od401}
\end{center}
\end{figure}

The periodogram of the noise (increments) $\Delta x_t =x_{t+1}-x_{t}$ has the form analogous to (\ref{rown:periodogr}),
where $x_t$ is replaced by $\Delta x_t$ and $T$ by $T-1$. Two characteristic plots of periodograms of noises are
presented in Figures \ref{figure:Periodogram_noise_WIG_do400} and \ref{figure:Periodogram_noise_WIG_od401}.

The result presented in Figure \ref{figure:Periodogram_noise_WIG_od401} makes possible to estimate the Hurst
exponent $H$ by using the Geweke and Porter-Hudak (GPH) method \cite{GPH,RW}. According to this method a simple
linear regression
\begin{eqnarray}
\ln I(\omega _k)=const - (H-0.5)\ln \left(4\sin ^2(\omega _k/2)\right)
\label{rown:regres}
\end{eqnarray}
should be valid at low frequencies $\omega _k,\; k=1,\ldots ,K\leq T/2$.

Several authors suggested that the length $K$ of part of periodogram covered by power law obeys the following
inequalities $Int\left[L^{0.2}\right]\leq K\leq Int\left[L^{0.5}\right] $. In our case (where the
length of time series $L=400$ and $K=14$) we obtain $K=Int[L^{0.44}]$ in accordance with these inequalities.

Furthermore, from fit of Relation (\ref{rown:regres}) into the empirical data within the range not larger than $K$ we
obtain $-2(H-0.5)=-3.02\Longleftrightarrow H=-1.01$. Recalling that stationary increment-increment correlation function
decays in time-lag according to power-law driven by exponent $2(1-H)$ \cite{RK}, we obtain as a result that the value
of this exponent equals 4.02.

On the other side, already simplified approach\footnote{The approach is simplified because more refined on-sided
Zipf-law should be considered here.} gives the consistence between this value of $H$ and exponents supplied by
power-laws shown in Figures \ref{figure:Histogram_noise_WIG} and \ref{figure:Periodogram_noise_WIG_od401}.
This approach is based on the scaling law given by Equations (26) - (28) in \cite{RK}. Hence, we have
\begin{eqnarray}
\mbox{Histogram}(\Delta x)\propto \frac{1}{\Delta x^{1-2/\eta }}=\frac{1}{\Delta x^{1.99}},
\label{rown:Histoscale}
\end{eqnarray}
where $\eta /2=H=-1/0.99=-1.01$. This histogram was already applied in Sect. \ref{section:noise}.

As is apparent, few independent approaches (considered in this Section and Section \ref{section:noise}) give the same
value of estimator of the Hurst exponent (burdened by statistical errors no greater than $5\%$).

\section{Qualitative explanation}\label{section:This}

\begin{figure}
\begin{center}
\includegraphics[width=130mm,angle=0,clip]{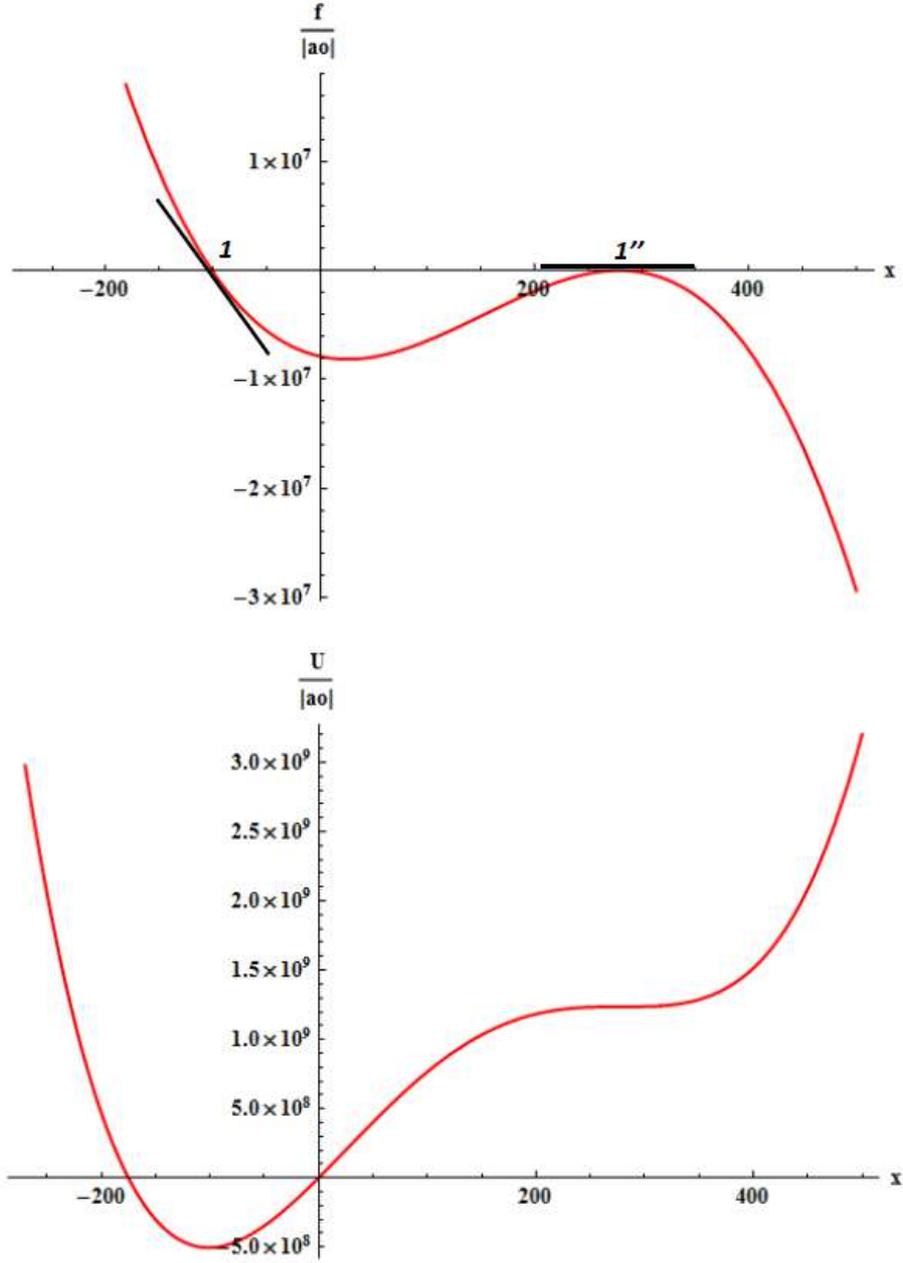}
\caption{Two complementary views at the threshold of the catastrophic bifurcation transition. Obviously, both curves
were plotted for the same values of the relative coefficients $a_1/a_0=-456.67,\; a_2/a_0=21359.70$ and
$a_3/a_0=7.87066\times 10^6$ derived in the main text. The curve $f/\mid a_0\mid $ vs. $x$ in the upper plot has single
twofold root $x_{1'}^*=x_{1''}^*=278.92$. This root, being the second tipping point, was denoted in Figure
\ref{figure:fixedpoint_DAX} by character x and placed in the nearest vicinity of the vertical dashed line. The first
root $x_1^*=-101.17$ is given directly by the empirical point placed on the vertical dashed line in this figure.
Apparently, the potential curve $U(x;P)/\mid a_0\mid $ vs. $x$ (presented in the lower plot) has positive asymmetry
in the vicinity of the equilibrium point $1''$. This results in the well seen positive asymmetry of the cumulative
skewness shown in the middle plot in Figure \ref{figure:Skosnosc_col} in the nearest vicinity of the vertical dashed
line.}
\label{figure:AtCatBif}
\end{center}
\end{figure}

In this section, we  explain how linear and nonlinear indicators (or early warnings) arise when system approaches the
regime shift or catastrophic bifurcation transition (threshold). The linear early warnings such as variance, recovery
time, reddened power spectra and related quantities can be derived in the frame of linearized time series defined by
the latter equation in (\ref{rown:fodeqx}). In contrast, nonlinear indicators (such as a non-vanishing skewness)
require an approach based on the nonlinear and asymmetric part of force $f(x;P)$ (present in Equation
(\ref{rown:fodeq}) in Section \ref{section:CSDgpv}) and on its asymmetric potential $U(x;P)$ (defined by Equation
(\ref{rown:potent})), both in the nearest vicinity of the regime shift (cf. plots in Figure \ref{figure:AtCatBif}
concerning the case at the catastrophic bifurcation threshold). This is one of the simplest viewpoint considered,
for instance, in article \cite{GJ}. Following this article it is possible to give herein at least a qualitative
explanation based on the mechanical like picture of the ball moving in the potential landscape.

\begin{figure}
\begin{center}
\includegraphics[width=140mm,angle=0,clip]{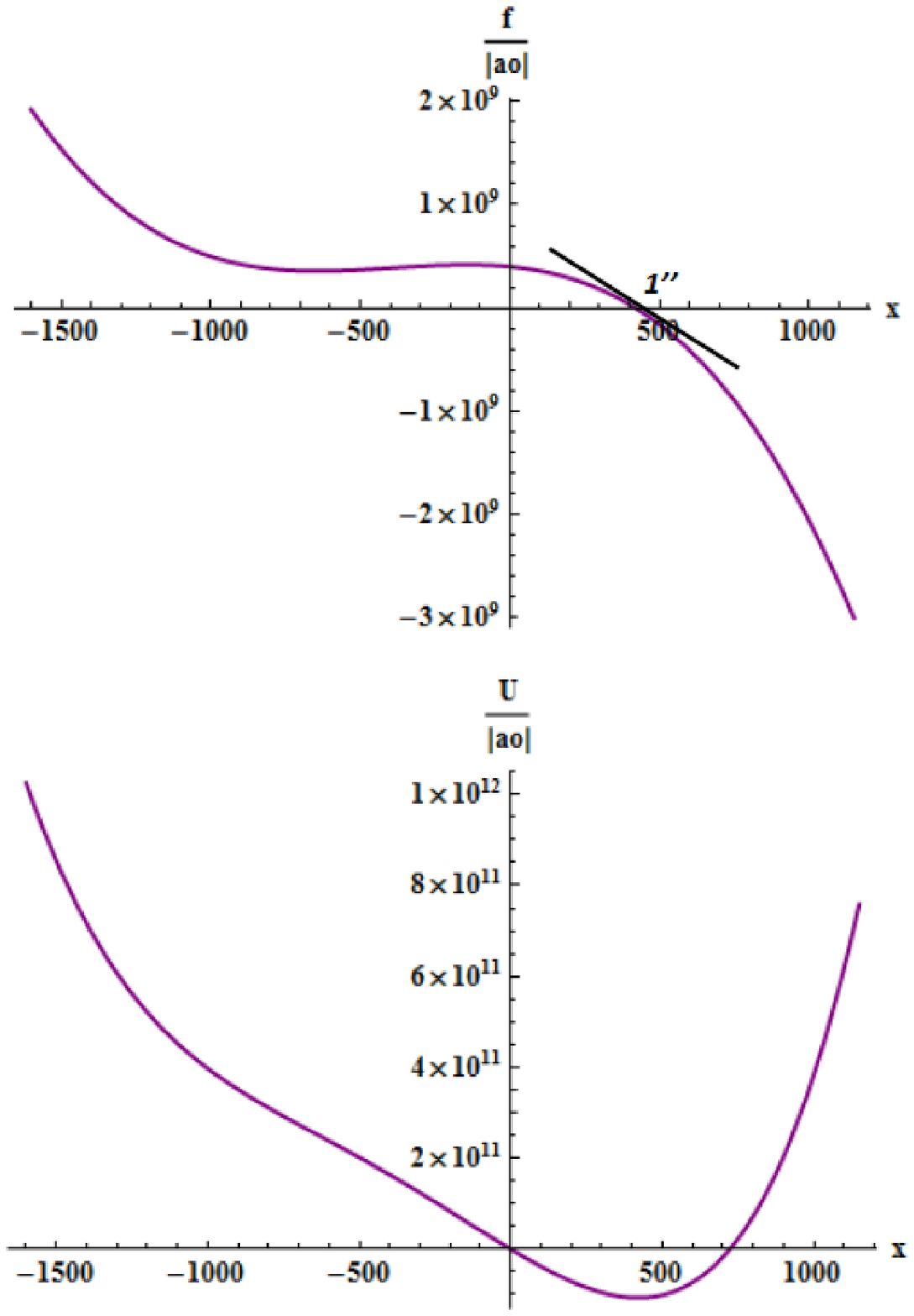}
\caption{Two complementary views of the situation before the catastrophic bifurcation region hidden
behind the highest peak of the variance (the spike) shown in Fig. \ref{figure:wariancja_MA_WIG_DAX_DJIA}. The upper
plot shows dependence of the force $f(x;P)$ (present in Equation (\ref{rown:fodeq})) vs. variable $x$ for the following
values of the relative coefficients $a_1/a_0=1179.81,\; a_2/a_0=278390$ and $a_3/a_0=-4.00948\times 10^8$ obtained in
the main text. Required mechanical empirical equilibrium point $1''(=x_{1''}^*)=421.009$, being the root of equation
$f(x;P)=0$, was taken from the empirical data (or backward folded curve) presented, for instance, for DAX in
Figure \ref{figure:fixedpoint_DAX}. The empirical equilibrium point $1''$ relates to ordinate equals
$665\, [td-11000]$ in this Figure. In the lower plot the corresponding potential $U(x;P)$ (related to force $f(x:P)$
by Equation (\ref{rown:potent})) is presented. Apparently, the equilibrium point $1''$ is a minimum of this potential.}
\label{figure:BeforeBeforeCatBif}
\end{center}
\end{figure}

\begin{figure}
\begin{center}
\includegraphics[width=115mm,angle=0,clip]{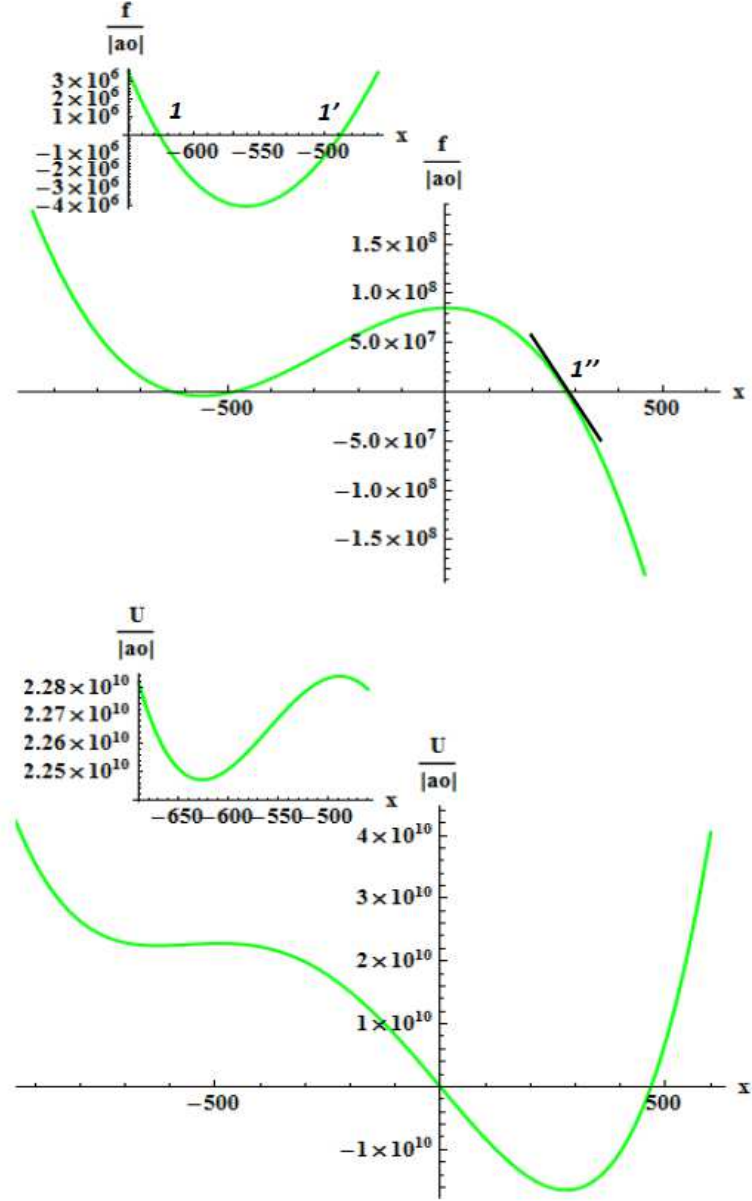}
\caption{Two complementary views of the situation inside bifurcation (bistable) region and before the catastrophic
bifurcation threshold hidden behind
the highest peak of the variance (the spike) shown in Fig. \ref{figure:wariancja_MA_WIG_DAX_DJIA}. The upper plot
shows dependence of the force $f(x;P)$ (present in Eq. (\ref{rown:fodeq})) vs. variable $x$ for some values of the
relative coefficients $a_1/a_0=835.861,\; a_2/a_0=-5022.94$ and $a_3/a_0=-8.53249\times 10^7$ obtained in the main text.
The mechanical equilibrium points $1(=x_1^*)=278.920,\, 1'(=x_{1'}^*)=-488.308$ and $1''(=x_{1''}^*)=-626.473$, being
roots of equation $f(x;P)=0$, were taken from empirical data (or backward folded curve) presented, for instance, for
DAX in Figure \ref{figure:fixedpoint_DAX}. The ordinates of these points are 669, 670, and 671, respectively. In the
lower plot the corresponding potential $U(x;P)$ is presented, where points $1$ and $1''$ are stable equilibrium
while $1'$ is the unstable one. The inset plots better visualize the behavior of $f$ and $U$ in the restricted region
containing points $1$ and $1'$.}
\label{figure:BeforeCatBif}
\end{center}
\end{figure}

\begin{figure}
\begin{center}
\includegraphics[width=130mm,angle=0,clip]{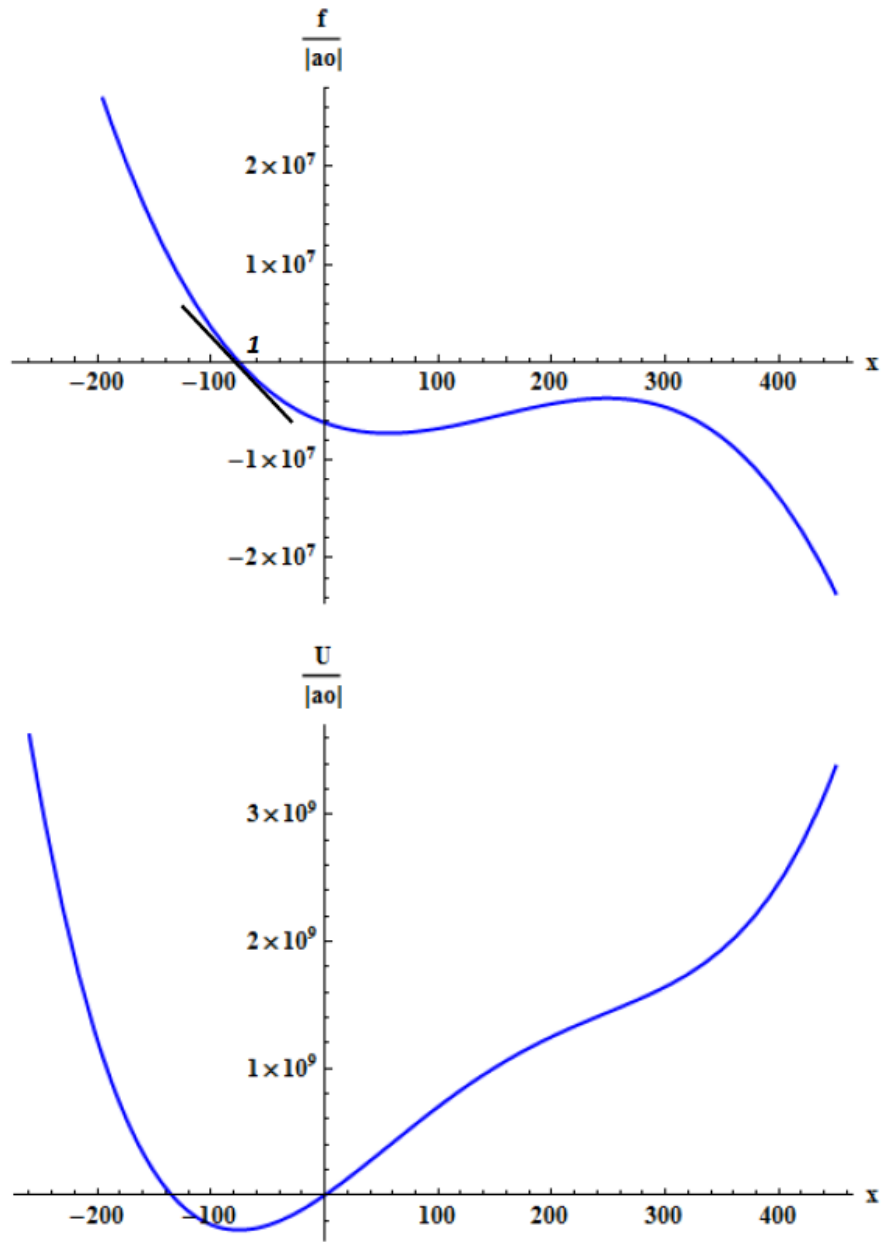}
\caption{Two complementary plots after the catastrophic bifurcation threshold. Obviously, both curves were obtained
in the main text for the same values of the relative coefficients $a_1/a_0=-456.67,\; a_2/a_0=41709.50$ and
$a_3/a_0=6.1682\times 10^6$. The mechanical empirical equilibrium point $1(=x_1^*)=-75,3875$, being the root of equation
$f(x;P)=0$, was taken from the empirical data (or backward folded curve) presented, for instance, for DAX in Figure
\ref{figure:fixedpoint_DAX}. The empirical equilibrium point $1$ relates to ordinate equals $?\, [td-11000]$ in this
Figure. This is well seen in the lower plot that potential $U(x;P)/\mid a_0\mid $ is (almost) symmetric
around the equilibrium point $x_1$, which results in the vanishing of the cumulative skewness (cf. Figure
\ref{figure:Skosnosc_col}) after the catastrophic bifurcation transition.}
\label{figure:AfterCatBif}
\end{center}
\end{figure}

\begin{figure}
\begin{center}
\includegraphics[width=160mm,angle=0,clip]{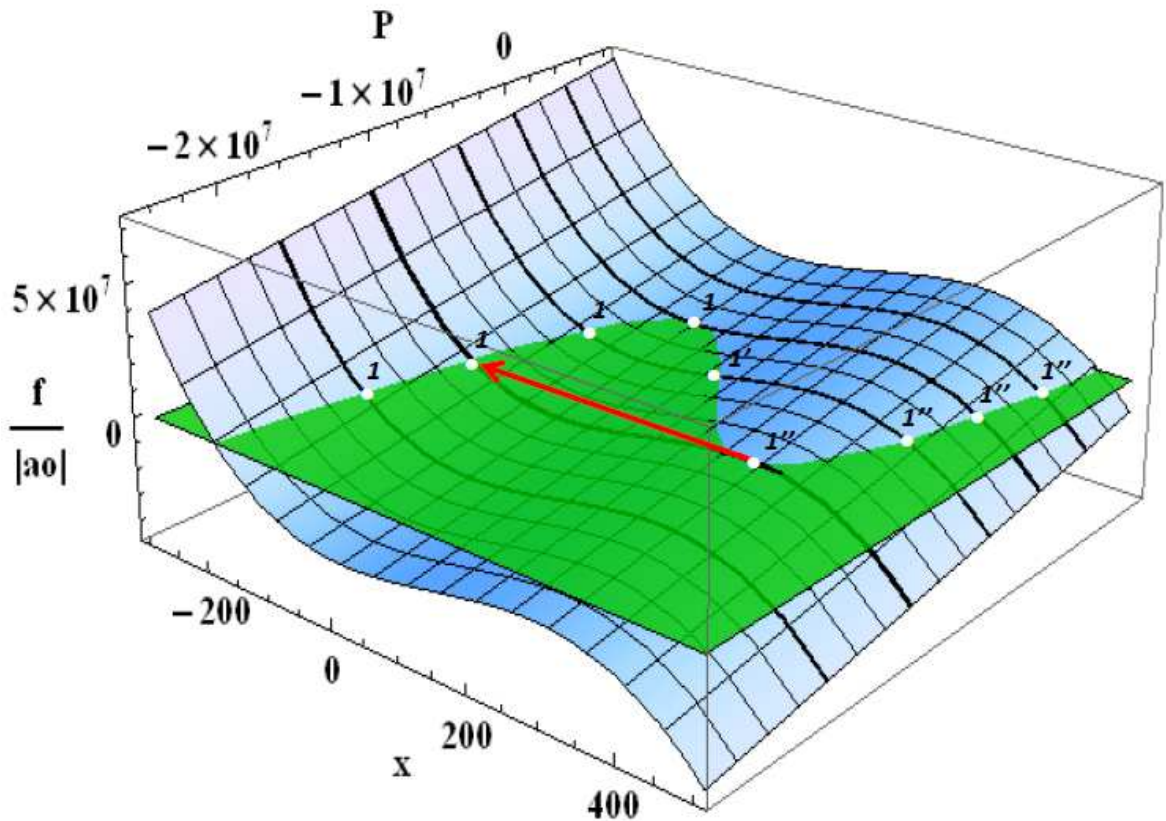}
\caption{A comprehensive three-dimensional schematic view showing the origin of the flat backward folded curve
$x^*$ vs. $P$ placed on a green (semi-transparent) plane. This backward folded curve originated as a section of the
green plane with the blue wavy surface. Points denoted by 1 and 1'' are stable mechanical equilibriums located on the
left and right segments of this curve, respectively. Points denoted by 1' are unstable mechanical equilibriums located
on the backward folded segment of this curve. The catastrophic bifurcation transition from equilibrium state
$1''$ to $1$ one is marked by the long red arrow. These particular points are placed on the catastrophic bifurcation
curve (being thicker than all other curves) located on the blue wavy surface. Note that the singular behavior of the
schematic backward folded curve in the vicinity of the catastrophic bifurcation threshold (cf. Figures
\ref{figure:fixedpoint_WIG} - \ref{figure:fixedpoint_DJIA}) is absent herein. Furthermore, the impact of noise
$\eta _t$ on states $x_t$ and $x^*_t$ is not visualized herein.}
\label{figure:Zbiorczy_CatBif}
\end{center}
\end{figure}

For instance, we consider a schematic snapshot pictorial views of four different modes (states) of the system on the
pathway to regime shift (cf. a sequence of the corresponding Figures
\ref{figure:BeforeBeforeCatBif}-\ref{figure:Zbiorczy_CatBif}). This pathway is defined by dependence $x^*=x^*(P)$
where driving (hidden) parameter $P$ is, by definition, a monotonically increasing function of time. The point $x^*(P)$
is a root of function $f(x;P)$, see point $1''(=x_{1''})$ shown in the upper plot in Figure
\ref{figure:BeforeBeforeCatBif}, points $1(=x_1),\, 1'(=x_{1'}),\, 1''$ in the upper  plot in Figure
\ref{figure:BeforeCatBif}, points $1,\, 1''$ shown in Figure \ref{figure:AtCatBif} and point $1$ in Figure
\ref{figure:AfterCatBif} as well as a sequence of points $1,\ 1'$ and $1''$ present in
the summary Figure \ref{figure:Zbiorczy_CatBif}. Indeed, any equilibrium state (stable or unstable) of the system
originates from Equation (\ref{rown:fodeq}) as a root of function $f(x;P)$.

\begin{figure}
\begin{center}
\includegraphics[width=140mm,angle=0,clip]{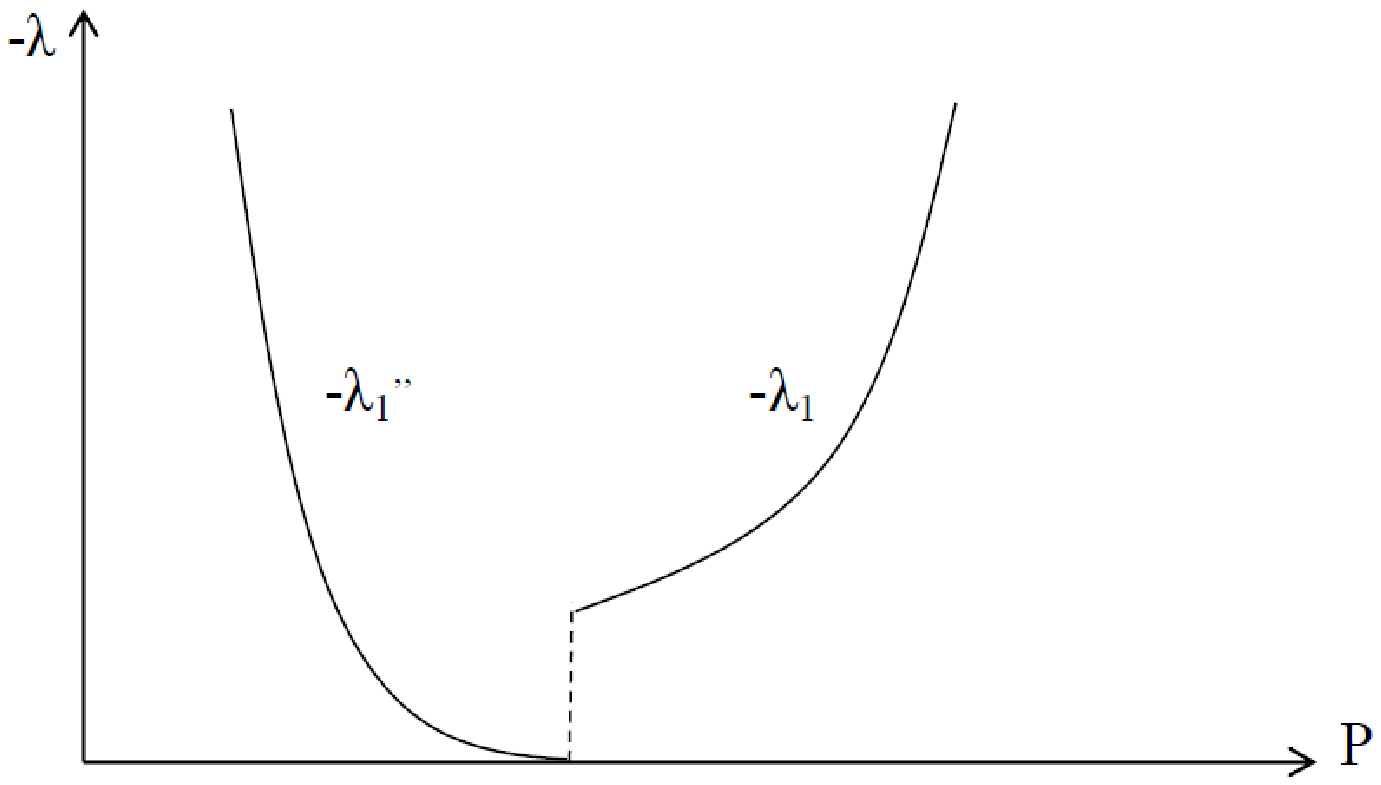}
\caption{The discontinuity of $-\lambda $ at the catastrophic bifurcation transition; this transition (between
points $x_{1''}^*$ and $x_1^*$) is marked in comprehensive Figure \ref{figure:Zbiorczy_CatBif} by the long arrow.}
\label{figure:Lambda_Uskok}
\end{center}
\end{figure}

The increase of driving parameter $P$ leads to the shifting of curve $f$ vs. $x$ in the downward direction. This is
well seen in the sequence of upper plots. The change of the shape of this curve is also observed although the change
is better seen in the corresponding swings of the curve $U$ vs. $x$ shown in lower plots.
As a consequence, roots representing stable equilibrium states that is, point $1''$ in Figure
\ref{figure:BeforeBeforeCatBif}, points $1,\, 1''$ present in Figures \ref{figure:BeforeCatBif} and point $1$ well
seen in Figure \ref{figure:AfterCatBif}, shift to the left while other, representing unstable equilibrium state that
is, point $1'$ shown in Figure \ref{figure:BeforeCatBif}, shifts to the right until disappear. Moreover, point $1''$
present in Figure \ref{figure:AtCatBif} also disappear (cf. next Figure \ref{figure:AfterCatBif}).
Furthermore, other roots (concerning other catastrophic bifurcation transition) can appear during the system evolution
and above given scenario can (at some time) repeats (cf. Figure \ref{figure:Scenario}).

Notably, two segments of the folded backward curve $x^*(P)$ containing points $1$ and $1''$ (schematically shown in
Figure \ref{figure:Zbiorczy_CatBif}), represent stable equilibria, while the third backward segment, containing points
$1'$, the unstable one. If the system is driven slightly away from the stable equilibrium it will return to this state
with relaxation time $\tau (P)$ (cf. considerations in Section \ref{section:dvequ}). Otherwise, the system driven from
unstable equilibrium will move away (to one of a stable equilibrium). In fact, the backward part of the curve $x^*(P)$
represent border or repelling threshold between the corresponding basins of attraction of two alternative stable
states (on the lower and upper branches of folded backward curve, marked by solid lines).

In this work we focus mainly on the analysis of stable equilibria. Some of them are tipping points at which a tiny
perturbation (spontaneous or systematic) can produce a large sudden transition (marked, e.g. for the second tipping
point, by long arrow in Figure \ref{figure:Zbiorczy_CatBif}). It should be noted that only in the vicinity of stable
equilibria that is, for points placed on the lower or upper branches of the folded curve, the variance of detrended
signal diverges according to power-law (cf. Expression (\ref{rown:variance}) in Section \ref{section:kp1oats}).
This is
a direct consequence of the critical slowing down, which can be detected well before a critical transition occurring.
This divergence can be intuitively understood as follows: as the return time diverges (cf. Sections
\ref{section:dvequ} and \ref{section:Dcbp}) the impact of a shock does not decay and its accumulating effect
increases the variance. Hence, CSD reduces the ability of the system to follow the fluctuations \cite{SBBB}.

It should be emphasized that observed increase of the skewness within the range of the spike of the variance
cannot be explained in the frame of linearized theory. Instead, the nonlinear asymmetric potential in the vicinity of
point 1 is needed to have deal with asymmetric distribution (cf. Equation (\ref{rown:eqP})).


\section{Conclusions}\label{section:conclus}

Following the Thomas Lux supposition concerning the possibility of existence of the catastrophic bifurcation
transitions on financial markets \cite{ThL}, we studied several linear and non-linear indicators
of such transitions on the stock exchanges of small and middle to large capitalization. All these indicators show
self-consistently that thresholds presented in Figures \ref{figure:WIG_C_24_09_2010} - \ref{figure:WariancjaL_WDD},
\ref{figure:Autocorrel}-\ref{figure:fixedpoint_DJIA}, and \ref{figure:Skosnosc_col} should be identified as
catastrophic bifurcation ones. There is a remarkable surprise that the catastrophic bifurcation
threshold itself is so distinct for daily empirical data received from various stock exchanges.
As it is seen, such a threshold (related to the kink-antikink pair) was noticed for several months before the global
crash. Notably that directly before and after the appearing of the kink-antikink pair the
variance of the signal is much less violent.

The basic results of this work was {\color{blue} {\bf the well established observation that: (i) $\lambda $ is
a negative quantity and (ii) recovery rate $\mid \lambda \mid $ vanishes when system approaches the catastrophic
bifurcation threshold (cf. Figure \ref{figure:Recovery}).}} This vanishing (together with result mentioned below)
permit us to formulate {\color{blue} {\bf the hypothesis that we deal herein with a catastrophic (but not critical)
slowing down.}} This result is significant the more so as $\lambda $ is a fundamental quantity which constitutes all
other linear indicators and partially participates in the non-linear ones.

Besides $\lambda $ we found also the shift parameter $b$ (cf. Fig. \ref{figure:Szum_sz_541_560} and, in particular,
the inside
figure present there). Hence, {\color{red}{\bf we were able to plot the empirical trajectory consisting of fixed points
$x^*$ plotted vs. trading time $t$ and directly observed the catastrophic bifurcation transition preceded by flickering
phenomenon (cf. Figures \ref{figure:fixedpoint_WIG}-\ref{figure:fixedpoint_DJIA}). This means that the Maxwell
construction can be broken here.}} This is the research highlight
of our work. Furthermore, we found that each catastrophic bifurcation transition is directly preceded by singularity,
which seems to be a superextreme event.

\begin{figure}
\begin{center}
\includegraphics[width=140mm,angle=0,clip]{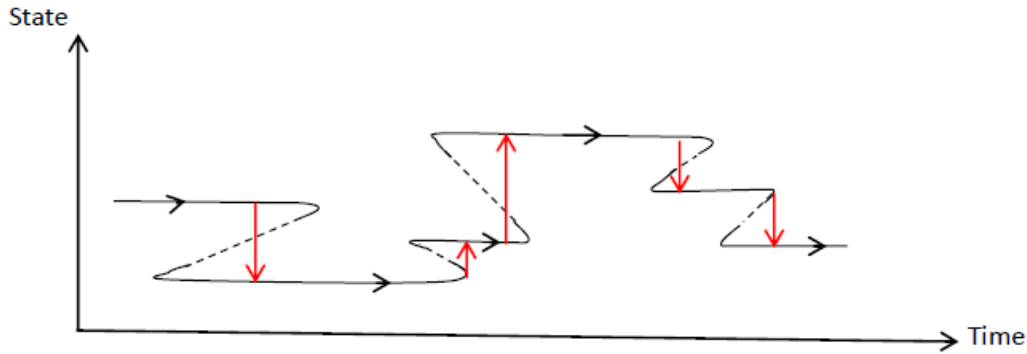}
\caption{The possible schematic scenario of a system evolution possesing several different catastrophic and
subcatastrophic bifurcation transitions.}
\label{figure:Scenario}
\end{center}
\end{figure}

\begin{figure}
\begin{center}
\includegraphics[width=150mm,angle=0,clip]{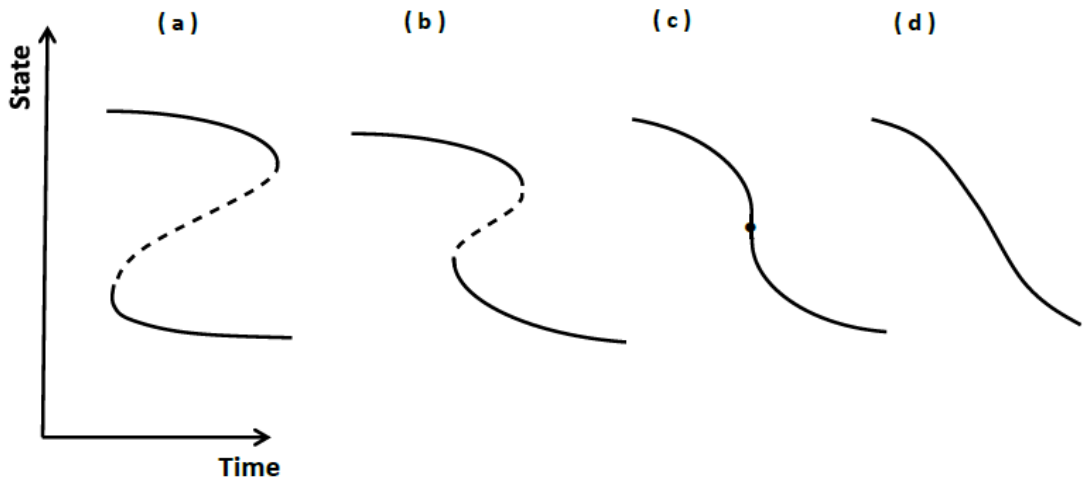}
\caption{The possible schematic scenario of a system evolution from likely subcatastrophic and catastrophic
bifurcation breakdowns (both (a) and (b)) to possible critical (c) and post-critical (d) transitions. Solid curves
denote stable equilibrium states. Unstable equilibrium states are marked by dashed curves in plots (a) and (b).
The critical point was schematically denoted by dot in plot (c). Note that such a sequence of curves can lead,
in our case, to 'pitchfork bifurcation' which arms and stem of the pitchfork are stable.}
\label{figure:Evolution}
\end{center}
\end{figure}

Finally, our study makes possible to formulate two kinds of hypothetical scenarios of financial market evolution shown
schematically in Figures \ref{figure:Scenario} and \ref{figure:Evolution}. Plot in Figure \ref{figure:Scenario} shows
that from time to time evolution of financial market can change abruptly according to subcatastrophic or
catastrophic bifurcation transition of different spread between stable regions. However, sometimes the sequence of
bifurcated curves presented in Figure \ref{figure:Evolution} can lead (in our case) to 'pitchfork bifurcation'
\cite{JJH} which arms and stem are stable. Maybe that some of them are so significant that become crashes.


We can conclude that linear and nonlinear indicators used in conjunction in this work provide reliable prediction
of regime shift on financial market.
However, more systematic verification of relation between the global crash (i.e. the time when hossa is changing
to bessa) and preceding it the corresponding catastrophic bifurcation transitions still leave a challenge.

\begin{acknowledgments}
We are grateful Piotr Suffczy\'nski for stimulating discussions.
Three of us (T. G., T. R. W., and R. K.) acknowledge partial financial support from the Polish Grant No. 119
awarded within the First Competition of the Committee of Economic Institute, organized by the National Bank of
Poland.
\end{acknowledgments}

\appendix

\section{Detrending procedure}\label{section:MLF}

We decided to use herein as a trend the following (deterministic) function of time $t$:
\begin{eqnarray}
X(\mid t-t_c\mid )&=&(X_0-X_1)E_{\alpha }\left(-\left(\frac{\mid t-t_c\mid }{\tau }\right)^{\alpha }\right) \nonumber \\
&-&X_1\cos(\omega \mid t-t_c\mid )\cos(\Delta \omega \mid t-t_c\mid ),\; X_0, \alpha , \tau , t_c >0,
\label{rown:MLFosc}
\end{eqnarray}
valid separately both for the bullish and bearish sides of a given well formed peak. In this work we have
$\omega,\, \Delta \omega \ll 1$ (as its is required). If additionally $\Delta \omega \ll \omega $ we deal with a beat.
The Mittag-Leffler function $E_{\alpha }(\ldots )$ is defined as follows \cite{MK}:
\begin{eqnarray}
E_{\alpha }\left(-\left(\frac{\mid t-t_c\mid }{\tau }\right)^{\alpha }\right)=\sum_{n=0}^{\infty }
\frac{(-1)^n}{\Gamma (1+\alpha n)}\left(\frac{\mid t-t_c\mid }{\tau }\right)^{\alpha n},
\label{rown:MLF}
\end{eqnarray}
where $t_c$ is the localization of the turning point (which change market state from bullish to bearish one),
$\alpha $ is the shape exponent and $\tau $ plays the role of the relaxation time. The trend function
(\ref{rown:MLFosc}) was already derived in our earlier work \cite{KK}. The parameters and coefficients describing
this function for indexes WIG, DAX, and DJIA hossas and bessas were presented in the corresponding tables placed below.
The trend function obtained on this way exhibits (as it is required) an unsustainable super-exponential (stretched
exponential) growth preceding the speculation-induced crash.

Remarkable that the trend function was derived in the frame of our Rheological Model of Fractional Dynamics of Financial
Markets. This model introduces the hypothesis that stock markets response like visco-elastic biopolimer. That is, they
are elastic (i.e., immediately response) if an external force impact on a stock market is sufficiently impetuous and
they are more like a liquid (plastic) material in the opposite case (of a weak external force). That is, the financial
markets can behave as a non-Newtonian liquid.

Notably, among fit parameters and coefficients for a given index (see tables \ref{table:par1L_WIG} -
\ref{table:par2R_DJIA}) there exists always at least one which
is burdened by large standard deviation. Indeed, this way the system is protected against an arbitrage. Other protection
is that time $t_c$ (the location of turning point from hossa to bessa) is (to some extend) random.

\begin{table}
\begin{center}
\caption{Values of fit parameters of the trend for WIG hossa ($R^2=0.9986$)}\label{table:par1L_WIG}
\vspace*{0.2in}
\begin{tabular}{|c||c|c|c|}
\hline
Parameter & Value & Standard deviation \\
\hline \hline
$t_c$ & $892\; [td]$ & $73\; [td]$ \\
$\tau $ & $105\; [td]$ & $420\; [td]$ \\
$\alpha $ & $0.57$ & $0.23$ \\
$\omega $ & $0.0041\; [td^{-1}]$ & $0.0005\; [td^{-1}]$ \\
$\Delta \omega $ & $0.0$ & $0.0$ \\
\hline
\end{tabular}
\end{center}
\end{table}

\begin{table}
\begin{center}
\caption{Values of fit coefficients of the trend for WIG hossa ($R^2=0.9986$)}\label{table:par2L_WIG}
\vspace*{0.2in}
\begin{tabular}{|c||c|c|c|}
\hline
Parameter & Value [p]& Standard deviation [p]\\
\hline \hline
$X_0+X_1$ & $60081$ & $85273$ \\
$X_1$ & $-8659$ & $2352$ \\
\hline
\end{tabular}
\end{center}
\end{table}

\begin{table}
\begin{center}
\caption{Values of fit parameters of the trend for WIG bessa ($R^2=0.9985$)}\label{table:par1R_WIG}
\vspace*{0.2in}
\begin{tabular}{|c||c|c|c|}
\hline
Parameter & Value & Standard deviation \\
\hline \hline
$t_c$ & $810\; [td]$ & $0\; [td]$ \\
$\tau $ & $272\; [td]$ & $20\; [td]$ \\
$\alpha $ & $1.562$ & $0.025$ \\
$\omega $ & $0.0431\; [td^{-1}]$ & $0.0005\; [td^{-1}]$ \\
$\Delta \omega $ & $0.0065$ & $0.0004$ \\
\hline
\end{tabular}
\end{center}
\end{table}

\begin{table}
\begin{center}
\caption{Values of fit coefficients of the trend for WIG bessa ($R^2=0.9985$)}\label{table:par2R_WIG}
\vspace*{0.2in}
\begin{tabular}{|c||c|c|c|}
\hline
Parameter & Value [p]& Standard deviation [p]\\
\hline \hline
$X_0+X_1$ & $41963$ & $334$ \\
$X_1$ & $-2528$ & $269$ \\
\hline
\end{tabular}
\end{center}
\end{table}

\begin{table}
\begin{center}
\caption{Values of fit parameters of the trend for DAX hossa ($R^2=0.9985$)}\label{table:par1L_DAX}
\vspace*{0.2in}
\begin{tabular}{|c||c|c|c|}
\hline
Parameter & Value & Standard deviation \\
\hline \hline
$t_c$ & $969\; [td]$ & $1\; [td]$ \\
$\tau $ & $426\; [td]$ & $391\; [td]$ \\
$\alpha $ & $0.52$ & $0.03$ \\
$\omega $ & $0.00362\; [td^{-1}]$ & $0.00004; [td^{-1}]$ \\
$\Delta \omega $ & $0.0065$ & $0.0004$ \\
\hline
\end{tabular}
\end{center}
\end{table}

\begin{table}
\begin{center}
\caption{Values of fit coefficients of the trend for DAX hossa ($R^2=0.9985$)}\label{table:par2L_DAX}
\vspace*{0.2in}
\begin{tabular}{|c||c|c|c|}
\hline
Parameter & Value [p]& Standard deviation [p] \\
\hline \hline
$X_0+X_1$ & $4698$ & $82$ \\
$X_1$ & $-763$ & $35$ \\
\hline
\end{tabular}
\end{center}
\end{table}

\begin{table}
\begin{center}
\caption{Values of fit parameters of the trend for DAX bessa ($R^2=0.9977$)}\label{table:par1R_DAX}
\vspace*{0.2in}
\begin{tabular}{|c||c|c|c|}
\hline
Parameter & Value & Standard deviation \\
\hline \hline
$t_c$ & $968\; [td]$ & $0\; [td]$ \\
$\tau $ & $426\; [td]$ & $72\; [td]$ \\
$\alpha $ & $1.12$ & $0.03$ \\
$\omega $ & $0.0089\; [td^{-1}]$ & $0.0001; [td^{-1}]$ \\
$\Delta \omega $ & $0.0246$ & $0.0001$ \\
\hline
\end{tabular}
\end{center}
\end{table}

\begin{table}
\begin{center}
\caption{Values of fit coefficients of the trend for DAX bessa ($R^2=0.9977$)}\label{table:par2R_DAX}
\vspace*{0.2in}
\begin{tabular}{|c||c|c|c|}
\hline
Parameter & Value [p]& Standard deviation [p] \\
\hline \hline
$X_0+X_1$ & $5464$ & $70$ \\
$X_1$ & $-847$ & $36$ \\
\hline
\end{tabular}
\end{center}
\end{table}

\begin{table}
\begin{center}
\caption{Values of fit parameters of the trend for DJIA hossa ($R^2=0.9996$)}\label{table:par1L_DJIA}
\vspace*{0.2in}
\begin{tabular}{|c||c|c|c|}
\hline
Parameter & Value & Standard deviation \\
\hline \hline
$t_c$ & $627\; [td]$ & $3\; [td]$ \\
$\tau $ & $333\; [td]$ & $38\; [td]$ \\
$\alpha $ & $1.29$ & $0.02$ \\
$\omega $ & $0.0107\; [td^{-1}]$ & $0.0002; [td^{-1}]$ \\
$\Delta \omega $ & $0.0220$ & $0.0002$ \\
\hline
\end{tabular}
\end{center}
\end{table}

\begin{table}
\begin{center}
\caption{Values of fit coefficients of the trend for DJIA hossa ($R^2=0.9996$)}\label{table:par2L_DJIA}
\vspace*{0.2in}
\begin{tabular}{|c||c|c|c|}
\hline
Parameter & Value [p]& Standard deviation [p] \\
\hline \hline
$X_0+X_1$ & $3486$ & $40$ \\
$X_1$ & $-332$ & $28$ \\
\hline
\end{tabular}
\end{center}
\end{table}

\begin{table}
\begin{center}
\caption{Values of fit parameters of the trend for DJIA bessa ($R^2=0.9971$)}\label{table:par1R_DJIA}
\vspace*{0.2in}
\begin{tabular}{|c||c|c|c|}
\hline
Parameter & Value & Standard deviation \\
\hline \hline
$t_c$ & $640\; [td]$ & $0\; [td]$ \\
$\tau $ & $165\; [td]$ & $191\; [td]$ \\
$\alpha $ & $1.938$ & $0.575$ \\
$\omega $ & $0.030\; [td^{-1}]$ & $0.070; [td^{-1}]$ \\
$\Delta \omega $ & $0.040$ & $0.070$ \\
\hline
\end{tabular}
\end{center}
\end{table}

\begin{table}
\begin{center}
\caption{Values of fit coefficients of the trend for DJIA bessa ($R^2=0.9971$)}\label{table:par2R_DJIA}
\vspace*{0.2in}
\begin{tabular}{|c||c|c|c|}
\hline
Parameter & Value [p]& Standard deviation [p] \\
\hline \hline
$X_0+X_1$ & $4010$ & $110$ \\
$X_1$ & $-866$ & $81$ \\
\hline
\end{tabular}
\end{center}
\end{table}

\section{Linear indicators of a critical slowing down from bifurcation point of view}\label{section:CSDgpv}

In this section we consider linear indicators of the critical slowing down or regime shift such as a variance,
recovery time and rate as well as reddened power spectra. The nonlinear indicator is defined by an asymmetric
skewness.

Let us suppose that detrended time-dependent signal, $x_t\stackrel{\rm def.}{=}X(t)-\mbox{Trend}(t)$,
obeys the first order difference equation of the stochastic dynamics
\begin{eqnarray}
x_{t+1}-x_t=f(x_t;P)+\eta _{\, t}
\label{rown:fodeq}
\end{eqnarray}
where $P$ is a driving (control, in general vector) parameter and the additive noise
$\eta _{\, t},\; t=0,1,2,\ldots ,$ is a $\delta $-correlated\footnote{Herein, the notation $\delta $ means the
Kronecker delta while $t$ numbers, in our case, trading days within a given trading month (which consists of 20
trading days). The trading month is our time window where $\lambda $ is constant.} $(0,\sigma ^2)$
random variable. In the upper plots in Figures \ref{figure:BeforeCatBif}-\ref{figure:AfterCatBif} the schematic
behaviors of function $f$ vs. variable $x$ were shown for different values of parameter $P$. In the comprehensive
Figure \ref{figure:Zbiorczy_CatBif} these dependences are collected in the form of the three-dimensional plot.
It is seen that function $f$ could be, for instance, a third-order polynomial having $P$-dependent coefficients (cf.
Section \ref{section:Example}).

\subsubsection{Continuous-time formulation}

Simultaneously, we consider the differential formulation of Equation (\ref{rown:fodeq}),
which basic ingredient is the Langevin dynamics \cite{DS0}, \cite{NGvK}.
This differential formulation takes the form of the massless stochastic dynamic equation
\begin{eqnarray}
\frac{\partial x_t}{\partial t}=-\frac{\partial U(x_t;P)}{\partial x_t}+\eta _t,
\label{rown:TDGL}
\end{eqnarray}
where $U$ plays the role of a mechanic potential while $\eta _t$ of a stochastic force.
This equation is equivalent to the quasi-linear (according to van Kampen terminology \cite{NGvK}) Fokker-Planck
equation
\begin{eqnarray}
\frac{\partial {\cal P}(x,t)}{\partial t} = - \frac{\partial j(x,t)}{\partial x}
\label{rown:FPeq}
\end{eqnarray}
which has the form of continuity equation (a conservation law) for probability density ${\cal P}(x,t)$,
where the current density is given by constitutive equation
\begin{eqnarray}
j(x,t)=f(x;P)\, {\cal P}(x,t)-\frac{\sigma ^2}{2}\frac{\partial {\cal P}(x,t)}{\partial x}.
\label{rown:jeq}
\end{eqnarray}

The equilbrium solution of Equation (\ref{rown:FPeq}) (obtained directly from requirement that no current is
present in the system, i.e. by assuming that $j(x,t)=0$ in Equation (\ref{rown:jeq})) is given by
\begin{eqnarray}
{\cal P}^{eq}(x)\sim \frac{2}{\sigma ^2}\exp\left(-\frac{2}{\sigma ^2}\, U(x;P)\right),
\label{rown:eqP}
\end{eqnarray}
where potential $U(x;P)$ already appeared in Equation (\ref{rown:TDGL}), while
\begin{eqnarray}
f(x;P)=-\frac{\partial U(x;P)}{\partial x}
\label{rown:potent}
\end{eqnarray}
is (in this terminology) a mechanical force. Both Formulas (\ref{rown:eqP}) and (\ref{rown:potent}) play an important
role in our explanations. That is, a positive asymmetry of the potential $U$ in the vicinity of fixed point $1''$ at
the bifurcation threshold (cf. the middle plot in Figure \ref{figure:AtCatBif}) could be a cause of large positive
cumulative skewness in this vicinity.

Taking into account the spirit of the Time Dependent Ginzburg-Landau theory of phase transition
\cite{DS0}, we can assume that potential $U(x;P)$ is a polynomial of the fourth
order (hence, force $f$ is a polynomial of the third order, cf. Section \ref{section:Example}). It is a technical task
to establish required relations between coefficients of the polynomial when system tends to the critical bifurcation.

In this Section our considerations consist of few stages: (i) the analysis of the linear stability, (ii) the
analysis of the non-equilibrium critical dynamics next to a catastrophic bifurcation, (iii) consideration, for
instance, the case of polynomial form of $f$, and (iv) derivation of the useful properties of the first order
autoregressive time series.

\subsection{Analysis of the linear stability}\label{section:dvequ}

In this stage we study, in fact, the linear stability of an equilibrium that is, we consider the relaxation of the
system which was slightly knocked out from its equilibrium \cite{CW}. The equilibrium of the system is defined by the
root (or fixed point) of function $f$ at corresponding value of the driving parameter. This definition does not exclude some
fluctuations supplied by noise $\eta $. For instance, in Figure \ref{figure:BeforeCatBif} the third root is placed
at $x=x_{1''}^*(P_{1''})$. Indeed, our expansion is made in this root and extended, herein, only to the nearest
(linear) surroundings. Moreover, we assume that this root is located next to the catastrophic bifurcation point
(cf. Figures \ref{figure:BeforeCatBif} and \ref{figure:AtCatBif}).

The linear expansion of $f$, e.g. at fixed point $1''$, gives
\begin{eqnarray}
y_{t+1}-y_t&=&f(x_{1''}^*(P_{1''});P_{1''})+\lambda \, y_t+\eta _{\, t}=\lambda \, y_t+\eta _{\, t} \nonumber \\
&\Leftrightarrow & y_{t+1}=(1+\lambda )\, y_t+\eta _{\, t}
\label{rown:fodeqx}
\end{eqnarray}
as, by definition of a root, $f(x_{1''}^*(P_{1''});P_{1''})$ vanishes
Herein, we used the notation: (i) for the
displacement from an equilibrium\footnote{The set of variables $y_t,\; t=0,1,2,\ldots ,$ is also called the first
order autoregressive time series.} or the (unnormalized) order parameter
$y_t=x_t-x_{1''}^{*}(P_{1''})\, ,\; \; t=0,1,2,\ldots ,$
and (ii) for rate $\lambda (P)= \frac{\partial f(x;P)}{\partial x}\mid _{x=x^*(P)}$.

The latter equation in (\ref{rown:fodeqx}), rewritten in the form
\begin{eqnarray}
x_{t+1}=(1+\lambda )\, x_t + b + \eta _{\, t},\; b = -\lambda \, x_{1''}^*
\label{rown:fodeqex}
\end{eqnarray}
makes possible to obtain $\lambda $ and $x_{1''}^*$ from the fit to empirical data, such as shown
in Figure \ref{figure:Szum_sz_541_560}. The related quantities were presented in Figures
\ref{figure:Autocorrel}-\ref{figure:fixedpoint_DJIA}. However, to make such a fit we assume that
coefficients $\lambda $ and $b$ are utmost a slowly varying function of time. Moreover, we assume that they are
piecewise functions of time.

The solution of Equation (\ref{rown:fodeqx}) is
\begin{eqnarray}
y_t=(1+\lambda )^ty_0+(1+\lambda )^{t-1}\sum_{\tau =0}^{t-1}\eta _{\tau }(1+\lambda )^{-\tau }
\approx \exp(\lambda t)\left[y_0+\int_0^t\eta _{\tau }\exp(-\lambda \tau )d\tau \right],
\label{rown:solut}
\end{eqnarray}
where the first equality is valid for $t\ge 1$. The second equality in (\ref{rown:solut}) has an exponential form
because we confined our considerations only to the case $\mid \lambda \mid \ll 1$ and $t\gg 1$ that is, to the nearest
vicinity of the threshold for sufficiently long time. However, in all our calculations we did not include the
flickering phenomenon, considering only the nearest vicinity of an equilibrium point (stable or unstable). We suppose
that by taking into account this phenomenon we could obtain the significant increase of the variance within the
bifurcation region.

>From Equation (\ref{rown:solut}) follows that given equilibrium state is stable
if\footnote{We observed that for our empirical data always inequality $-1<\lambda <0$ is obeyed.} $-2<\lambda <0$
otherwise, it is unstable. Hence, e.g. points $1$ and $1''$ in Figure \ref{figure:BeforeCatBif} define
stable equilibria. In general, stable equilibrium points (or fixed points) are placed on the solid segments
of the folded backward curves, while unstable ones are placed on the dashed-dotted segment (cf. all bottom plots in
Figures \ref{figure:BeforeCatBif}-\ref{figure:AfterCatBif} and plot in Figure \ref{figure:Zbiorczy_CatBif}).

For the stable equilibrium the relaxation (recovery or return) time $\tau (P)=1/\mid \lambda (P)\mid $ is well
defined that is it is a positive quantity. Particularly interesting are stable equilibrium points
$x_1^*$ and $x_{1''}^*$ shown in Figure \ref{figure:AtCatBif} as they are border points of two corresponding
bifurcation regions (attraction basins) therefore called the catastrophic bifurcation points (or tipping points).
They are the most significant states of the system considered in this work.

\subsection{Generic properties of the first order autoregressive time series}\label{section:kp1oats}


It is well known \cite{WAF,BD} that particularly useful quantities, i.e. variance, covariance and autocorrelation
function as well as power spectrum, are related. We calculate them by exploiting an exact solution
given by the first equality in (\ref{rown:solut}).

Firstly, we calculate the covariance,
\begin{eqnarray}
Cov(y_t\, y_{t+h})&=&\left<y_t\,y_{t+h}\right>-\left<y_t\right>\left<y_{t+h}\right>=
(1+\lambda )^{\mid h\mid }\, Var(y_t)= \nonumber \\
Cov(x_t\, x_{t+h})&=&(1+\lambda )^{\mid h\mid }\, Var(x_t) \nonumber \\
&\Leftrightarrow & ACF(h)=\frac{Cov(y_t\,y_{t+h})}{Var(y_{\, t})}=\frac{Cov(x_t\,x_{t+h})}{Var(x_{\, t})}=(1+\lambda )^{\mid h\mid }
\nonumber \\
&\Rightarrow &ACF(h)\approx \exp(\lambda \mid h\mid ), \; h=0,\pm 1,\pm 2,\ldots ,
\label{rown:covar}
\end{eqnarray}
where $\lambda $ is expressed by Equation (\ref{rown:lambda}) and the variance $Var(y_t) $ is given
(after straightforward calculations) by the significant formula
\begin{eqnarray}
Var(y_{\, t})=\left<y_t^2\right>-\left<y_t\right>^2=Var(x_{\, t}) =Var(y_0)(1+\lambda )^{2t}
-\frac{1}{\lambda  (2+\lambda )}\left[1-(1+\lambda )^{2t}\right]\sigma ^2,
\label{rown:variance}
\end{eqnarray}
where $Var(y_t)=Var(x_t)$ and notation $\left<\ldots \right>$ denotes an average over the noise and the initial
condition (within the statistical ensemble of solutions $y_t$). The latter equality in (\ref{rown:covar}) is obeyed
for $\mid \lambda \mid \ll 1$. Therefore, it further simplifies into the form
\begin{eqnarray}
Var(y_{\, t})\approx Var(y_0)(1+2\lambda t)+t\sigma^2\approx Var(y_0)+\sigma ^2 t,
\label{rown:varsimp}
\end{eqnarray}
where the latter approximate equality was obtained for $2t\ll \mid \lambda \mid ^{-1}$. We suppose that by taking
into account the flickering phenomenon we could obtain the significant increase of the variance within the bifurcation
region. However, the analytical calculation of such a variance requires the solution of nonlinear Equation
(\ref{rown:fodeq}) for $f$ given by the polynomial (\ref{rown:fpolxP}), which is still a challenge.

The coefficient $1+\lambda $ (present in (\ref{rown:covar})) is the lag-1 autocorrelation function, which can be find
directly from empirical data (cf. Fig. \ref{figure:Autocorrel}). Apparently, it does not depend on the variance.


Now, as we have an exact simple formula for the stationary autocorrelation function, $ACF(h)$, we can obtain power
spectrum $PS(\omega )$ in an exact analytical form. Namely, from the Wiener-Khinchine theorem \cite{KTH} we have
\begin{eqnarray}
PS(\omega )&=&\lim_{H\rightarrow \infty }\sum _{h=-H}^{H}ACF(h)\, \exp(-i \, h\, \omega)
=\lim_{H\rightarrow \infty }\sum _{h=-H}^{H}(1+\lambda )^{\mid h\mid }\, \exp(-i \, h\, \omega) \nonumber \\
&=&-\lim_{H\rightarrow \infty }\frac{\lambda (2+\lambda ) - 2(1+\lambda )^H(1+\lambda -\cos (\omega ))}
{2+\lambda (2+\lambda )-2(1+\lambda )\cos (\omega )}.
\label{rown:PSomega}
\end{eqnarray}
Obviously, for $-1<\lambda <0$ (which is our case) the term containing  factor $(1+\lambda )^H$ in
the numerator vanishes. Hence, we obtain a useful formula
\begin{eqnarray}
PS(\omega )=-\frac{\lambda (2+\lambda )}
{2+\lambda (2+\lambda )-2(1+\lambda )\cos (\omega )}.
\label{rown:PSomeg}
\end{eqnarray}

For vanishing $p_j$ the variance and covariance diverge according to power-law while autocorrelation
function tends then to $1$. Besides, all odd moments of variable $y_t$ vanish. Hence and from Equation
(\ref{rown:variance}) we find that in the frame of linear theory the skewness also vanishes.

Furthermore, it can be easily verified (again by using solution (\ref{rown:solut})) that the excess kurtosis
vanishes if variables $y_0$ and $\eta _{\, t}$ are drawn from some Gaussian distributions.
That is, in the frame of the linear theory (i.e. in the vicinity of the threshold) the distribution of variable
$y_t$ is the Gaussian, which has the variance given by Expression (\ref{rown:variance}) and centered around the mean
value $\left<y_t\right>=y_0(1+\lambda )^t$.

The above given considerations clearly show that indeed {\color{red}{\bf $ACF(1)$ or $AR(1)$ coefficient is a basic
quantity to study the catastrophic bifurcation, possible for direct determination from empirical data}}.

\subsection{Nonlinear time series analysis}

As linear term in the linear time series can be relatively easy masked by the noise term, we expanded $f$ in Expression
(\ref{rown:fodeq}) one step further up to the quadratic term (see also Equation (\ref{rown:fodeqx})). Hence, we deal
with the nonlinear time series. For time series of this type several methods were developed to determine the level
of the noise and to make the noise reduced.
\cite{KS}.

\subsection{Catastrophic dynamics next to catastrophic bifurcation transition}\label{section:Dcbp}

In the second stage, we more subtle consider the non-equilibrium dynamics in the vicinity of particular equilibrium
(fixed) points \cite{CW}. That is, the dynamics next to catastrophic bifurcation point is considered herein.

Let us make the Taylor expansion at point $x_j^*(P_j)$:
\begin{eqnarray}
f(x;P)=f(x_j^*(P_j)+y_j;P_j+p_j)=\beta \, p_j+\alpha \, y_j^2+\ldots ,
\label{rown:secstag}
\end{eqnarray}
herein $j=1''$ means the only catastrophic bifurcation point, i.e., the catastrophic point placed on the catastrophic
bifurcation curve (cf. Figure \ref{figure:Zbiorczy_CatBif}). For sufficiently small $\mid p_j\mid $ and
$\mid y_j\mid $ only first two terms in (\ref{rown:secstag}) play an essential role. We used the following
notation: for the displacement from catastrophic bifurcation point $y_j=x-x_j^*(P_j)$, negative
parameter's deviation $p_j=P-P_j$, negative coefficient
$\beta =\frac{\partial f(x;P)}{\partial P}\mid _{x=x_j^*(P_j),\, P=P_j}$ and for negative coefficient
$\alpha =\frac{1}{2}\frac{\partial ^2 f(x;P)}{\partial x^2}\mid _{x=x_j^*(P_j),\, P=P_j}$. Note that the sign of each
quantity can be easily recognized by comparing curves $f(x)$ vs. $x$ plotted in Figure \ref{figure:Zbiorczy_CatBif}.
We used also helpful properties: $f(x_j^*(P_j);P_j)=0$ and $\lambda (P_j)=0$ (cf. Figure
\ref{figure:AtCatBif} and the catastrophic curve in Figure \ref{figure:Zbiorczy_CatBif}), where the former defines
a fixed point while the latter the local maximum at the catastrophic bifurcation point. Due to these properties
we get below the searched scaling relations.

By substituting in Equation (\ref{rown:secstag}) variable $x=x^*(P)$, we get the scaling relation (as
the left-hand side of (\ref{rown:secstag}) vanishes then)
\begin{eqnarray}
y_j^*=\pm \left[-\left(\frac{\beta \, p_j}{\alpha }\right)\right]^{\mu },
\label{rown:propert}
\end{eqnarray}
where displacement $y_j^*=x^*(P)-x_j^*(P_j)$ is, in this case, a positive quantity (therefore, sign '+'
standig outside the square brackets should be taken into account) while exponent $\mu =1/2$ plays
the role of the \emph{critical index} (or \emph{citical exponent}).

{\color{red} {\bf The relation (\ref{rown:propert}) shows that the displacement vanishes according to the
power-law, when system reaches a catastrophic bifurcation point. This is a signature of phase transition; in our
case the discontinuous one that is, the first-order phase transition.}}

Moreover, from Equations (\ref{rown:secstag}), (\ref{rown:propert}) and definition of $\lambda $ we also
obtain\footnote{Formula (\ref{rown:lambda}) was derived from (\ref{rown:secstag}) by making the derivative over
$x$ variable of its both sides at $x=x^*(P)$ .} a {\color{red}{\bf significant scaling relation
\begin{eqnarray}
\lambda =2\alpha \, y_j^*=\pm 2\alpha \left[-\left(\frac{\beta \, p_j}{\alpha }\right)\right]^{\mu }.
\label{rown:lambda}
\end{eqnarray}
}}
The plus sign in the second equality in above equation (outside square brackets) defines the stable equilibrium
(as then $\lambda $ is negative), which is our case ($j=1''$). The relation (\ref{rown:lambda}) shows
that rate $\lambda $ also vanishes according to power-law when system reaches a catastrophic bifurcation point,
which is a succeeding signature of phase transition. {\color{red} {\bf Nevertheless, there is a challenge how
$\lambda $ scales with time when it is closed to the transition instant.}}

\subsection{Simplest example: approximation of force $f$ by the third-order polynomial}\label{section:Example}

Let us assume that potential $U$, present in Eq. (\ref{rown:potent}), is defined by the fourth-order polynomial
\begin{eqnarray}
U(x;P)=A_0x^4+A_1x^3+A_2x^2+A_3x+A_4,
\label{rown:UpolxP}
\end{eqnarray}
where $A_j,\, j=0,1, \ldots ,4$, are its real coefficients somehow related to (combined) parameter $P$; this relation
is considered further in this section. According to Eq. (\ref{rown:potent}), force $f$ is a polynomial
of one order of magnitude lower
\begin{eqnarray}
f(x;P)=a_0x^3+a_1x^2+a_2x+a_3,
\label{rown:fpolxP}
\end{eqnarray}
where coefficients $a_{4-j}=-j\, A_{4-j},\, j=1,\ldots ,4$ and herein we assume $a_0<0$ as it is indicated by empirical
data. Below, we consider three characteristic cases: (a) the catastrophic bifurcation transition, (b) the transition
before the catastrophic bifurcation transition, and (c) the transition after it. For instance, we consider situations
(common for all three cases), where smallest root of polynomial (\ref{rown:fpolxP}) is negative. That is, we consider
situations presented in Figures \ref{figure:fixedpoint_WIG} (and placed on the left-hand side of the threshold marked
by the straight vertical dashed line) and also \ref{figure:fixedpoint_DAX}, where the latter is our working example.

The generic aim of this section is to express coefficients of polynomial (\ref{rown:fpolxP}) in terms of roots of this
polynomial.

\subsubsection{Case of the catastrophic bifurcation transition}\label{section:casecbt}

Let us focus on the case (a) (presented in Figure \ref{figure:AtCatBif}) concerning the catastrophic
bifurcation transition. This means that coefficients $a_j,\, j=0, \ldots ,3$, make such a parametrisation, which gives
curve $f(x;P)$ vs. $x$ indeed in the form shown in Figure \ref{figure:AtCatBif} (we relate these coefficients to
parameter $P$ at the end of this section).

Now, we can define the goals of this case. There are as follows:
\begin{itemize}
\item[(i)] the derivation of the roots $x_1^*$ and $ x_{1''}^*$ of polynomial (\ref{rown:fpolxP}) and hence
calculation, for instance, the catastrophic bifurcation jump, $\Delta x_{1'',1}^* = x_1^*-x_{1''}^*$, as a function
of polynomial coefficients and
\item[(ii)] the solution of the inverse problem that is, the derivation of relative parameters
$a_1/a_0,\, a_2/a_0$ and $a_3/a_0$ by means of roots $x_1^*$ and $ x_{1''}^*$, which can be found from empirical
data (cf. Figures \ref{figure:fixedpoint_WIG}-\ref{figure:fixedpoint_DJIA} in Section \ref{section:AR}).
\end{itemize}

Notably, the catastrophic bifurcation transition $1''\Rightarrow 1$ (cf. the upper plot in Figure
\ref{figure:AtCatBif}) begins at point $1''$, which is not only the largest twofold root of polynomial $f$ but also
it is its local maximum that is, it is an inflection point on curve $U$ vs. $x$ (cf. both plots in Figure
\ref{figure:AtCatBif}). Hence,
\begin{eqnarray}
\frac{\partial f(x;P)}{\partial x}\mid _{x=x_0,\, x_{1''}^*}\,= 0\Leftrightarrow
3\,x_{0,1''}^{*\, 2}+2\, \frac{a_1}{a_0}\, x_{0,1''}^*+\frac{a_2}{a_0}=0,
\label{rown:minimum}
\end{eqnarray}
where $0(=x_0)$ is the first inflection point on curve $U$ vs. $x$ (see again the lower plot in Figure
\ref{figure:AtCatBif}) and it is the local minimum of curve $f$ vs. $x$ (see again the upper plot in Figure
\ref{figure:AtCatBif}).

>From Equation (\ref{rown:minimum}) we obtain
\begin{eqnarray}
x_{0,1''}^*=x_{ip}\mp \frac{1}{3}\, \sqrt{D},\, \, D\stackrel{\rm def.}{=}\left(\frac{a_1}{a_0}\right)^2-
3\, \frac{a_2}{a_0},
\label{rown:x10}
\end{eqnarray}
where sign $-$ concerns minimum $x_0^*$, sign $+$ concerns root $x_{1''}^*$, while
\begin{eqnarray}
x_{ip}=-\frac{1}{3}\frac{a_1}{a_0},
\label{rown:x101}
\end{eqnarray}
and we assumed $D>0$ as two real roots of Eq. (\ref{rown:minimum}) should exist.

We can easily derive (from vanishing of the second derivative $f$ over $x$) that $x_{ip}$, given by
Eq. (\ref{rown:x101}), is the inflection point (cf. the upper plot in Figure \ref{figure:AtCatBif}),
\begin{eqnarray}
\frac{\partial ^2 f(x;P)}{\partial x^2}\mid _{x=x_{ip}}\,=0 \Rightarrow  x_{ip}=-\frac{1}{3}\frac{a_1}{a_0}.
\label{rown:xip}
\end{eqnarray}
As it follows from Equation (\ref{rown:x10}), both extremes $x_0^*$ and $x_{1''}^*$ are located symmetrically on
either sides of the inflection point $x_{ip}$ that is, minimum $x_0^*$ is placed on the left-hand side and $x_{1''}^*$
on the right-hand side.

By the way, from Equation (\ref{rown:minimum}) and definition of coefficient $\alpha $ (given in Section
\ref{section:Dcbp}) follow that
\begin{eqnarray}
\alpha =\frac{1}{2}\left(6x_{1''}+2\, \frac{a_1}{a_0}\right)=-\sqrt{D}.
\label{rown:alphaxt}
\end{eqnarray}

We can postulate the complementary properties of coefficients and roots suggested by the empirical data points
and corresponding backward folded hypothetical curve shown, for instance, in Figure \ref{figure:fixedpoint_DAX},
that is
\begin{itemize}
\item[(i)] we assume inequalities for the roots $x_1^*<0$ and $x_{1''}^*>0$,
\item[(ii)] the root $x_{1''}^*$ is a twofold one of polynomial (\ref{rown:fpolxP}), where local maximum of this
polynomial is located.
\end{itemize}

>From item (ii) and by looking for such values of variable $x$ and parameter $P$ for which $f(x;P)$ vanishes
in Equation (\ref{rown:fpolxP}), we easily obtain (with help of Eq. (\ref{rown:x10})) the third root
\begin{eqnarray}
x_1^*=x_{ip}-\frac{2}{3}\, \sqrt{D}.
\label{rown:root11}
\end{eqnarray}

>From Equations (\ref{rown:x10}), (\ref{rown:alphaxt}) and (\ref{rown:root11}) we get the catastrophic bifurcation
jump
\begin{eqnarray}
\Delta x_{1'',1}^*=-\sqrt{D}=\alpha ,
\label{rown:jump}
\end{eqnarray}
being function of only two relative parameters $a_1/a_0$ and $a_2/a_0$. This equation gives significant
interpretation of parameter $\alpha $ making possible to easily detect it from empirical plots presented in Figures
\ref{figure:fixedpoint_WIG} - \ref{figure:fixedpoint_DJIA}.

>From Egs. (\ref{rown:x10}), (\ref{rown:xip}) and (\ref{rown:root11}) we easily find the solution
of the inverse problem in the form,
\begin{eqnarray}
\frac{a_1}{a_0}=-(2x_{1''}^*+x_1^*)\leq 0, \nonumber \\
\frac{a_2}{a_0}=x_{1''}^* \left(x_{1''}^*+2x_1^*\right)\geq 0,
\label{rown:inverse}
\end{eqnarray}
together with the constraint for the relative free parameter
\begin{eqnarray}
\frac{a_3}{a_0}=-x_1^*\, (x_{1''}^*)^2\geq 0;
\label{rown:twoconstr}
\end{eqnarray}
the latter relation makes the above procedure self-consistent. The above given three inequalities were suggested
by empirical data. Following for the data we can assume helpful inequalities: $a_1>0,\, a_2\leq 0,\, a_3<0$.

Apparently, having tipping points $x_1^*$ and $x_{1''}^*$, taken from empirical data, we can obtain searched relative
parameters $a_1/a_0,\, a_2/a_0$ and $a_3/a_0$. Here, we have $x_1^*=-101.17$ and $x_{1''}^*=278.92$ which gives
$a_1/a_0=-456.67,\, a_2/a_0=21359.70$ and $a_3/a_0=7.87066\times 10^6$.

Notably, the combined parameter $P$ consists of only two independent components, e.g., $a_1/a_0$ and $a_2/a_0$.
The first component could be, for example, the volume averaged over month
while the second component could be the basic rate of the central bank (which is, obviously, a piecewise function
of time counted in month scale). We can say that two components of combined parameter $P$ are already
sufficient to perform some simulations at catastrophic transition.

Furthermore, the schematic bottom plots shown
in Figures \ref{figure:BeforeCatBif}-\ref{figure:Zbiorczy_CatBif} tacitly assumed that combined parameter $P$
monotonically depends on time (counted in month time scale) at least in the vicinity of BCT.


\subsubsection{Case before the catastrophic bifurcation transition}

The case considered here (i.e., the case represented by Figure \ref{figure:BeforeCatBif}) is a generalization of the
one discussed in Section \ref{section:casecbt}. That is, we consider variable $x$ placed inside
the bifurcation region, where three different real roots exist (cf. backward folded curves shown in Figures
\ref{figure:fixedpoint_WIG} - \ref{figure:fixedpoint_DJIA} and schematically shown in Figure
\ref{figure:Zbiorczy_CatBif}).

The goal of this section is analogous to that considered in Section \ref{section:casecbt}, i.e., to find coefficients
of polynomial (\ref{rown:fpolxP}) by using its roots found from empirical data (shown in the above mentioned figures).
By assuming that polynomial (\ref{rown:fpolxP}) has three real different roots we obtain, by comparing
Eq. (\ref{rown:fpolxP}) with its multiplicative form
$f(x;P)=\left(x-x_1^*\right)\left(x-x_{1'}^*\right)\left(x-x_{1''}^*\right)$, the searched relations for coefficients
of the polynomial
\begin{eqnarray}
\frac{a_1}{a_0}=-(x_{1''}^*+x_{1'}^*+x_1^*), \nonumber \\
\frac{a_2}{a_0}=x_{1'}^*x_{1''}^*+x_1^*x_{1''}^*+x_1^*x_{1'}^*, \nonumber \\
\frac{a_3}{a_0}=-x_1^*\, x_{1'}^*\, x_{1''}^*.
\label{rown:3rootcoeff}
\end{eqnarray}
Apparently, above equations are generalization of the corresponding Eqs. (\ref{rown:inverse}) and
(\ref{rown:twoconstr}), as we obtain the latter equations by setting in Eqs. (\ref{rown:3rootcoeff})
$x_{1'}^*=x_{1''}^*$.

Notably, Figure \ref{figure:BeforeCatBif} was constructed by using coefficients obtained from
Eqs. (\ref{rown:3rootcoeff}) by introducing into their right-hand sides the following empirical values of roots:
$x_1^* = 278.92,\, x_{1'}^* = -488.308,\, x_{1''}^* = -626.473$, taken, for instance, from the backward folded curve
shown in Figure \ref{figure:fixedpoint_DAX}. According to (\ref{rown:3rootcoeff}), this leads to $a_1/a_0= 835.861,\,
a_2/a_0=-5022.94,\, a_3/a_0= -8.53249\times 10^8$.

\subsubsection{Case after the catastrophic bifurcation transition}\label{section:casec}

For this case (represented by Figure \ref{figure:AfterCatBif}) we have insufficient empirical data
for unique solution as only single real root $x_1^*$ we can find from them (roots $x_{1'}^*$ and $x_{1''}^*$ are
the complex conjugated). Hence, we deal only with a single relation between coefficients of polynomial
\begin{eqnarray}
f(x_1^*;P)=-\left(x_1^*\right)^3-\frac{a_1}{a_0}\left(x_1^*\right)^2-x_1^*\frac{a_2}{a_0}-\frac{a_3}{a_0}=0,
\label{rown:a1a2a3}
\end{eqnarray}
which makes the ratio of parameters (e.g., $a_1/a_0$) dependent on other two ratios (herein, on $a_2/a_0$ and
$a_3/a_0$). For instance, in Figure \ref{figure:AfterCatBif} we show plots for root $x_1^*= -75.3875$ as well as
ratios of parameters $a_3/a_0= 6.1682\times 10^6,\, a_2/a_0= 41709.50$ and $a_1/a_0= -456.67$.

With analogous situation we deal if $x$ is placed before CBT and outside of the bifurcation region. Then, we
deal again with a single real root, here $x_{1''}^*=421.009$, while roots $x_1$ and $x_{1'}^*$ are complex conjugated.
In Figure \ref{figure:BeforeCatBif} we show plots for ratios of parameters
$a_3/a_0=-4.00948\times 10^8,\, a_2/a_0=278390$ and $a_1/a_0=1179.81$.


Finally one can state, that the approach developed in Sections \ref{section:casecbt} - \ref{section:casec} is
sufficiently flexible that is, can be reformulated for the situation where the smallest root of
Eq. (\ref{rown:fpolxP}), $x_1^*$, become positive. Then, obviously, signs and values of coefficients
$a_0,\, a_1,\, a_2,\, a_3$ should correspondingly changed.



\section{Scaling relations}\label{section:scalrels}

Following Equations (26)-(28) in \cite{RK} (and reference [28] therein), we can propose an approximate expression
of the noise (increments) distribution in the form
\begin{eqnarray}
P(\Delta x,\Delta t)={\cal B}\Delta t^{-\eta /2}{\cal F}(\xi )
\label{rown:propgatorsd}
\end{eqnarray}
for large values of the scaling variable $\xi =\mid \Delta x\mid /\Delta t^{\eta /2}$, where the scaling function
\begin{eqnarray}
{\cal F}(\xi )=\xi ^{\bar \nu }\exp\left(-\frac{1}{4{\bar {\cal D}}}\, \xi ^{\nu }\right).
\label{rown:scalfunctF}
\end{eqnarray}
The coefficients are defined as
\begin{eqnarray}
{\cal B}=\frac{1}{2}\frac{1}{\sqrt{\pi (2-\eta )}}\left(\frac{\eta }{2\sqrt{{\cal D}}}\right)^{\bar \nu },\; \;
{\bar {\cal D}}=\frac{1}{4}\left(\frac{2\sqrt{{\cal D}}}{\eta }\right)^{\nu }\left(\frac{\eta }{2-\eta }\right)
\label{rown:coeffs}
\end{eqnarray}
and the auxiliary exponents
\begin{eqnarray}
\nu = \frac{2}{2-\eta }, \; \; {\bar \nu }=\frac{\eta-1}{\eta -2},
\label{rown:coeffss}
\end{eqnarray}
all for basic scaling exponent $\eta <2$. The normalization of distribution (\ref{rown:propgatorsd}) can be easily
verified.

Note that only for $\eta =1$ distribution (\ref{rown:propgatorsd}) reduces into the Gaussian form. Besides, the
fractional Brownian motion (fBm) is never kept by distribution (\ref{rown:propgatorsd}).

The distribution (\ref{rown:propgatorsd}) has properties which are helpful for our considerations in Section
\ref{section:Pons}. The first one, concerning the border distribution,
\begin{eqnarray}
P(\Delta x)=\int_0^{\infty }P(\Delta x,\Delta t)\, d(\Delta t)=\frac{const}{\mid \Delta x\mid^{1-2/\eta }},
\label{rown:border}
\end{eqnarray}
where
\begin{eqnarray}
const = \frac{2{\cal B}}{\eta }\int _0^{\infty } \frac{1}{\xi ^{2/\eta }}F(\xi )\, d\xi .
\label{rown:constb}
\end{eqnarray}
The direct, left-sided empirical representation of this distribution is shown in Figure
\ref{figure:Histogram_noise_WIG} for the power-law (here Zipf-law) decay driven by L\'evy exponent equals
$-2/\eta =0.99$. This leads to negative exponent $\eta =-2.02$.

Another, very helpful property concerns the second moment
\begin{eqnarray}
\left<[\Delta x(\Delta t)]^2\right>={\cal B}\int_{-\infty }^{\infty }\Delta t^{-\eta /2}F(\xi )\Delta x^2 d(\Delta x)=
const\, \Delta t^{\eta },
\label{rown:moment}
\end{eqnarray}
where
\begin{eqnarray}
const=2{\cal B}\int_{0}^{\infty }\xi ^2F(\xi )d\xi .
\label{rown:constx}
\end{eqnarray}
Here, the second moment given above is the monotonically decreasing function of $\Delta t$, which is a consequence of
a scaling form of distribution (\ref{rown:propgatorsd}). Relation (\ref{rown:constx}) is a basis for our derivation
of the autocorrelation function and hence power-spectra in subsection below.

\subsection{Autocorrelation and power spectrum}\label{section:AcorrPsectr}

Provided that one observes the signal at times $t_1$ and $t_2$ such that the observation time $t=t_2-t_1$ is short
compared with the time elapsed since the process began (i.e., $t\ll t_1$), one can evaluate the unnormalized
autocorrelation function (or autocovariance in econometric terminology) as a stationary one. For this evaluation we
use Equation (\ref{rown:moment}) in a more generic notation
\begin{eqnarray}
\left<[\Delta x(t)]^2\right>=\left<[x_{t_2}-x_{t_1}]^2\right>&\propto &|t_2-t_1|^{\eta } \Rightarrow \nonumber \\
\left<x_{t_1}\, x_{t_1+t}\right>&\propto &|t_1|^{\eta }+|t|^{\eta }-|t_1+t|^{\eta } \Rightarrow \nonumber \\
\left<x_{t_1}\, x_{t_1+t}\right>&\propto &-(\eta -1)|t|^{\eta }, \nonumber \\
\left<\Delta x(t_1)\Delta x (t_1+t)\right>&\propto & \eta (\eta -1)\frac{1}{\mid t\mid ^{2-\eta }},
\label{rown:genermoment}
\end{eqnarray}
where $t\stackrel{\rm def.}{=}t_2-t_1$ and we assumed $\left<x_0^2\right>=0$. Note that above given results are valid for arbitrary
(real) value of $\eta $ (and not only $\eta <2$). Furthermore, only for $\eta >1$ the signal is negatively correlated
(cf. the third equation in (\ref{rown:genermoment})).

However, as in our case $\eta = -2.02$ (cf. considerations in Sections \ref{section:noise}, \ref{section:Pons} and
\ref{section:AcorrPsectr}) the question arises whether the power spectrum, $S_x(f)$, of detrended
(time-dependent) signal $x_t$ (related to above given unnormalized signal autocorrelation function) properly
reproduces the empirical frequency ($f$) dependence of the corresponding periodogram
(shown in Figure \ref{figure:Power_spectrum_sygnal_400_859_polowa}) for small $f$ (i.e., for $1\le j\le 14$).

By making the Laplace transformation, ${\cal L}(s)$, of both sides of the third equation in
(\ref{rown:genermoment}) and by using the definition of power spectrum, we obtain for detrended signal
\begin{eqnarray}
S_x(f)\stackrel{\rm def.}{=}2Re\, {\cal L}(s)|_{s=2\pi f i }\; \propto \frac{1}{|f|^{\eta +1}}
\label{rown:ssff}
\end{eqnarray}
where $i$ is an imaginary unit. Notably, this result is well confirmed by plot shown in Figure
\ref{figure:Power_spectrum_sygnal_400_859_polowa}. That is, the slope of the periodogram for small frequencies is
directed there by increasing power law with positive slope equals $-(1+\eta )=1.02$.

By using the fourth equation in (\ref{rown:genermoment}) and performing an analogous calculation as for $S_x(f)$,
we obtain for related power spectrum of the noise
\begin{eqnarray}
S_{\Delta x}(f)\; \propto \frac{1}{|f|^{\eta -1}},
\label{rown:dssff}
\end{eqnarray}
which is confirmed by the plot in Figure \ref{figure:Periodogram_noise_WIG_od401}. That is, the slope of the
periodogram for small frequencies is directed there by increasing power law with positive slope almost equals
$-\eta +1=3.02$.

Below, we outline the derivation of Expression (\ref{rown:ssff}) and hence (\ref{rown:dssff}).

Let us assume that function $C(t)\propto (t+t_0)^{\eta }$, where $t_0$ is a small (insignificant from a physical point
of view) positive quantity calculated below. We can derive its Laplace transform in the form \cite{GHW,AE}
\begin{eqnarray}
{\cal L}(s)=\int_0^{\infty }C(t)\exp(-st)dt\propto \frac{1}{s^{1+\eta }}\int_{y_0}^{\infty }y^{\eta }\exp(-y)dy=
\frac{\exp(y_0)\Gamma _{up}(1+\eta ,y_0)}{s^{1+\eta }},
\label{rown:SfCt}
\end{eqnarray}
where $|y_0|=|st_0|\ll 1$ and $\Gamma _{up}(1+\eta , y_o)$ is an upper incomplete gamma function of variable $1+\eta $.
Indeed, Expression (\ref{rown:ssff}) was directly obtained from (\ref{rown:SfCt}) by simply putting there
$s=2\pi f i $ and next taking a real part of (\ref{rown:SfCt}).

The derivation of Expression (\ref{rown:dssff}) is analogous to the derivation of Expression (\ref{rown:ssff}) if
exponent $\eta $ in (\ref{rown:SfCt}) is formally replaced by exponent $\eta -2$.

\bibliographystyle{unsrt}
\bibliography{EmpiricalSymptoms}

\end{document}